\documentclass[twocolumn,showpacs,preprintnumbers,amsmath,amssymb]{revtex4}
\usepackage{epsfig}
\usepackage{dcolumn}
\usepackage{bm}

\newcommand{\remove}[1]{}

\def\be{\begin{equation}}
\def\ee{\end{equation}}

\newcommand{\beq}{\begin{equation}}
\newcommand{\eeq}{\end{equation}}
\newcommand{\beqa}{\begin{eqnarray}}
\newcommand{\eeqa}{\end{eqnarray}}

\newcommand{\vv}{{\bf v}}

\newcommand{\vk}{{\bf k}}

\newcommand{\bea}{\begin{array}}
\newcommand{\ea}{\end{array}}

\begin{document}

\title{Self-acceleration in scalar-bimetric theories}

\author{Philippe Brax}
\affiliation{Institut de Physique Th\'eorique,\\
CEA, IPhT, F-91191 Gif-sur-Yvette, C\'edex, France\\
CNRS, URA 2306, F-91191 Gif-sur-Yvette, C\'edex, France}
\author{Patrick Valageas}
\affiliation{Institut de Physique Th\'eorique,\\
CEA, IPhT, F-91191 Gif-sur-Yvette, C\'edex, France\\
CNRS, URA 2306, F-91191 Gif-sur-Yvette, C\'edex, France}
\vspace{.2 cm}

\date{\today}
\vspace{.2 cm}

\begin{abstract}

We describe scalar-bimetric theories where the dynamics of the Universe are governed by
two separate metrics, each with an Einstein-Hilbert term. In this setting, the baryonic
and dark matter components of the Universe couple to metrics which are constructed
as functions of these two gravitational metrics. More precisely, the two metrics coupled
to matter are obtained by a linear combination of their vierbeins, with scalar-dependent
coefficients.
The scalar field, contrary to dark-energy models, does not have a potential whose role
is to mimic a late-time cosmological constant. The late-time acceleration of the expansion
of the Universe can be easily obtained at the background level in these models by appropriately
choosing the coupling functions appearing in the decomposition of the vierbeins for the baryonic
and dark matter metrics. We explicitly show how the concordance model can be retrieved
with negligible scalar kinetic energy.
This requires the scalar coupling functions to show variations of order unity during the
accelerated expansion era.
This leads in turn to deviations of order unity for the effective Newton constants and
a fifth force that is of the same order as Newtonian gravity, with peculiar features.
The baryonic and dark matter self-gravities are amplified although  the gravitational force between
baryons and dark matter is reduced and even becomes repulsive at low redshift.
This slows down the growth of baryonic density perturbations on cosmological scales,
while dark matter perturbations are enhanced.
These scalar-bimetric theories have a perturbative cutoff scale of the
order of one astronomical unit (a.u.), which prevents a precise comparison with Solar System data. On the other hand, we can deduce
strong requirements on putative Ultra Violet (UV) completions by analyzing the stringent constraints in the Solar System.
Hence, in our local environment, the upper bound on the time evolution of Newton's constant requires an efficient screening
mechanism that both damps the fifth force on small scales and decouples the local value
of Newton constant from its cosmological value.
This cannot be achieved by a quasistatic chameleon mechanism, and requires going beyond
the quasistatic regime and probably using derivative screenings, such as Kmouflage
or Vainshtein screening, on small scales.

\end{abstract}

 \vskip2pc

\maketitle

\section{Introduction}
\label{sec:Introduction}

A very common way of reproducing the late-time acceleration of the expansion of the Universe \cite{Perlmutter:1998hx,Riess:1998cb} is to add a scalar-field energy density, which would mimic
a cosmological constant at small redshifts \cite{Copeland:2006wr}. Recently, it has been proposed that the acceleration could be an illusion due to the different metrics coupled
to either the baryons or dark matter \cite{Berezhiani:2016dne}. This was achieved by considering that baryons couple to a metric that can be constructed from both the metric felt by dark matter
and the velocity field of the dark matter particles. In the same vein, it has been known for some time that  conformally coupled models with a single metric
and screening properties, thus evading the local tests of gravity, cannot generate the late-time acceleration of the Universe \cite{Wang:2012kj}. In this paper, we generalize the latter approach by
introducing two gravitational metrics, with an Einstein-Hilbert term each, and we consider that the baryons and dark matter couple to different dynamical metrics. These metrics are obtained
by taking linear combinations of the two gravitational vierbeins, with each of the coefficients dependent on a scalar field. Contrary to dark-energy models (even coupled), we do not require
that the scalar field should play any explicit role in generating an effective cosmological constant at late time. Quite the contrary, the scalar is only a free and massless scalar, with
positive pressure. The role of the scalar is to provide a time-dependent mapping
and transform the deceleration of the two gravitational metrics into an acceleration
for the baryonic metric.

Our approach is inspired by the construction of doubly coupled bigravity models \cite{deRham:2010kj,Hassan:2011zd} where
the late-time acceleration is due to an explicit cosmological constant, albeit related to the mass
of the massive graviton \cite{Volkov:2011an,vonStrauss:2011mq}, and  matter couples to a  combination (with constant coefficients) of
the two dynamical metrics \cite{deRham:2014naa}. Here, we remove the potential term of massive gravity and introduce
scalar-dependent coupling functions, as our goal is to build self-accelerated solutions.
As expected, we find that this leads to major difficulties, as a self-acceleration implies
effects of order unity on cosmological scales. This generically gives rise to effective Newton
constants that evolve on Hubble timescales and a fifth force of the same order as Newtonian
gravity.

The models that we construct have no ghost in Minkowski space, where they correspond simply to two copies
of General Relativity. When matter is introduced, only the diagonal diffeomorphism invariance is preserved.
The order parameter of the symmetry  breaking, from two copies of diffeomorphism invariance to the diagonal one,
is the Hubble parameter induced by the matter sectors. When performing a St\"uckelberg analysis of the breaking pattern and
introducing the corresponding scalar  Goldstone mode, we find that the absence of ghosts
associated with an Ostrogradsky instability is guaranteed below the cutoff scale
$\Lambda_{\rm cut}= (H^3 M_{\rm Pl})^{1/4}$. This energy scale is smaller than the strong coupling
scale $\Lambda_3 = (H^2 M_{\rm Pl})^{1/3}$.
Thus, $\Lambda^{-1}_{\rm cut} \sim 1 $ a.u whereas $\Lambda_3^{-1} \sim 1000$ km
in the late-time Universe, but $\Lambda^{-1}_{\rm cut}$ remains much below cosmological
scales.
Down to the scale $\Lambda^{-1}_{\rm cut}$ and around compact objects in the weak gravitational regime, the scalar
Goldstone mode is decoupled from matter and no Vainshtein mechanism is at play. For scales below $\Lambda_{\rm cut}^{-1}$, it is very likely that the models should be altered. Notice, too, the analogy with
doubly coupled bigravity, where the order parameter of diffeomorphism breaking is given by the graviton mass $m$ as appearing in the potential term of massive gravity and
the strong coupling is given by $\Lambda_3= (m^2 M_{\rm Pl})^{1/3}$. In this case, too, a ghost is known to be present at higher energy and the models should also be completed.

Our analysis of the presence of ghosts and the existence of a cutoff scale has been performed perturbatively around Friedmann-Lemaitre-Robertson-Walker backgrounds. We have found that, at energies higher
than the perturbative cutoff scale, a ghost is likely to exist due to the mixing between the tensor modes and higher derivatives in the St\"uckelberg field. It is quite likely that
a nonperturbative analysis along the lines of \cite{Hassan:2011zd,Heisenberg:2014rka,Hassan:2018mbl}  would unravel the existence of nonperturbative effects which would lower the cutoff scale and reduce the domain of validity of the scalar-bimetric models.
This is left for future work. Here, we only focus on the perturbative cutoff scale and treat the corresponding range as the one where the scalar-bimetric models are well defined.

We mostly focus on the late-time Universe, in the matter and dark-energy eras.
However, doubly coupled bigravity theories suffer from instabilities in the radiation era
\cite{Comelli:2015pua,Gumrukcuoglu:2015nua,Brax:2016ssf} for tensor and vector
modes. We briefly rederive these behaviors for our models.
Tensor modes have a tachyonic regime that implies an anomalous growth in the early
Universe. This has some effect on the Cosmic Microwave Background B-modes
which may be amplified \cite{Brax:2017pzt} in models where there is a nonlinear coupling
between the metrics, such as in bigravity theories. Similarly, the vector modes that are
decoupled from  matter suffer from a gradient instability which could pose serious
problems for the viability of the models. However, in our case, these instabilities
only affect ``hidden'' modes that are not seen by the matter metrics (at the linear
level).

In this paper, we do not perform detailed comparisons with cosmological and astrophysical
data, as our goal is only to distinguish which families of solutions offer a realistic
framework, which may deserve further investigations.
Indeed, imposing a $\Lambda$-CDM expansion history for the cosmological background
(which ensures consistency with cosmological data at the background level),
we find that the tight constraint on the velocity of gravitational waves
\cite{Abbott2017a}
already provides significant constraints on the model.
Moreover, we find that a nonlinear screening mechanism \cite{Brax:2013ida} must come
into play on small scales, to ensure convergence to General Relativity in the Solar
System. This follows from  the upper bound on the local time dependence of Newton's constant
\cite{Williams:2004qba}, which would have to be obeyed by any UV completion on scales below the cutoff of order one a.u.
This must go beyond the quasistatic approximation and probably rely
on derivatives of the scalar field (as in Kmouflage
\cite{Babichev:2009ee,Brax:2012jr, Brax:2014a} or Vainshtein mechanisms
\cite{Vainshtein:1972sx}),
while quasistatic chameleon screening \cite{Khoury:2003aq,Khoury:2003rn} cannot occur.
We leave the analysis of this regime for future work.

This article is organized as follows.
We first define the bimetric model in section~\ref{sec:definition} and next provide the
equations of motion in section~\ref{sec:Eqs-of-motion}.
We describe the cosmological background in section~\ref{sec:background}.
We show how to construct solutions that mimic a $\Lambda$-CDM expansion and
discuss both the simplified cases where all metrics have the same conformal time
and the cases where they have different conformal times.
We turn to linear perturbations in section~\ref{sec:linear}, for both baryonic and matter
density fluctuations. We then describe in section \ref{sec:freedom-linear} how linear perturbations behave beyond the quasistatic approximation. We consider the possible presence of ghosts in section \ref{sec:Stuckelberg}.
We then compare our results to doubly coupled bigravity in section \ref{sec:dou}.
We discuss consistency with small-scale tests of General Relativity in
section~\ref{sec:cancel} and conclude in section~\ref{sec:conclusion}. Several appendices are dedicated to more technical details.

\section{Scalar-Bimetric models}
\label{sec:definition}

\subsection{Defining the models}
\label{sec:model-definition}

In the following, we focus on models where the dynamics are driven by two independent metrics coupled to a scalar field. We do not add any nontrivial
dynamics for the scalar field, which we choose to be massless with a canonical kinetic term.
We consider models with the scalar-bimetric action
\beq
S = S_{\rm grav} + S_{\rm mat} ,
\label{S-def}
\eeq
with
\beq
S_{\rm grav} = \int d^4x \frac{M_{\rm pl}^2}{2} \left[ \sqrt{-g_1} R_1
+ \sqrt{-g_2} R_2 \right] ,
\label{Sgrav-def}
\eeq
and
\beq
S_{\rm mat} = \int d^4x \left[ \sqrt{-g_{\rm d}} \,
{\cal L}_{\rm d}(\varphi,\psi_{\rm dm}^i;g_{\rm d}) + \sqrt{-g_{\rm b}} \,
{\cal L}_{\rm b}(\psi_{\rm b}^i;g_{\rm b}) \right] .
\label{Smatter-def}
\eeq
The gravitational action $S_{\rm grav}$ contains two Einstein-Hilbert terms for the
two gravitational metrics $g_{1\mu\nu}$ and $g_{2\mu\nu}$.
The matter action $S_{\rm mat}$ contains the dark sector Lagrangian
${\cal L}_{\rm d}$, which includes dark matter fields $\psi_{\rm dm}^i$ and an
additional scalar field $\varphi$, and the baryonic Lagrangian ${\cal L}_{\rm b}$,
which includes the ordinary particles of the standard model,
both matter and radiation (photons) components.
These two matter Lagrangians involve two associated dynamical metrics, $g_{{\rm d}\mu\nu}$
and $g_{{\rm b}\mu\nu}$.
In the following, we will usually omit the subscript ``b'', as this is the main sector
that is probed by observations and experiments.

We split the dark sector Lagrangian in  its scalar-field and dark matter components,
\beq
{\cal L}_{\rm d} = {\cal L}_{\varphi}(\varphi;g_{\rm d})
+ {\cal L}_{\rm dm}(\psi_{\rm dm}^i;g_{\rm d}) ,
\label{Ldark-def}
\eeq
and for simplicity, we only keep the kinetic term in the scalar-field Lagrangian:
\beq
{\cal L}_{\varphi}(\varphi) = - \frac{1}{2} \, g^{\mu\nu}_{\rm d} \partial_{\mu}\varphi
\partial_{\nu}\varphi ,
\label{Lphi-def}
\eeq
as  we wish to recover the late-time acceleration of the expansion of the
Universe through a dynamical mechanism, rather than through an effective cosmological
constant associated with a nonzero minimum of the scalar-field potential.

The dark and baryonic metrics are functions of the two gravitational metrics
$g_{1\mu\nu}$ and $g_{2\mu\nu}$. We write this relationship in terms of the vierbeins
of these four metrics.
Thus, introducing the vierbeins $e^a_{1\mu}$ and $e^a_{2\mu}$ of the metrics
$g_{1\mu\nu}$ and $g_{2\mu\nu}$,
\beq
g_{1\mu\nu} = e^a_{1\mu} e^b_{1\nu} \eta_{ab} , \;\;\;
g_{2\mu\nu} = e^a_{2\mu} e^b_{2\nu} \eta_{ab} ,
\label{vierbein-12-def}
\eeq
we define the dark and baryonic metrics as
\beq
g_{{\rm d}\mu\nu} = e^a_{{\rm d}\mu} e^b_{{\rm d}\nu} \eta_{ab} , \;\;\;
g_{\mu\nu} = e^a_{\mu} e^b_{\nu} \eta_{ab} ,
\label{gd-g-def}
\eeq
with
\beqa
e^a_{{\rm d}\mu} & = & s_{{\rm d}1}(\varphi) e^a_{1\mu} + s_{{\rm d}2}(\varphi) e^a_{2\mu} ,
\nonumber \\
e^a_{\mu} & = & s_{1}(\varphi) e^a_{1\mu} + s_{2}(\varphi) e^a_{2\mu} .
\label{vierbein-db-def}
\eeqa
Thus, both dynamical matter metrics are a combination of the two gravitational
metrics that depend on the scalar field $\varphi$. This leads to nonminimal couplings
between the matter sectors and the scalar field.

\subsection{Number of components and degrees of freedom}
\label{sec:degrees-of-freedom}

Although we have defined the model in terms of the vierbeins of the two dynamical metrics,
we treat the theory as a metric theory, which is a function of the two metrics
$g_{\ell \mu\nu},\ i=1,2$.
The two Einstein-Hilbert terms are invariant under two copies of the diffeomorphism group.
The coupling to matter, which involves the diagonal subgroup, reduces the diffeomorphism
invariance to one diagonal copy. The two metrics involve 20 components,
which can be reduced to 16 when the diagonal gauge invariance under reparameterization
of coordinates has been used. The vierbeins are four-by-four matrices, which involves
$2\times 16=32$ components. This is redundant even when the diagonal diffeomorphism
invariance has been used, reducing to 28 the number of components.
The two vierbeins have two copies of the local Lorentz symmetry group as
an invariance group.
Again, this is broken to the diagonal Lorentz group by the coupling to matter.
This removes six components, bringing it down to 22. This is still more than
the 16 components of the metric description. This can be made to coincide
by imposing the symmetric condition \cite{Deffayet:2012zc}
\be
Y_{\mu\nu}= Y_{\nu\mu}
\label{Ymunu-sym}
\ee
where we have introduced the tensor
\be
Y_{\mu\nu}= \eta_{ab} e^a_{1\mu} e^b_{2\nu} =  e^a_{1\mu} e_{2a\nu} .
\ee
The $4\times 4$ tensor $Y_{\mu\nu}$ can be decomposed into ten symmetric components
and six antisymmetric ones which are imposed to be vanishing. This brings the number
of vierbein components down to 16, matching the ones for a bimetric theory.

The consequences of the symmetric conditions are well known; let us recall some salient
features here. First of all, let us define
\be
X^\mu_{2\nu}= g^{\mu\lambda  }_{2} Y_{\lambda\nu} = e^{\mu}_{2a} e^a_{1\nu} , \;\;\;
X^\mu_{1\nu}= g^{\mu\lambda  }_{1} Y_{\lambda\nu} = e^{\mu}_{1a} e^a_{2\nu} ,
\label{Xmunu-def}
\ee
then we have that
\be
X^\theta_{2\nu} X^\nu _{2\lambda} = g_2^{\theta \nu}g_{1\nu\lambda} , \;\;\;
X^\theta_{1\nu} X^\nu _{1\lambda} = g_1^{\theta \nu}g_{2\nu\lambda} , \;\;\;
\ee
which implies that in matrix notation
\be
Y = g_2 (g_2^{-1} g_1)^{1/2} = g_1 (g_1^{-1} g_2)^{1/2} ,
\ee
with an appropriate definition for the square root of a matrix \cite{Deffayet:2012zc}.
As a result, $Y_{\mu\nu}$ becomes a function of the two metrics $g_{\ell \mu\nu}$,
which implies that the dark and baryonic metrics,
\begin{eqnarray}
&& g_{\mu\nu}= s_1^2(\varphi) g_{1\mu\nu} +2 s_1(\varphi)s_2(\varphi) Y_{\mu\nu}
+ s_2^2 (\varphi) g_{2\mu\nu} , \nonumber \\
&& g_{{\rm d}\mu\nu}= s_{{\rm d}1}^2(\varphi) g_{1\mu\nu} +2 s_{{\rm d}1}(\varphi)
s_{{\rm d}2}(\varphi) Y_{\mu\nu} + s_{{\rm d}2}^2(\varphi) g_{2\mu\nu} , \nonumber \\
\end{eqnarray}
are simply functions of the two metrics, too.

Not all the components become physical degrees of freedom. For instance, when no matter
is present, our models reduce to two copies of General Relativity (GR), and as such only carry two copies of massless gravitons, i.e. $2\times 2$ physical degrees of freedom.
When matter is present, in particular cosmologically, the Hubble expansion rate of one of the two types of matter becomes the order parameter of the symmetry breaking
pattern $({\rm diff}_1 \times {\rm diff}_2)/{\rm diff}_{\rm diag}$, where the two copies of diffeomorphism invariance are broken down to the diagonal subgroup. As such, we could
expect that four Goldstone bosons $\xi^\mu$ could become physical. In fact, we find that out of the divergenceless vector and the two scalars associated with
$\xi^\mu$, only the two independent components of the vector are dynamical. The validity of the model can be
probed by a St\"uckelberg analysis, where we focus on the scalar $\xi^\mu=\partial^\mu \pi$, and we show in section~\ref{sec:Stuckelberg} below
that no ghost appears below the cutoff scale of order $\Lambda_{\rm cut}= (H^3 M_{\rm Pl})^{1/4}$.
This is the physical regime we analyze in this paper. In particular, it applies to cosmology as the horizon scale $H^{-1}$ is always much larger than $\Lambda_{\rm cut }^{-1}$ since the very early Universe.
Only in the Solar System, as the cutoff scale is of order 1 a.u., shall we be prevented from strong conclusions for want of explicit UV completions.

\subsection{Equations of motion}
\label{sec:Eqs-of-motion}

\subsubsection{Einstein's equations}
\label{sec:Einstein}

We cannot obtain the Einstein equations by requiring the functional derivatives
of the action with respect to the vierbeins $e^a_{1\mu}$ and $e^a_{2\mu}$ to vanish.
Indeed, because of the symmetry condition (\ref{Ymunu-sym}), which reduces the
number of components to those of $g_{1\mu\nu}$ and $g_{2\mu\nu}$,
the vierbeins are correlated and constrained by Eq.(\ref{Ymunu-sym}).
This means that we must take the variations along the directions that span the subspace
defined by the constraint (\ref{Ymunu-sym}). If we vary the metric $g_1$ while keeping
$g_2$ fixed, hence we vary $e^a_{1\mu}$ at fixed $e^a_{2\mu}$, the symmetric constraint
(\ref{Ymunu-sym}) reads as
\beq
\delta e^a_{1\mu} e_{2a\nu} = \delta e^a_{1\nu} e_{2a\mu} \;\;\;
\mbox{for all} \;\;\; \{ \mu, \nu \} .
\eeq
We can check that these constraints are satisfied if the variations $\delta e^a_{1\mu}$
are of the form
\beq
\delta e^a_{1\mu} = \delta Z_{1\mu\nu} e_2^{a\nu} ,
\label{deltaZ-def}
\eeq
where $\delta Z_{1\mu\nu}$ is an arbitrary infinitesimal symmetric matrix,
$\delta Z_{1\mu\nu}=\delta Z_{1\nu\mu}$.
As expected, the matrix $\delta Z_{1\mu\nu}$ provides the same number of components
as the metric $g_{1\mu\nu}$. This also gives
$\delta g_{1\mu\nu} = \delta Z_{1\mu\lambda} X^{\lambda}_{2\nu}
+ \delta Z_{1\nu\lambda} X^{\lambda}_{2\mu}$.
Then, the Einstein equations follow from the variation of the action with respect to
$\delta Z_{1\mu\nu}$.
We can write
\beq
\frac{\delta e^a_{1\lambda}}{\delta Z_{1\mu\nu}} = \Theta_{\mu\nu} \left(
\delta^{\mu}_{\lambda} e^{a\nu}_2 + \delta^{\nu}_{\lambda} e^{a\mu}_2 \right) ,
\label{de-dZ}
\eeq
where $\Theta_{\mu\nu}=1$ if $\mu<\nu$ and $\Theta_{\mu\nu}=1/2$ if $\mu=\nu$.
Here we restrict to $\mu\leq\nu$ as $\delta Z_{1\mu\nu}=\delta Z_{1\nu\mu}$
so that $\delta Z_{1\mu\nu}$ and $\delta Z_{1\nu\mu}$ are not independent.
This gives
\beq
\mbox{for all} \;\;\; \{ \mu, \nu \} : \;\;\;
\frac{\delta S}{\delta e^a_{1\mu}} e^{a\nu}_2
+ \frac{\delta S}{\delta e^a_{1\nu}} e^{a\mu}_2 = 0 .
\label{Einstein-e-a-mu}
\eeq
This provides the expected 16 symmetric Einstein equations (hence 10 equations, before
we use diffeomorphism invariance), which read as
\beqa
&& \!\!\!\!\! M_{\rm Pl}^2 \sqrt{-g_1} \left[ G_1^{\mu\sigma} X^{\nu}_{2\sigma}
+ G_1^{\nu\sigma} X^{\mu}_{2\sigma} \right] = \nonumber \\
&& s_1 \sqrt{-g} \left[ T^{\mu\sigma}
(s_1 X^{\nu}_{2\sigma} \!+\! s_2 \delta^{\nu}_{\sigma}) + T^{\nu\sigma}
(s_1 X^{\mu}_{2\sigma} \!+\! s_2 \delta^{\mu}_{\sigma}) \right] \nonumber \\
&& + s_{\rm d1} \sqrt{-g_{\rm d}} \left[ T^{\mu\sigma}_{\rm d}
(s_{\rm d1} X^{\nu}_{2\sigma} \!+\! s_{\rm d2} \delta^{\nu}_{\sigma}) + T_{\rm d}^{\nu\sigma}
(s_{\rm d1} X^{\mu}_{2\sigma} \!+\! s_{\rm d2} \delta^{\mu}_{\sigma}) \right] \;\; \nonumber \\
&&
\label{Einstein-e1}
\eeqa
The Einstein equations with respect to the second metric $g_2$ are obtained by exchanging
the indices $1 \leftrightarrow 2$
\footnote{Had we taken the variations with respect to all the vierbeins as independent
variables, we would have obtained the non-symmetric form of the Einstein equations
which does not guarantee the symmetry of the Einstein tensor unless the matter contents of the Universe is particularly tuned \cite{Hinterbichler:2015yaa}
$
M_{\rm Pl}^2 \sqrt{-g_{\ell}} G^{\mu\nu}_{\ell} e^a_{\ell\nu} =
s_{\ell} \sqrt{-g} T^{\mu\nu} e^a_{\nu}
+ s_{{\rm d}\ell} \sqrt{-g_{\rm d}} T^{\mu\nu}_{\rm d} e^a_{{\rm d}\nu} .
$
}.
Here, $T^{\mu\nu}$ and $T^{\mu\nu}_{\rm d}$ are the baryonic and dark-energy energy-momentum
tensors, defined with respect to their associated metrics,
\beq
T_{\mu\nu} = \frac{-2}{\sqrt{-g}}
\frac{\delta(\sqrt{-g} {\cal L}_{\rm b})}{\delta g^{\mu\nu}} , \;\;\;
T_{{\rm d}\mu\nu} = \frac{-2}{\sqrt{-g_{\rm d}}}
\frac{\delta(\sqrt{-g_{\rm d}} {\cal L}_{\rm d})}{\delta g^{\mu\nu}_{\rm d}} .
\eeq
We recover the standard Einstein equations when the two metrics are identical
(case of a single-metric model) with $s_1+s_2=1$ and $s_{\rm d1}+s_{\rm d2}=1$,
as it yields $X^{\mu}_{2\nu} = X^{\mu}_{1\nu} = \delta^{\mu}_{\nu}$.

When the metrics are diagonal, that is, we have
\[
g_{*\mu\nu} \subset \delta_{\mu\nu} , \;\;
e^a_{*\mu} \subset \delta^a_{\mu} , \;\;
Y_{\mu\nu} \subset \delta_{\mu\nu} , \;\;
X^{\mu}_{\ell\nu} \subset \delta^{\mu}_{\nu} ,
\]
where $*=\{1,2,{\rm d,b}\}$ and $g_{*\mu\nu} \subset \delta_{\mu\nu}$
means that $g_{*\mu\nu}=0$ for $\mu\neq\nu$,
the Einstein equations (\ref{Einstein-e1}) simplify along the diagonal as
(no summation over $\mu$)
\[
M_{\rm Pl}^2 \sqrt{-g_1} G_1^{\mu\mu} e^{a}_{1\mu} = s_1 \sqrt{-g}
T^{\mu\mu} e^a_{\mu} + s_{{\rm d}1} \sqrt{-g_{\rm d}} T^{\mu\mu}_{\rm d}
e^a_{{\rm d}\mu}
\]
with $a=\mu$. This coincides with the Einstein equations that would have been
obtained by taking derivatives with respect to the vierbeins without taking care
of the symmetric constraint (\ref{Ymunu-sym}). However, the off-diagonal Einstein
equations remain modified.

\subsubsection{Scalar-field equation of motion}
\label{sec:scalar field eom}

The dependence of the matter metrics $g_{{\rm d}\mu\nu}$ and $g_{\mu\nu}$ on the scalar
field $\varphi$, through Eq.(\ref{vierbein-db-def}), gives rise to source terms in
the Klein-Gordon equation that governs the scalar-field dynamics,
\beqa
&& \partial_{\mu} \left[ \sqrt{-g_{\rm d}} g^{\mu\nu}_{\rm d} \partial_{\nu}\varphi \right]
+ \sqrt{-g_{\rm d}} T^{\mu\nu}_{\rm d} \sum_{\ell=1}^2
\frac{ds_{{\rm d}\ell}}{d\varphi} e^a_{\ell\mu} e^b_{{\rm d}\nu} \eta_{ab}
\nonumber \\
&& + \sqrt{-g} T^{\mu\nu} \sum_{\ell=1}^2
\frac{ds_{\ell}}{d\varphi} e^a_{\ell\mu} e^b_{\nu} \eta_{ab} = 0 .
\label{KG-1}
\eeqa

\subsubsection{Matter equations of motion}
\label{sec:matter eom}

The equations of motion of the dark and baryonic matter components take their standard
form in their Jordan frames,
\beq
\nabla_{{\rm d}\mu} T^{\mu}_{{\rm dm}\nu} = 0 , \;\;\;
\nabla_{\mu} T^{\mu}_{\nu} = 0 ,
\label{matter-eom}
\eeq
where $\nabla_{{\rm d}\mu}$ and $\nabla_{\mu}$ are the covariant derivatives with respect
to the metrics $g_{{\rm d}\mu\nu}$ and $g_{\mu\nu}$.

\section{Cosmological background}
\label{sec:background}

In this section, we investigate the cosmological backgrounds that can be achieved
in these scalar-bimetric scenarios. We show how we recover a standard cosmology
at high redshift, when the scalar field is almost constant and plays no role
and all metrics follow the same expansion, whereas a self-accelerated expansion
without a cosmological constant can be achieved at low redshift thanks to the running of the
scalar field, through the interplay between the matter and gravitational metrics.

\subsection{Friedmann's equations}
\label{sec:backg-Friedmann}

We consider diagonal metrics of the form
\beq
g_{_*\mu\nu}(\tau) = {\rm diag}(-b_*^2(\tau), a_*^2(\tau), a_*^2(\tau), a_*^2(\tau)) ,
\label{g-diag-def}
\eeq
where $*=\{1,2,{\rm d,b}\}$, with the vierbeins
\beq
e^{a}_{*\mu} = {\rm diag}(b_*,a_*,a_*,a_*) ,
\label{e-diag-def}
\eeq
and we denote by  ${\cal H}_{b_*}=d\ln b_*/d\tau$, ${\cal H}_{a_*}=d\ln a_*/d\tau$, the conformal
expansion rates of the time and spatial components.
We can choose to define the conformal time $\tau$ with respect to the baryonic
metric $g_{\mu\nu}$, so that
\beq
b=a , \;\;\; g_{\mu\nu}(\tau) = {\rm diag}(-a^2, a^2, a^2, a^2) ,
\label{b=a}
\eeq
and we use either $\tau$ or $\ln(a)$ as the time variable.
From the definitions (\ref{vierbein-db-def}) we obtain the constraints
\beqa
&& b=a= s_1 b_1 + s_2 b_2 , \;\;\; a= s_1 a_1 + s_2 a_2 , \nonumber \\
&& b_{\rm d}= s_{{\rm d}1} b_1 + s_{{\rm d}2} b_2 , \;\;\;
a_{\rm d}= s_{{\rm d}1} a_1 + s_{{\rm d}2} a_2 .
\label{a-ad-a1-a2}
\eeqa

The $(0,0)$ component of the Einstein equations (\ref{Einstein-e1}) reads
\beq
3 M_{\rm Pl}^2 a_{\ell}^3 b_{\ell}^{-2} {\cal H}_{a_{\ell}}^2 = s_{\ell} a^3
(\bar\rho+\bar\rho_{\gamma}) + s_{{\rm d}{\ell}} a_{\rm d}^3
( \bar\rho_{\rm dm}+\bar\rho_{\varphi} ) ,
\label{E1-(00)}
\eeq
while the $(i,i)$ components read
\beqa
&& M_{\rm Pl}^2 a_{\ell}^2 b_{\ell}^{-1} \left[ 2 {\cal H}'_{a_{\ell}} + 3 {\cal H}^2_{a_{\ell}} - 2
{\cal H}_{a_{\ell}} {\cal H}_{b_{\ell}} \right] = - s_{\ell} a^3 \frac{\bar\rho_{\gamma}}{3}
\nonumber \\
&& - s_{{\rm d}{\ell}} a_{\rm d}^2 b_{\rm d} \bar\rho_{\varphi}  .
\label{E1-(ii)}
\eeqa
Here, we assumed nonrelativistic matter components, $p_{\rm dm}=p=0$,
and we used $p_{\gamma}=\rho_{\gamma}/3$ and $\bar{p}_{\varphi}=\bar{\rho}_{\varphi}$
for the radiation and scalar pressure.

\subsection{Conservation equations}
\label{sec:backg-conserv}

The Jordan-frame equations of motion (\ref{matter-eom}) lead to the usual conservation
equations; hence,
\beq
\bar\rho_{\rm dm} = \frac{\bar\rho_{{\rm dm}0}}{a_{\rm d}^3} , \;\;\;
\bar\rho = \frac{\bar\rho_{0}}{a^3} , \;\;\;
\bar\rho_{\gamma} = \frac{\bar\rho_{\gamma 0}}{a^4} .
\label{rho-backg-0}
\eeq
We define the cosmological parameters associated with these characteristic densities
by
\beq
\Omega_{{\rm b}0} = \frac{\bar\rho_{\rm b 0}}{3 M_{\rm Pl}^2 H_0^2} , \;\;\;
\Omega_{\gamma 0} = \frac{\bar\rho_{\gamma 0}}{3 M_{\rm Pl}^2 H_0^2} , \;\;\;
\Omega_{{\rm dm}0} = \frac{\bar\rho_{\rm dm 0}}{3 M_{\rm Pl}^2 H_0^2} ,
\label{Omega-def}
\eeq
where $H_0$ is the physical expansion rate associated with the baryonic metric
today, at $a=1$. We also define the rescaled scalar-field energy density $\xi$ as
the ratio of the scalar field to dark matter energy densities
\beq
\xi(a) = \frac{a_{\rm d}^3 \bar\rho_{\varphi}}{3 M_{\rm Pl}^2 H_0^2}  , \;\;\;
\frac{\xi(a)}{\Omega_{{\rm dm}0}} = \frac{\bar\rho_{\varphi}(a)}{\bar\rho_{\rm dm}(a)} .
\label{xi-def}
\eeq
It is convenient to introduce the dimensionless combination
\beq
\ell=1, 2 : \;\;\;  \omega_{\ell} = a_{\ell}^3 b_{\ell}^{-2} \frac{{\cal H}_{a_{\ell}}^2}{H_0^2} .
\label{omega-i-def}
\eeq
Then, the Friedmann equations (\ref{E1-(00)})-(\ref{E1-(ii)}) simplify as
\beq
\omega_{\ell} = s_{\ell} \left( \Omega_{\rm b 0} + \frac{\Omega_{\gamma 0}}{a} \right)
+ s_{{\rm d} \ell} \left( \Omega_{\rm dm 0} + \xi(a) \right) ,
\label{Friedmann-00}
\eeq
\beq
\frac{b_{\ell}}{a_{\ell}} \frac{{\cal H}}{{\cal H}_{a_{\ell}}}
\frac{d\omega_{\ell}}{d\ln a} = - s_{\ell} \frac{\Omega_{\gamma 0}}{a}
- s_{{\rm d} \ell} \, 3 \frac{b_{\rm d}}{a_{\rm d}} \xi .
\label{Friedmann-ii}
\eeq
We recover the usual Friedmann equations of General Relativity with
$\xi=0$, $b_*=a_*$, ${\cal H} = {\cal H}_{a_{\ell}}$, $s_{\ell}=1$ and $s_{{\rm d}\ell}=1$.
In this case, we can check that the second Friedmann equation is a consequence
of the first Friedmann equation and of the conservation equations, as it is the derivative
of Eq.(\ref{Friedmann-00}) with respect to $\ln a$.

By taking the first derivative of Eq.(\ref{Friedmann-00}) and combining with
Eq.(\ref{Friedmann-ii}), we obtain the useful combinations
\beqa
&& \frac{ds_{\ell}}{d\ln a} \left( \Omega_{\rm b 0} + \frac{\Omega_{\gamma 0}}{a} \right)
+ \frac{ds_{{\rm d} \ell}}{d\ln a} \left( \Omega_{\rm dm 0} + \xi(a) \right) =
\nonumber \\
&& s_{\ell} \frac{\Omega_{\gamma 0}}{a} \left( 1 - \frac{a_{\ell}}{b_{\ell}}
\frac{{\cal H}_{a_{\ell}}}{\cal H} \right) - s_{{\rm d}\ell} \xi \left( \frac{d\ln\xi}{d\ln a}
+ 3 \frac{b_{\rm d} a_{\ell}}{a_{\rm d} b_{\ell}} \frac{{\cal H}_{a_{\ell}}}{\cal H} \right) .
\nonumber \\
&& \label{ds-dsd}
\eeqa
This shows that the evolutions of the baryonic and dark matter couplings
are correlated and related to the running of the scalar field ($\xi > 0$)
and the deviations between the different expansion rates ${\cal H}_*$.
In the absence of the  scalar field in the dynamics the relation (\ref{ds-dsd}) reduces
to the branch of solutions
\be
\frac{a_{\ell}}{b_{\ell}} {\cal H}_{a_{\ell}} = {\cal H} ,
\label{branch-2}
\ee
which appears in doubly coupled bigravity \cite{Brax:2016ssf}.

\subsection{Scalar-field equation of motion}
\label{sec:backg-scalar-field-eom}

The scalar-field energy density reads as
\beq
\bar\rho_{\varphi} = \frac{1}{2b_{\rm d}^2} \left( \frac{d\bar\varphi}{d\tau} \right)^2 .
\label{rho-phi-def}
\eeq
Then, we can check that the background Klein-Gordon equation (\ref{KG-1}) can be
written in terms of $\bar\rho_{\varphi}$. Using the rescaled scalar-field density $\xi$
of Eq.(\ref{xi-def}), this gives
\beqa
&& \frac{b_{\rm d}}{a_{\rm d}^3} \frac{d}{d\ln a} \left[ a_{\rm d}^3 \xi \right]
+ ( \Omega_{\rm dm 0}+\xi ) \sum_{\ell} \frac{ds_{{\rm d}\ell}}{d\ln a} b_{\ell}
\nonumber \\
&& - 3 \frac{b_{\rm d}}{a_{\rm d}} \xi \sum_{\ell}
\frac{ds_{{\rm d}\ell}}{d\ln a} a_{\ell}
+ \left( \Omega_{\rm b0}+\frac{\Omega_{\gamma 0}}{a} \right) \sum_{\ell}
\frac{ds_{\ell}}{d\ln a} b_{\ell} \nonumber \\
&& - \frac{\Omega_{\gamma 0}}{a} \sum_{\ell} \frac{ds_{\ell}}{d\ln a} a_{\ell}
 =  0 .
\label{KG-backg}
\eeqa

From $\xi(a)$, we obtain the evolution of the scalar field $\bar\varphi(a)$ by integrating
Eq.(\ref{rho-phi-def}). With the initial condition $\bar\varphi(0)= 0$, this gives
\beq
\bar\varphi(a) = M_{\rm Pl} \int_0^a \frac{da}{a} \frac{b_{\rm d} H_0}{{\cal H}}
\sqrt{ \frac{6\xi}{a_{\rm d}^3} } .
\label{phi-a-int}
\eeq

We can actually check that the Klein-Gordon equation (\ref{KG-backg}) is also a
consequence of the Friedmann equations (\ref{Friedmann-00})-(\ref{Friedmann-ii}),
supplemented with the constraints (\ref{a-ad-a1-a2}).
Therefore, as in General Relativity, the Friedmann equations and the conservation
equations are not independent. In General Relativity, it is customary to
work with the first Friedmann equation and the conservation equations of the various
matter components, leaving aside the second Friedmann equation that is their automatic
consequence.
In this paper, because we have two symmetric sets of Friedmann equations
(\ref{Friedmann-00})-(\ref{Friedmann-ii}), for $\ell=1,2$, and the Klein-Gordon equation
(\ref{KG-backg}) takes a complicated form with its new source terms,
we instead work with the four Friedmann equations and we discard the
Klein-Gordon equation (\ref{KG-backg}), which is their automatic
consequence.

\subsection{Einstein - de Sitter reference}
\label{sec:backg-EdS}

When the scalar field is a constant, it should not play any role and we expect to recover
a standard cosmology. Because we did not introduce any cosmological constant,
this must be an Einstein-de Sitter universe without late-time acceleration
(more precisely, a universe with only matter and radiation components).
In this reference universe, obtained within General Relativity with only one metric,
the Friedmann equations (\ref{Friedmann-00})-(\ref{Friedmann-ii}) read as
\beqa
&& \omega^{(0)} = a \frac{{\cal H}^{(0)2}}{H_0^2} = \Omega_{\rm dm 0}+\Omega_{\rm b 0}
+ \frac{\Omega_{\gamma 0}}{a} , \nonumber \\
&& \frac{d\omega^{(0)}}{d\ln a} = - \frac{\Omega_{\gamma 0}}{a} .
\label{Friedmann-ref-0}
\eeqa
Here and in the following, we denote with the superscript ``(0)'' quantities associated
with this Einstein-de Sitter reference Universe, which follows General Relativity.
As noticed above, here the second Friedmann equation is trivial as it is a mere
consequence of the first Friedmann equation and of the conservation equations,
which have already been used in the first Eq.(\ref{Friedmann-ref-0}).

We can recover the standard cosmology (\ref{Friedmann-ref-0}) within the bimetric
model (\ref{S-def}) by the simple solution
\beqa
&& a_{\ell}^{(0)} = s_{\ell}^{(0)} a , \;\;
a_{\rm d}^{(0)} = a , \;\; b_*^{(0)} = a_*^{(0)} ,
\;\; {\cal H}^{(0)}_* = {\cal H}^{(0)} ,
\nonumber \\
&& \omega_{\ell}^{(0)}= s_{\ell}^{(0)} \omega^{(0)} , \;\; \xi^{(0)} = 0 , \;\;
s_{{\rm d} \ell}^{(0)} = s_{\ell}^{(0)} ,
\label{EdS-0}
\eeqa
where the coefficients $s_{\ell}^{(0)}$ are constants that obey the condition
\beq
\left( s_1^{(0)} \right)^2 + \left( s_2^{(0)} \right)^2 = 1 .
\label{s12-s22-1}
\eeq
The scalar field $\varphi$ is also constant, as the derivatives in the source terms
of Eq.(\ref{KG-1}) vanish.
In this solution, all four metrics are essentially equivalent, as $b_*=a_*$ and all
scale factors $a_*$ are proportional. The common expansion rate ${\cal H}_*(a)$ follows
the standard Einstein-de Sitter reference ${\cal H}^{(0)}(a)$ of Eq.(\ref{Friedmann-ref-0}).

\subsection{$\Lambda$-CDM reference}
\label{sec:backg-LCDM}

To match observations, the expansion rate ${\cal H}(a)$ must deviate from the
Einstein-de Sitter reference (\ref{Friedmann-ref-0}) and remain close to the
concordance $\Lambda$-CDM cosmology.
To ensure that this is the case, in this paper, we constrain the baryonic expansion rate
${\cal H}(a)$ to follow exactly a reference $\Lambda$-CDM cosmology.
Of course, in practice, small deviations from the $\Lambda$-CDM limit are allowed
by the data, and we could also generalize the solutions that we consider in this paper
by adding small deviations. However, by definition, this would not significantly modify
the properties of these solutions. Besides, being able to recover a
$\Lambda$-CDM expansion rate is sufficient to show that the bimetric model can be made
consistent with data at the level of the cosmological background.

In the $\Lambda$-CDM cosmology, we add a cosmological constant to the components of the
Universe. The usual Friedmann equation reads as
\beq
\frac{{\cal H}^2}{H_0^2} = \frac{\Omega_{\rm dm 0}+\Omega_{\rm b 0}}{a}
+ \frac{\Omega_{\gamma 0}}{a^2} + \Omega_{\Lambda 0} a^2 ,
\label{Friedmann-LCDM}
\eeq
where $\Omega_{\Lambda 0}$ is the cosmological parameter associated with the
cosmological constant. In terms of the variable $\omega$ this gives
\beq
\omega(a) = \omega^{(0)}(a) + \Omega_{\Lambda 0} a^3 ,
\label{omega-LCDM}
\eeq
which explicitly shows the deviation from the Einstein-de Sitter reference
(\ref{Friedmann-ref-0}).
[Here, the Einstein-de Sitter reference is normalized with
$\Omega_{\rm dm 0}+\Omega_{\rm b 0}+\Omega_{\gamma 0} = 1-\Omega_{\Lambda 0} \neq 1$,
because we normalize the cosmological densities by $H_0$ instead of $H^{(0)}_0$.]

The bimetric solution with a constant scalar field, which was able to reproduce
the Einstein-de Sitter cosmology (\ref{Friedmann-ref-0}),
cannot mimic the $\Lambda$-CDM cosmology (\ref{omega-LCDM}) because of the extra term
on the right-hand side of the Friedmann equation [a constant scalar field implies
$\xi=0$ in Eq.(\ref{Friedmann-00})].
Therefore, to recover a $\Lambda$-CDM expansion rate, we must consider more general
solutions with a nonconstant scalar field.
In particular, even if the scale factors $a_i$ of the gravitational metrics
keep decelerating at late times, the baryonic scale factor $a=s_1 a_1 + s_2 a_2$
can accelerate at late times if $s_1$ or $s_2$ grows sufficiently fast.
Then, the acceleration experienced by the baryonic metric is a dynamical effect
due to the time-dependent relationship between this metric and the two
gravitational metrics.

On the other hand, at early times where data show that the dark-energy density
is negligible, we converge to the simple Einstein - de Sitter solution
(\ref{EdS-0}). This will be the common early-time behavior of all the solutions
that we consider in this paper.
We can check that the integral in Eq.(\ref{phi-a-int}) is indeed finite and
goes to zero for $a\to 0$, both in the radiation and dark matter eras, provided
\beq
\xi \to 0 \;\;\; \mbox{for} \;\;\; a \to 0 .
\label{xi-a-to-0}
\eeq
This also ensures that the dark-energy density is negligible as compared with the
dark matter density.
As we shall see below, the families of solutions that we build in this paper are
parametrized by $\xi(a)$, which is treated as a free function of the model.
Therefore, the condition (\ref{xi-a-to-0}) is easily satisfied, by choosing
functions $\xi(a)$ that exhibit a fast decay at high redshift.

\subsection{Solutions with common conformal time}
\label{sec:backg-common-conformal}

To illustrate how we can build bimetric solutions that follow a $\Lambda$-CDM
expansion rate, we first consider solutions with
\beq
a_* = b_* ,
\label{symm-ab}
\eeq
that is, the conformal time $\tau$ is the same for all metrics.
Then, at the background level, each metric is defined by a single scale factor $a_*$
and the two constraints in the first line in Eq.(\ref{a-ad-a1-a2}) reduce to one,
$a=s_1 a_1 + s_2 a_2$.
As all metrics are proportional, at the background level, this scenario is similar to a single gravitational
metric model, $\tilde{g}_{\mu\nu}$, where the baryonic and the dark matter metrics
are given by different conformal rescalings,
$g_{\mu\nu} = A^2(\varphi) \tilde{g}_{\mu\nu}$ and
$g_{\mu\nu} = A_{\rm d}^2(\varphi) \tilde{g}_{\mu\nu}$.

\subsubsection{Symmetric solution}
\label{sec:backg-symmetric}

We first consider a simple symmetric solution where we split the single constraint
$a=s_1 a_1 + s_2 a_2$ into two symmetric constraints:
\beq
\frac{s_1 a_1}{a} = \left( s_1^{(0)} \right)^2 , \;\;
\frac{s_2 a_2}{a} = \left( s_2^{(0)} \right)^2 .
\label{symm-s1-s2}
\eeq
This is consistent with the initial conditions defined by the early-time solution
(\ref{EdS-0})-(\ref{s12-s22-1}).
Then, Eq.(\ref{symm-s1-s2}) gives $s_{\ell}(a)$ as an explicit function of
$\{a,a_{\ell}(a)\}$, and we solve for the two sets
$\{a_{\ell}(a), \omega_{\ell}(a), s_{{\rm d} \ell}(a)\}$.
Thanks to the splitting (\ref{symm-s1-s2}), these two sets of variables can be solved
independently.
Then, the three functions $\{a_{\ell}(a), \omega_{\ell}(a), s_{{\rm d}\ell}(a)\}$
are determined by the two Friedmann equations (\ref{Friedmann-00})-(\ref{Friedmann-ii})
and the definition (\ref{omega-i-def}).
The definition (\ref{omega-i-def}) provides ${\cal H}_{\ell}$ at each time step,
hence $d\ln a_{\ell}/d\ln a$.
The second Friedmann equation (\ref{Friedmann-ii}) gives $d\omega_{\ell}/d\ln a$.
The first Friedmann equation (\ref{Friedmann-00}) provides $s_{{\rm d}\ell}(a)$.
The dark sector scale factor $a_{\rm d}$ is given by Eq.(\ref{a-ad-a1-a2}),
$a_{\rm d}(a)= s_{{\rm d}1} a_1 + s_{{\rm d}2} a_2$.
The scalar-field energy density $\xi(a)$ is an arbitrary function, which is a free
function of the bimetric model. It must be positive, and we only request that it vanishes
at early times to recover the high-redshift cosmology (\ref{EdS-0}).

This procedure provides a family of solutions that are parametrized by the initial
coefficients $s_{\ell}^{(0)}$ and the scalar energy density $\xi(a)$, and which follow
the $\Lambda$-CDM expansion history for ${\cal H}(a)$.
The latter enters the dynamical equations through the factors ${\cal H}(a)$
in Eqs.(\ref{Friedmann-ii}) and (\ref{omega-i-def}) [when we write
$d\ln a_{\ell}/d\tau = {\cal H} (d\ln a_{\ell}/d\ln a)$].
As the coefficients $s_{\ell}^{(0)}$ do not appear in these equations,
the two metrics are actually equivalent, with
\beq
\frac{a_1}{a_2} = \frac{s_1}{s_2} = \frac{s_1^{(0)}}{s_2^{(0)}} , \;\;\;
{\cal H}_1 = {\cal H}_2 .
\label{sym-H1-H2}
\eeq

\begin{figure*}
\begin{center}
\epsfxsize=8.5 cm \epsfysize=5.2 cm {\epsfbox{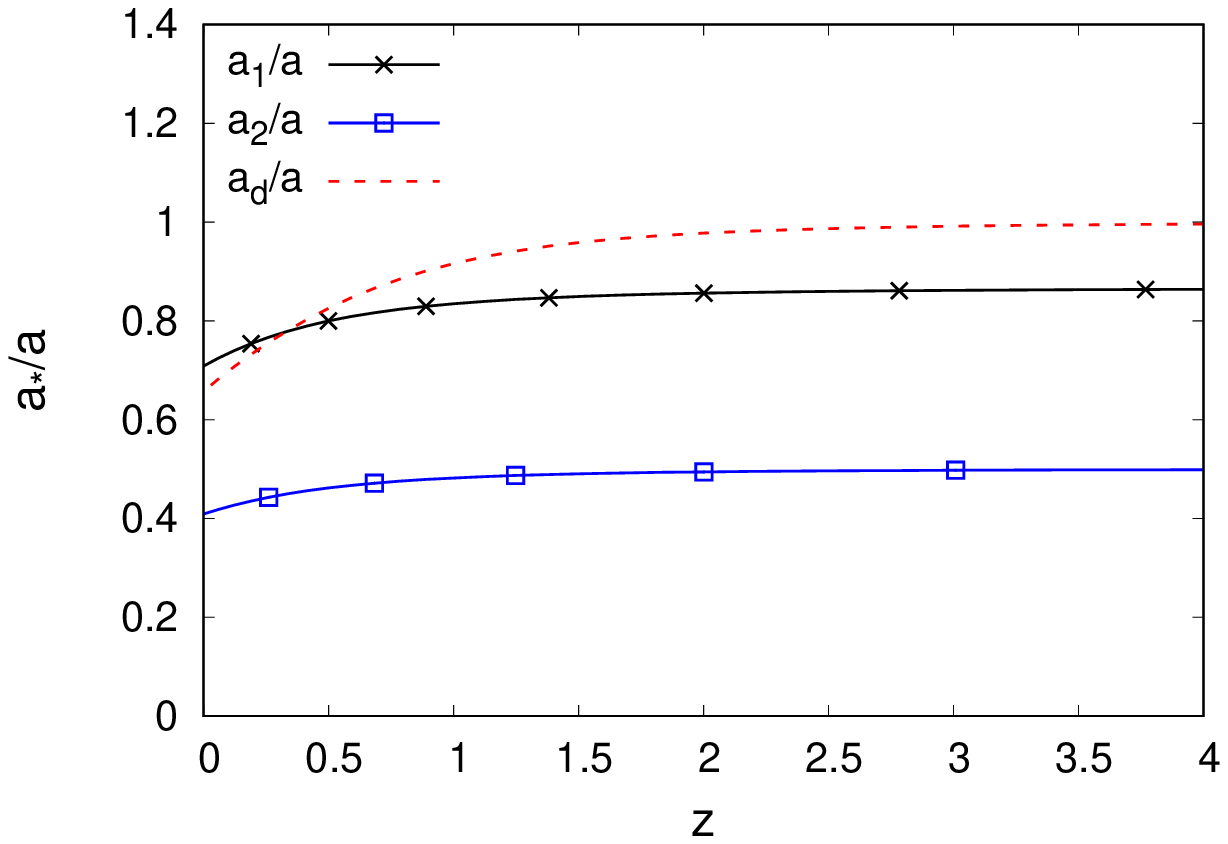}}
\epsfxsize=8.5 cm \epsfysize=5.2 cm {\epsfbox{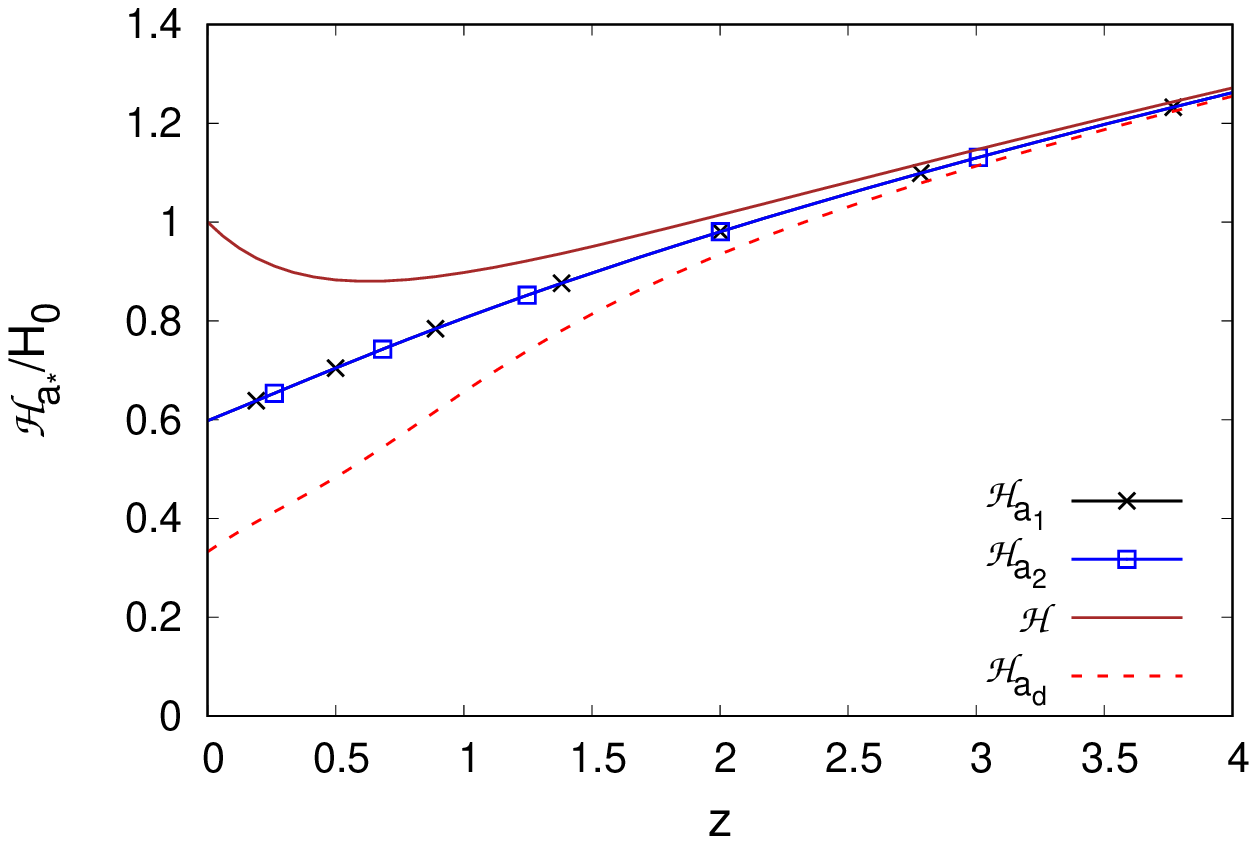}}\\
\epsfxsize=8.5 cm \epsfysize=5.2 cm {\epsfbox{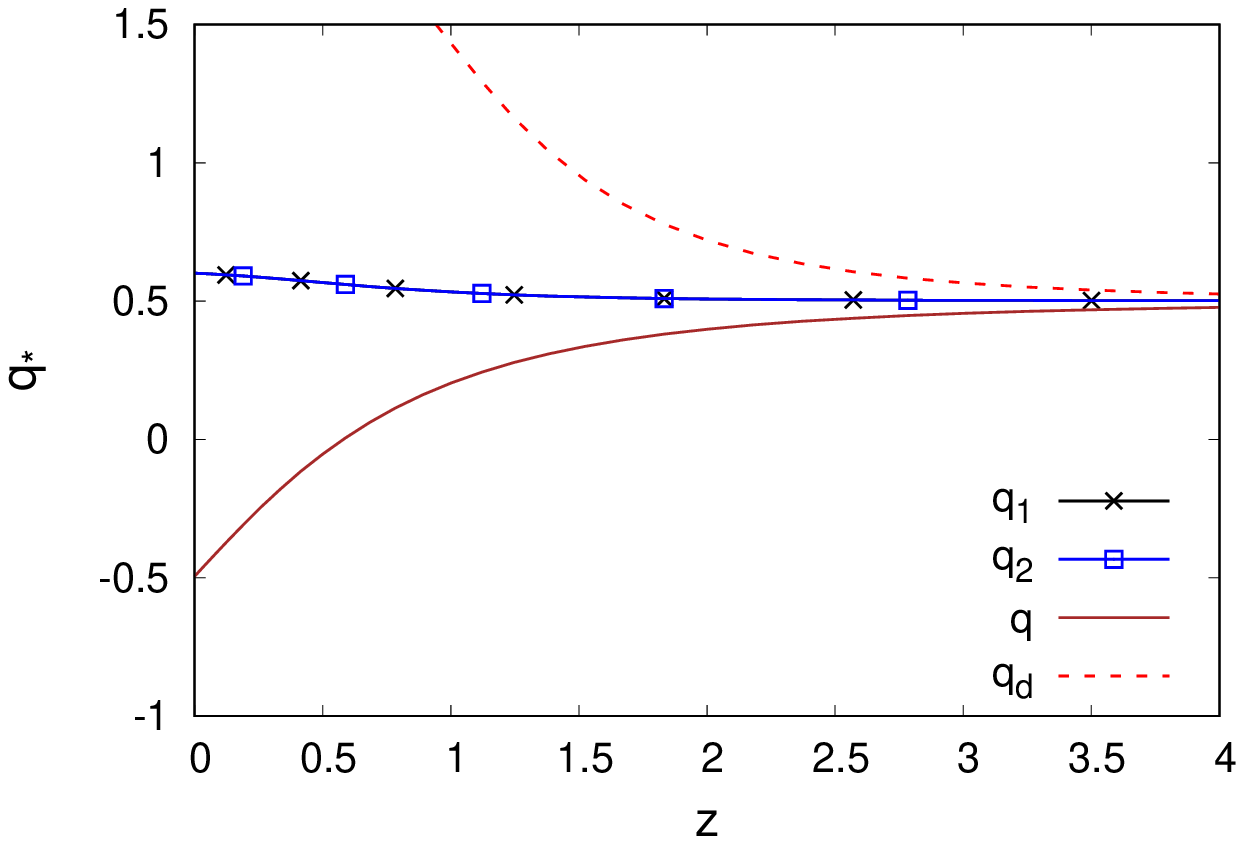}}
\epsfxsize=8.5 cm \epsfysize=5.2 cm {\epsfbox{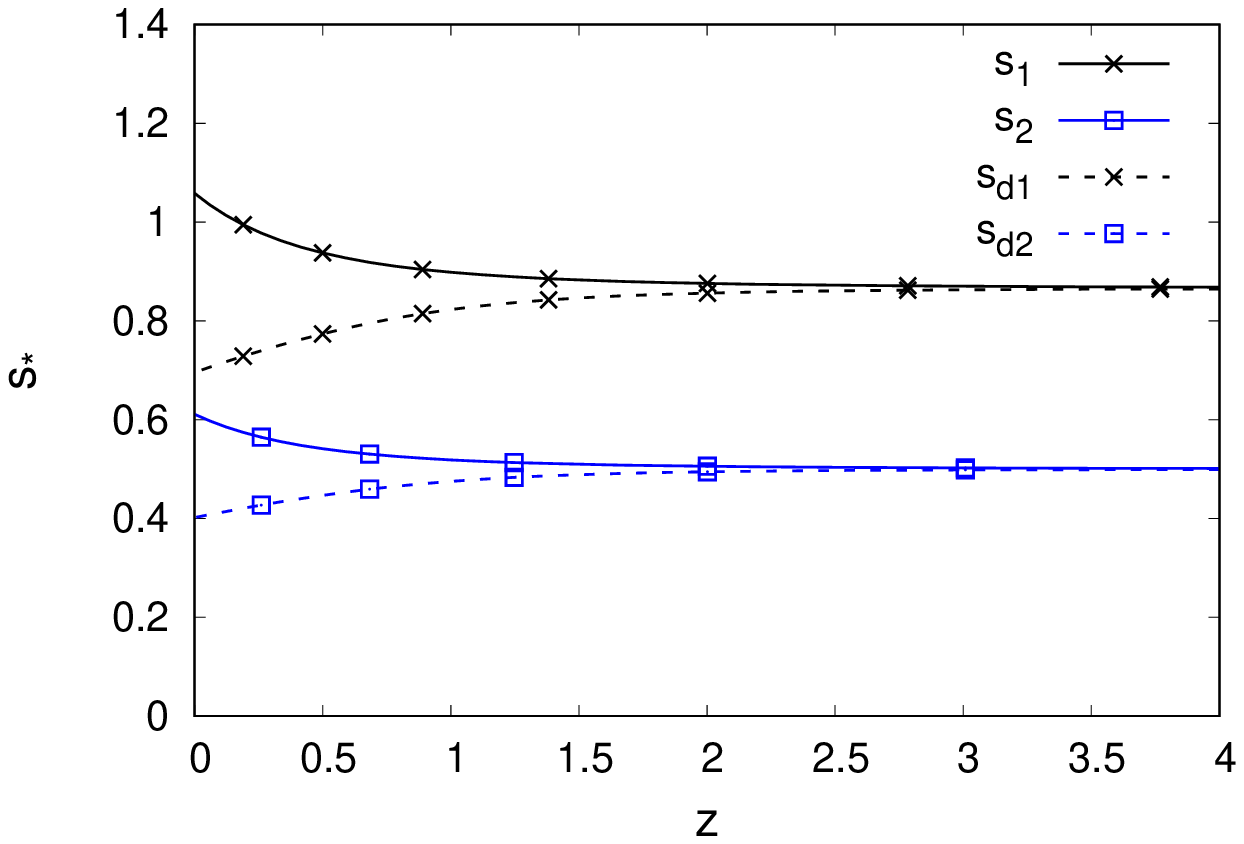}}
\end{center}
\caption{Background quantities for a symmetric solution of the form (\ref{symm-s1-s2}),
as a function of redshift.
{\it Upper left panel:} ratio of the various scale factors $a_*$ to the baryonic scale factor $a$.
{\it Upper right panel:} the various expansion rates ${\cal H}_*$ normalized to $H_0$.
{\it Lower left panel:} the various deceleration parameters $q_*$.
{\it Lower right panel:} coefficients $s_{\ell}$ and $s_{{\rm d}\ell}$ of Eqs.(\ref{vierbein-db-def})
and (\ref{a-ad-a1-a2}).}
\label{fig_sym_a1a2}
\end{figure*}

\begin{figure}
\begin{center}
\epsfxsize=8.5 cm \epsfysize=5.2 cm {\epsfbox{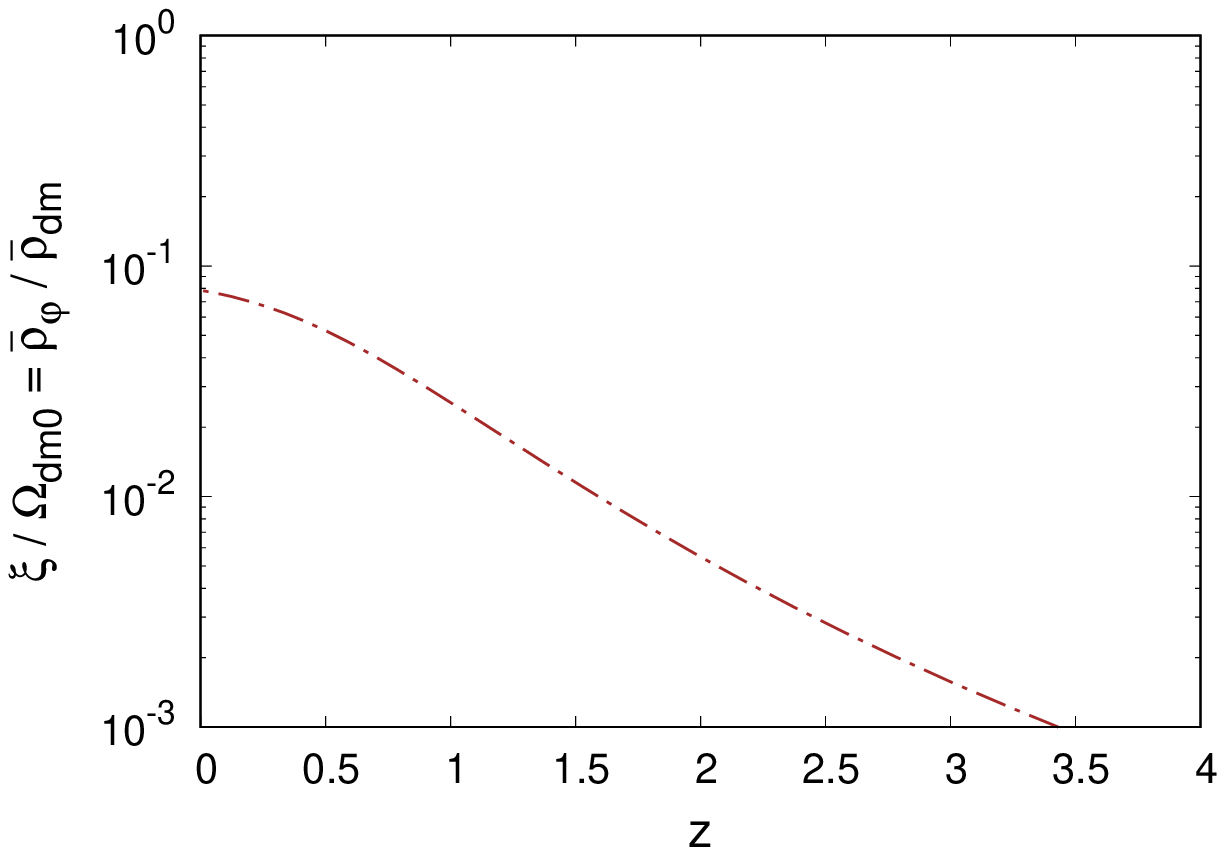}}
\epsfxsize=8.5 cm \epsfysize=5.2 cm {\epsfbox{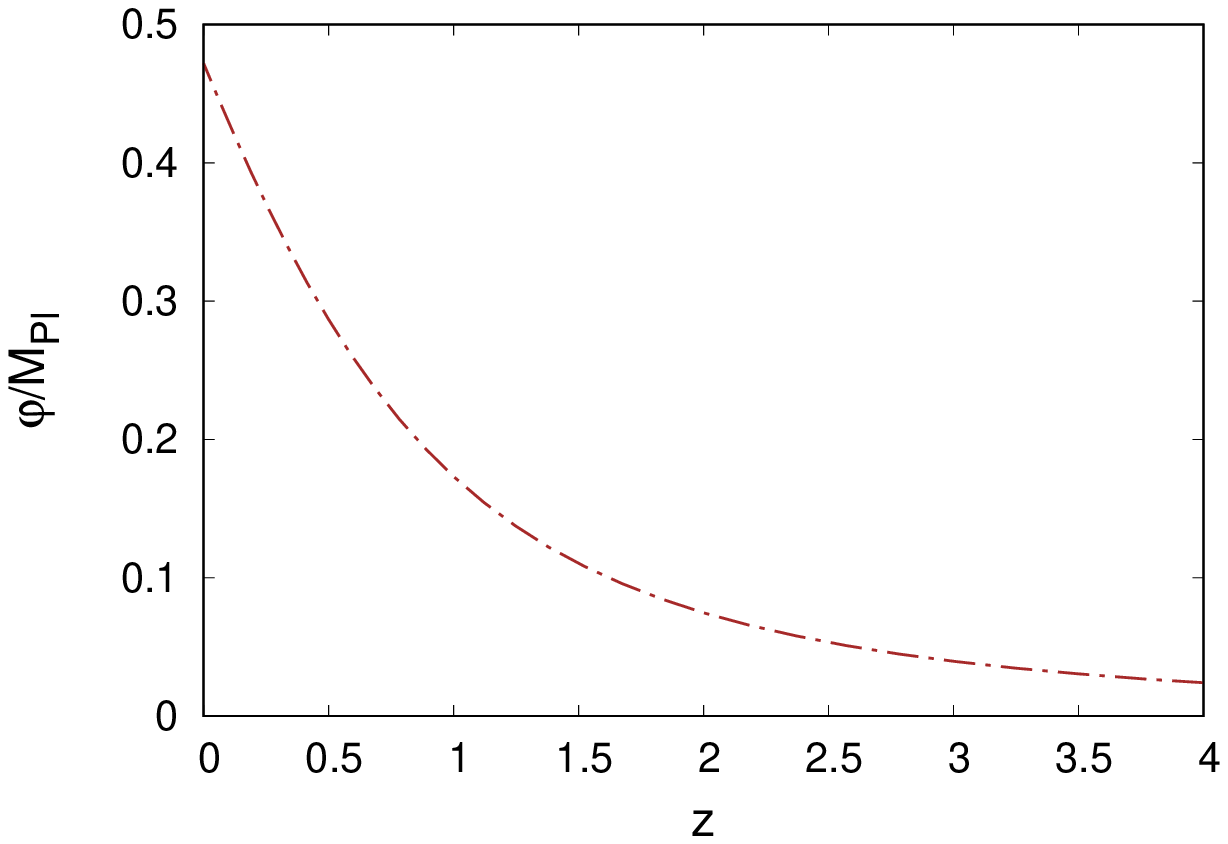}}
\epsfxsize=8.5 cm \epsfysize=5.2 cm {\epsfbox{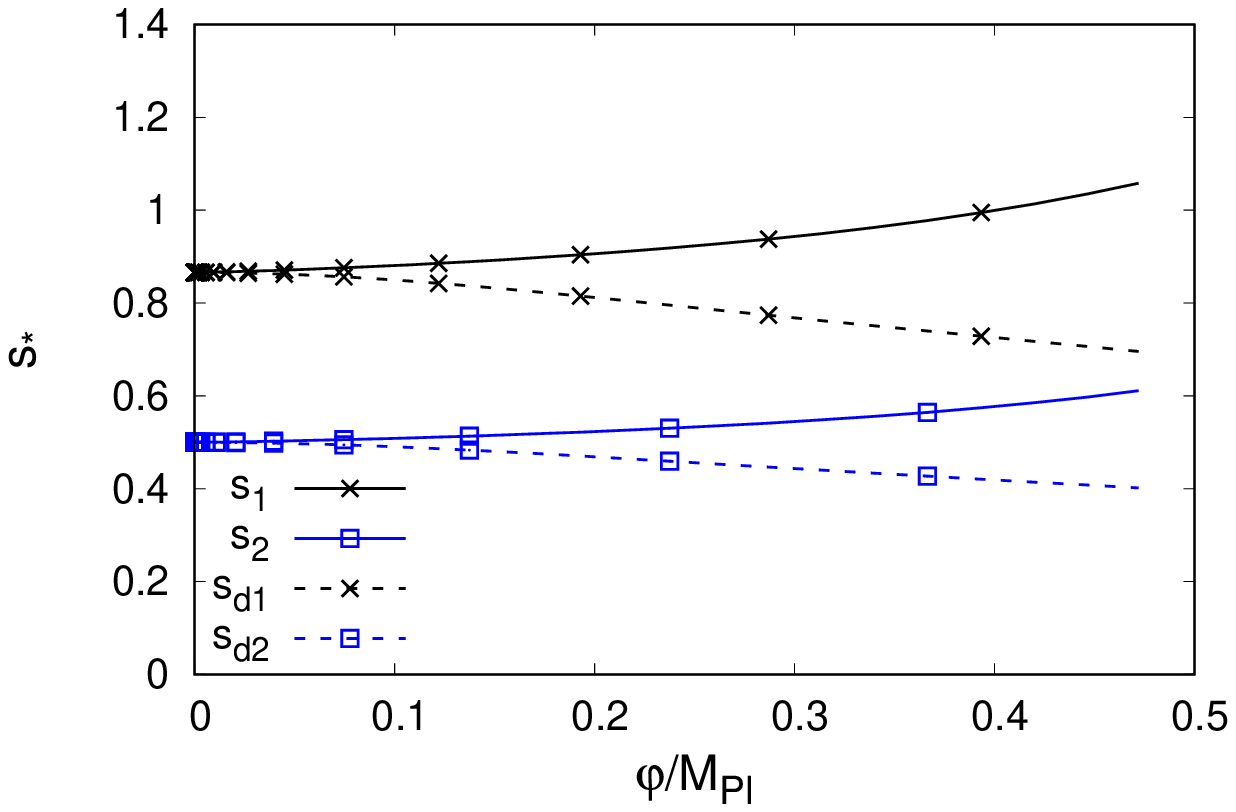}}
\end{center}
\caption{Background quantities for a symmetric solution of the form (\ref{symm-s1-s2}).
{\it Upper panel:} ratio of the scalar-field energy density to the dark matter
energy density.
{\it Middle panel:} value of the scalar field in Planck mass unit.
{\it Lower panel:} coefficients $s_{\ell}$ and $s_{{\rm d}\ell}$ as a function of the
scalar field.
}
\label{fig_sym2_a1a2}
\end{figure}

We show in Figs.~\ref{fig_sym_a1a2} and ~\ref{fig_sym2_a1a2}
the evolution with redshift of the main background quantities, in such a solution
with $s_1^{(0)}=\sqrt{3}/2, s_2^{(0)}=1/2$.
The scalar-field energy density $\xi(z)$ is chosen to vanish
at high $z$ and to remain much smaller than the dark matter energy density at all times,
$\xi \ll 1$.
More specifically, we use the simple form
\beq
\xi(a) \propto \frac{u^{3/2}}{1+u^{3/2}} , \;\;\;
u = \frac{\Omega_{\Lambda0}a^4}{\Omega_{\gamma 0}
+ (\Omega_{\rm dm 0}+\Omega_{\rm b 0})a} .
\label{xi-u-def}
\eeq
From Eq.(\ref{Friedmann-LCDM}), the quantity $u(a)$ is a natural measure of the
deviation of the $\Lambda$-CDM cosmological background from the Einstein-de Sitter
background. It is also the ratio of the effective dark-energy density to the matter and
radiation energy densities and we have $\omega=\omega^{(0)} (1+u)$.
In this paper, we write the free functions of the models in terms of powers of $u(a)$,
to ensure that we recover the Einstein-de Sitter reference of Section~\ref{sec:backg-EdS}
at early times. This also means that the effects of the scalar field only appear at
low redshifts, where the departure from the Einstein-de Sitter reference is associated
with a running of the scalar field.

We can see in Fig.~\ref{fig_sym_a1a2} that $a_1/a$ and $a_2/a$ decrease at low $z$
while $s_1$ and $s_2$ increase. Indeed, because of the absence of a cosmological constant,
the scale factors $a_i(\tau)$ of the gravitational metric tend to follow an
Einstein-de Sitter expansion rate, which falls below the $\Lambda$-CDM expansion rate
of $a(\tau)$. The latter manages to mimic the $\Lambda$-CDM history thanks to the
late-time growth of the factors $s_{\ell}$ in Eq.(\ref{a-ad-a1-a2}).
On the other hand, the dark factors $s_{{\rm d}\ell}$ decrease at low $z$,
in a fashion that is opposite to the baryonic factors $s_{\ell}$.
This follows from the relationship (\ref{ds-dsd}), which gives
\beq
\xi \ll 1 , \;\;\; \Omega_{\gamma 0} \ll 1 : \;\;\;
\frac{ds_{{\rm d}\ell}}{d\ln a} \simeq - \frac{\Omega_{\rm b 0}}{\Omega_{\rm dm 0}}
\frac{ds_{\ell}}{d\ln a} \;\;\; \mbox{for} \;\;\; a\sim 1.
\label{s-sd-low-z}
\eeq
Then, from Eq.(\ref{a-ad-a1-a2}), the dark sector scale factor $a_{\rm d}(\tau)$
grows even more slowly than the gravitational scale factors $a_{\ell}(\tau)$ at late times,
and we have ${\cal H}_{a_{\rm d}} < {\cal H}_{a_{\ell}} < {\cal H}$.

These different cosmic evolutions are clearly shown by the deceleration factors $q_*$,
defined for each metric with respect to its cosmic time $dt_* = b_* d\tau$ by
\beq
q_* = - \frac{\ddot a_* a_*}{\dot a_*^2}
= -\frac{d^2a_*}{dt_*^2} a_* \left( \frac{da_*}{dt_*} \right)^{-2} .
\eeq
Thus, we can see that the gravitational metrics $g_1$ and $g_2$ show no acceleration.
They keep behaving like an Einstein de Sitter cosmology, except for a slightly stronger
deceleration at low $z$. Only the baryonic metric shows an accelerated expansion with
$q<0$. Because of the opposite behavior of the dark sector coefficients $s_{{\rm d}\ell}$,
as compared with the baryonic coefficients $s_{\ell}$, the dark sector metric shows instead
a stronger deceleration at late times than the Einstein de Sitter cosmology.
This clearly shows that the apparent acceleration of the baryonic metric is not due
to a dark-energy component, associated for instance with the scalar field $\varphi$,
as the ``Einstein-frame'' metrics $g_1$ and $g_2$ do not accelerate. It is only due to
the time-dependent mapping (\ref{a-ad-a1-a2}) between these metrics and the baryonic
metric. Therefore, this provides a ``self-accelerated model'', in the sense that the acceleration
is not due to a hidden cosmological constant (e.g., the nonzero minimum of some potential,
or a dark-energy fluid with negligible kinetic energy).

As we wish to mimic a $\Lambda$-CDM cosmology, with $\Omega_{\Lambda 0} \simeq 0.7$,
the deviations from the Einstein-de Sitter cosmology are of order unity at low $z$.
This implies that the deviation of the coefficients $s_{\ell}$ and $s_{{\rm d}\ell}$
from their initial value is also of order unity at low $z$, while from
Eq.(\ref{phi-a-int}) we have $\bar\varphi \sim M_{\rm Pl} \sqrt{\xi}$,
\beq
z=0 : \;\; s_{\ell}-s^{(0)}_{\ell} \sim 1, \;\; s_{{\rm d}\ell}-s^{(0)}_{\ell} \sim 1 , \;\;
\frac{\bar\varphi}{M_{\rm Pl}} \sim \sqrt{\xi} .
\label{scaling-s-phi}
\eeq
As explained below, after Eq.(\ref{scaling-beta}), we cannot take $\xi$ too small
as this would give rise to a large fifth force. On the other hand, we wish to
keep the scalar-field energy density to be subdominant.
We choose for all the solutions that we consider in this paper the same scalar-field
energy density, shown in the upper panel in Fig.~\ref{fig_sym2_a1a2}.
It is of order $\Omega_{\rm dm 0}/10$ at $z=0$ and decreases at higher $z$.
The $u^{3/2}$ falloff of $\xi(a)$ is fast enough to make the scalar field subdominant
and to converge to the Einstein - de Sitter solution (\ref{EdS-0}). It is also slow enough
to enforce $ds_*/d\varphi \to 0$, as we have
$ds_*/d\varphi = (ds_*/d\ln a) / (d\varphi/d\ln a) \sim u {\cal H} \sqrt{a/\xi}$.
This yields vierbein coefficients $s_{\ell}(\varphi)$ that look somewhat more natural
than functions with a divergent slope at the origin. We can see in the lower panel
that the functions $s_*(\varphi)$ built by this procedure have simple shapes
and do not develop fine-tuned features.
The model chosen for $\xi(a)$ gives scalar-field excursions of about $M_{\rm Pl}/2$
at $z=0$.

\subsubsection{Nonsymmetric solution}
\label{sec:backg-non-symmetric}

We can also build nonsymmetric solutions, which do not obey Eq.(\ref{symm-s1-s2}).
Instead of splitting the constraint $a=s_1 a_1 + s_2 a_2$ into the two conditions
(\ref{symm-s1-s2}), we can add another condition, such as requiring the ratio $s_2/s_1$
to follow an arbitrary function of time $\kappa(a)$.
Then, the function $\kappa(a)$ parametrizes this extended family of solutions.
The symmetric solution of section~\ref{sec:backg-symmetric} corresponds to the
particular case $\kappa(a) = s_2^{(0)} / s_1^{(0)}$.
Because the effective Newton constant is given by $s_1^2+s_2^2$, in units of
${\cal G}_{\rm N}=1/8\pi M_{\rm Pl}^2$, as we shall see in Eq.(\ref{G-b}) below,
we choose instead to parametrize the solutions by the sum $s_1^2+s_2^2$,
as a function of redshift. Thus, we solve the system
\beq
s_1 a_1 + s_2 a_2 = a , \;\;\;  s_1^2+s_2^2 = \lambda(a) ,
\label{lambda-def}
\eeq
where $\lambda(a)$ is a new arbitrary function that parametrizes this extended family
of solutions.
These two equations now provide $\{s_1,s_2\}$ as a function of $\{a,a_1,a_2\}$,
\beqa
s_1 & = & \frac{a a_1 + \epsilon a_2 \sqrt{\lambda (a_1^2+a_2^2) - a^2}}{a_1^2+a_2^2} ,
\nonumber \\
s_2 & = & \frac{a a_2 - \epsilon a_1 \sqrt{\lambda (a_1^2+a_2^2) - a^2}}{a_1^2+a_2^2} ,
\label{s2-lambda}
\eeqa
where $\epsilon=\pm 1$.
Then, we can again solve for the two sets
$\{a_{\ell}(a), \omega_{\ell}(a), s_{{\rm d} \ell}(a)\}$
from Eqs.(\ref{omega-i-def}), (\ref{Friedmann-00}), and (\ref{Friedmann-ii}),
the only difference being that these two sets of variables are now coupled.

\begin{figure}
\begin{center}
\epsfxsize=8.5 cm \epsfysize=5.2 cm {\epsfbox{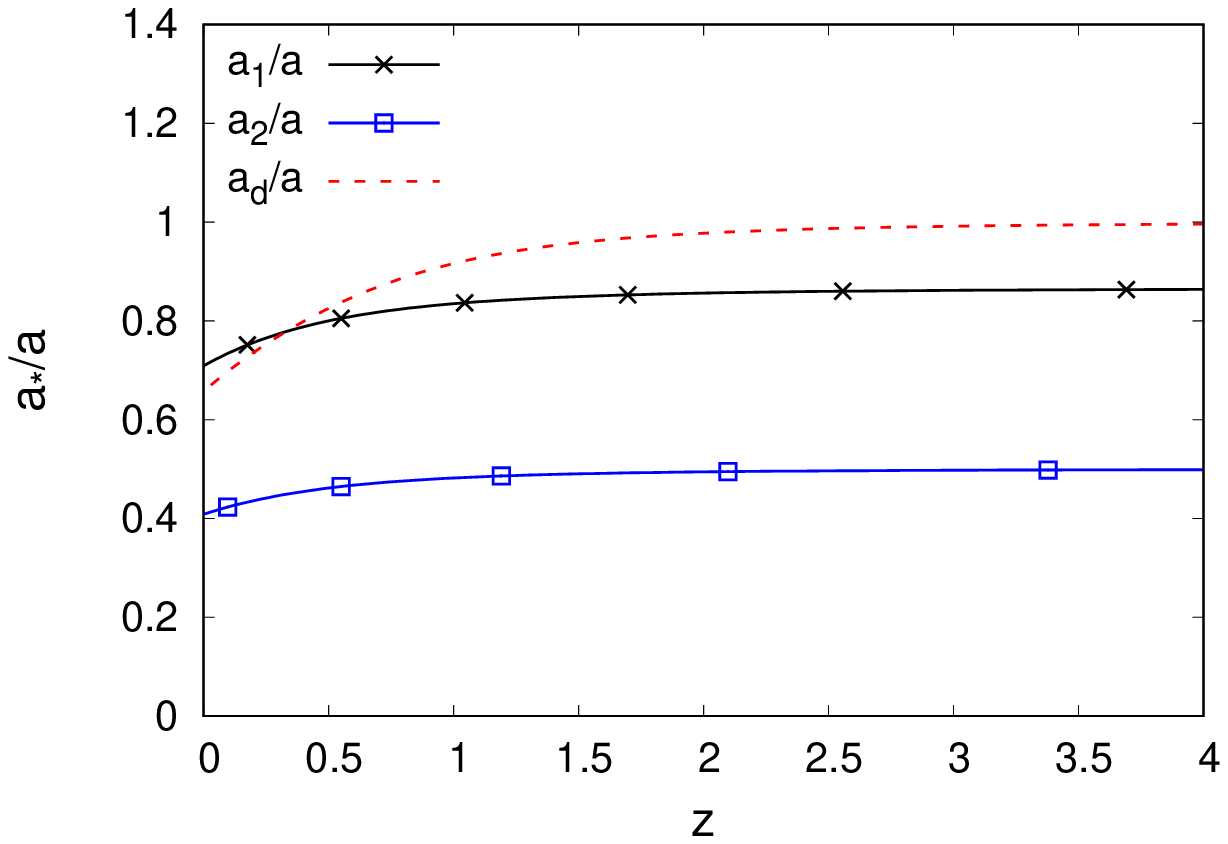}}\\
\epsfxsize=8.5 cm \epsfysize=5.2 cm {\epsfbox{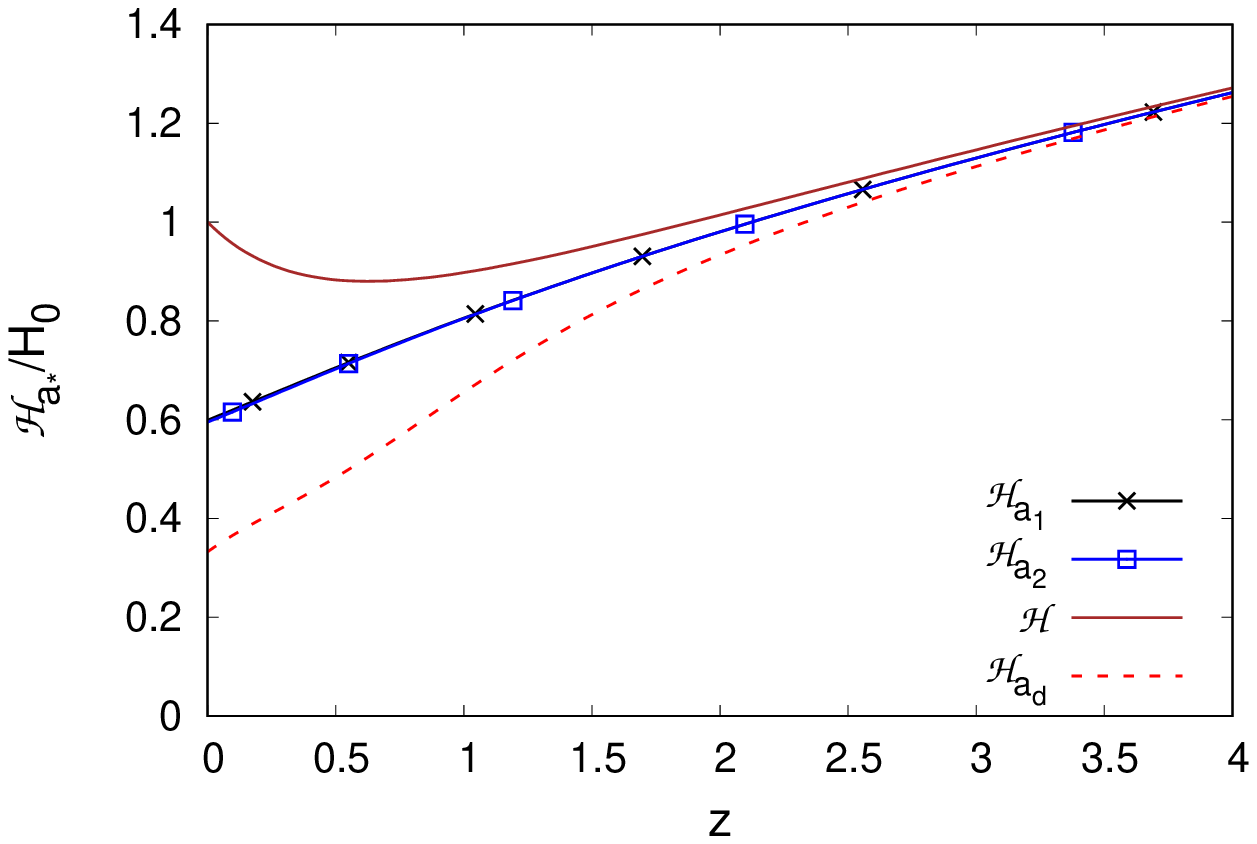}}\\
\epsfxsize=8.5 cm \epsfysize=5.2 cm {\epsfbox{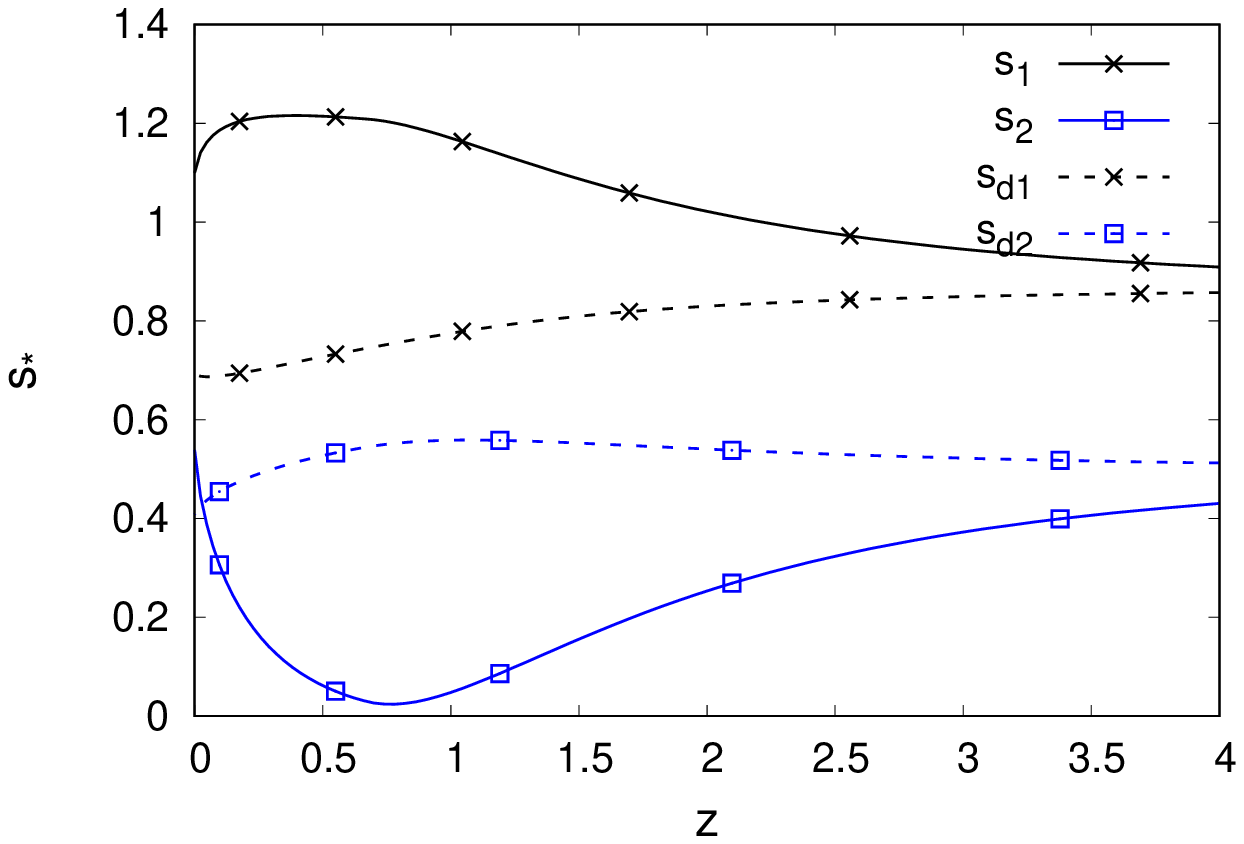}}\\
\end{center}
\caption{Background quantities for a solution of the form (\ref{lambda-def}),
where $s_2/s_1$ is not constant.}
\label{fig_lambda_a1a2}
\end{figure}

We show in Fig.~\ref{fig_lambda_a1a2} the evolution of the background quantities
for a solution of the form (\ref{lambda-def}),
where $s_2/s_1$ is no longer constant and we impose that $d\lambda/da=0$ at $z=0$.
Despite this difference, the scale factors and the Hubble expansion rates
are very close to those of Fig.~\ref{fig_sym_a1a2}.
This is because, at late times after the radiation-to-matter transition,
$a \gg a_{\rm eq}$, and for $\xi \ll 1$, the second Friedmann equation (\ref{Friedmann-ii})
reduces to
\beq
\frac{d\omega_{\ell}}{d\ln a} \sim - \frac{\Omega_{\gamma 0}}{a} - \xi , \;\;
\mbox{hence} \;\; \left| \frac{d\omega_{\ell}}{d\ln a} \right| \ll 1 .
\label{Friedmann-ii-low-z}
\eeq
Since the dark-energy era and the running of the scalar field occur much later than
the radiation-to-matter transition, we can actually see from the first Friedmann
equation Eq.(\ref{Friedmann-00}) that we must have
\beq
\omega_{\ell} \simeq  s_{\ell} \Omega_{\rm b 0} + s_{{\rm d}\ell} \Omega_{\rm dm 0}
\simeq s^{(0)}_{\ell} \Omega_{\rm b 0} + s^{(0)}_{\ell} \Omega_{\rm dm 0}  .
\label{Friedmann-00-low-z}
\eeq
Thus, we recover the relationship (\ref{s-sd-low-z}) and we also find that
for the general class of solutions with a common conformal time the quantities
$\omega_{\ell}$ are set by the initial conditions and show a negligible dependence
on the late-time evolution of the coefficients $s_{\ell}$ and $s_{{\rm d}\ell}$
and on the scalar field (as long as it remains subdominant).
This explains why we recover almost the same evolution for the scale factors $a_*$ and
the Hubble expansion rates ${\cal H}_*$, which are determined by the
definition (\ref{omega-i-def}).
Then, the deceleration parameters $q_*$ are also close to those obtained in
Fig.~\ref{fig_sym_a1a2}.
The change to the factors $s_{\ell}$ associated with different solutions is almost
fully compensated by the change to the dark coefficients $s_{{\rm d}\ell}$
that is implied by the constraint of recovering a $\Lambda$-CDM expansion rate
for the baryonic metric.
By the same mechanism, we also find that in these solutions, despite the
different behaviors of $s_1$ and $s_2$, the two gravitational metrics
are mostly equivalent, with again the same expansion rates
${\cal H}_1 \simeq {\cal H}_2$ up to negligible deviations.

\subsection{Solutions with different conformal times}
\label{sec:backg-different-conformal}

\begin{figure*}
\begin{center}
\epsfxsize=8.5 cm \epsfysize=5.2 cm {\epsfbox{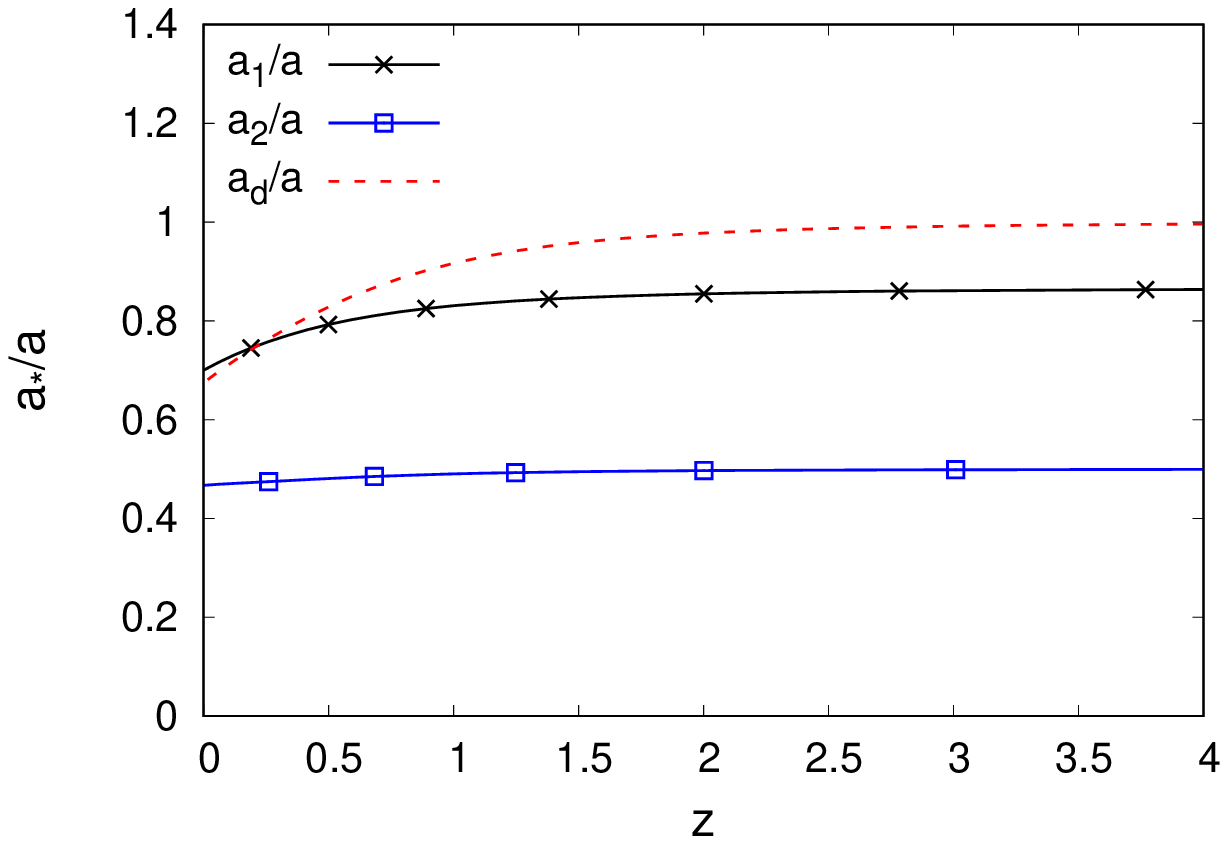}}
\epsfxsize=8.5 cm \epsfysize=5.2 cm {\epsfbox{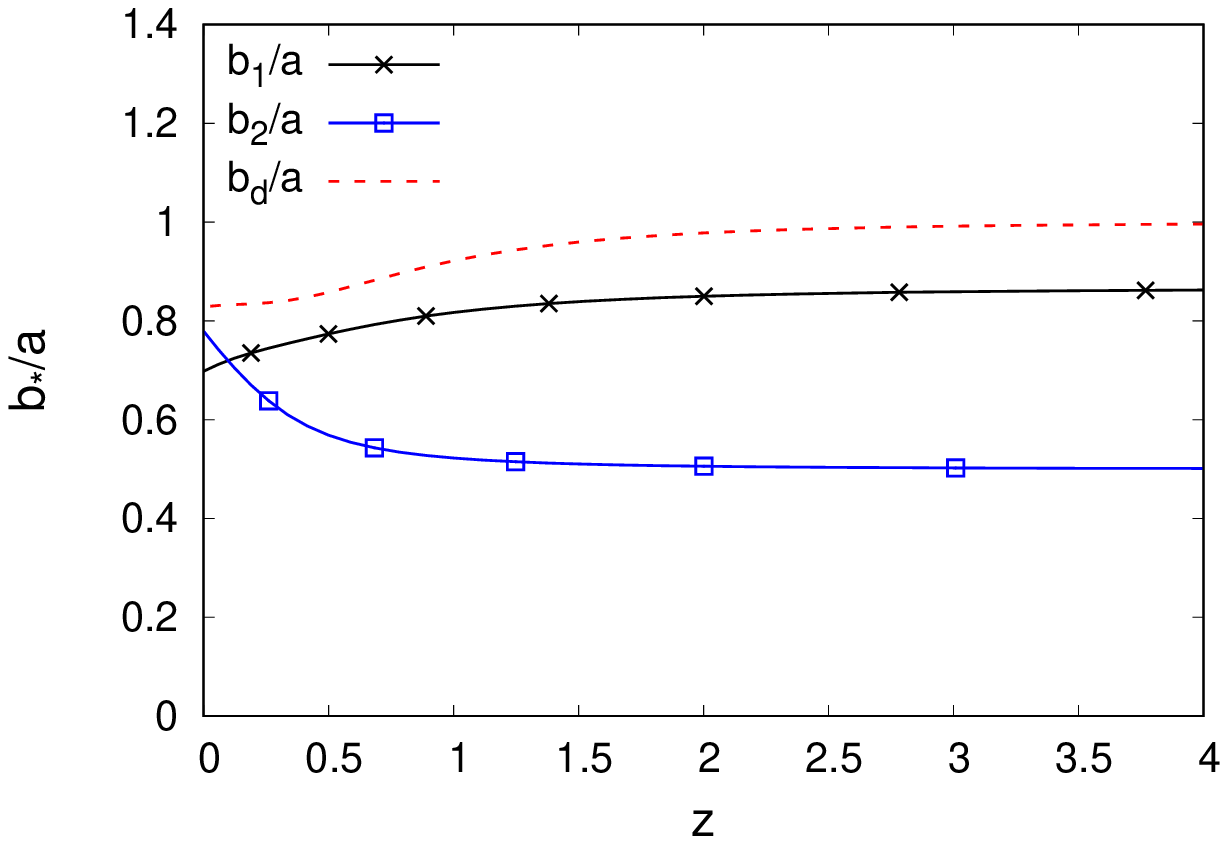}}\\
\epsfxsize=8.5 cm \epsfysize=5.2 cm {\epsfbox{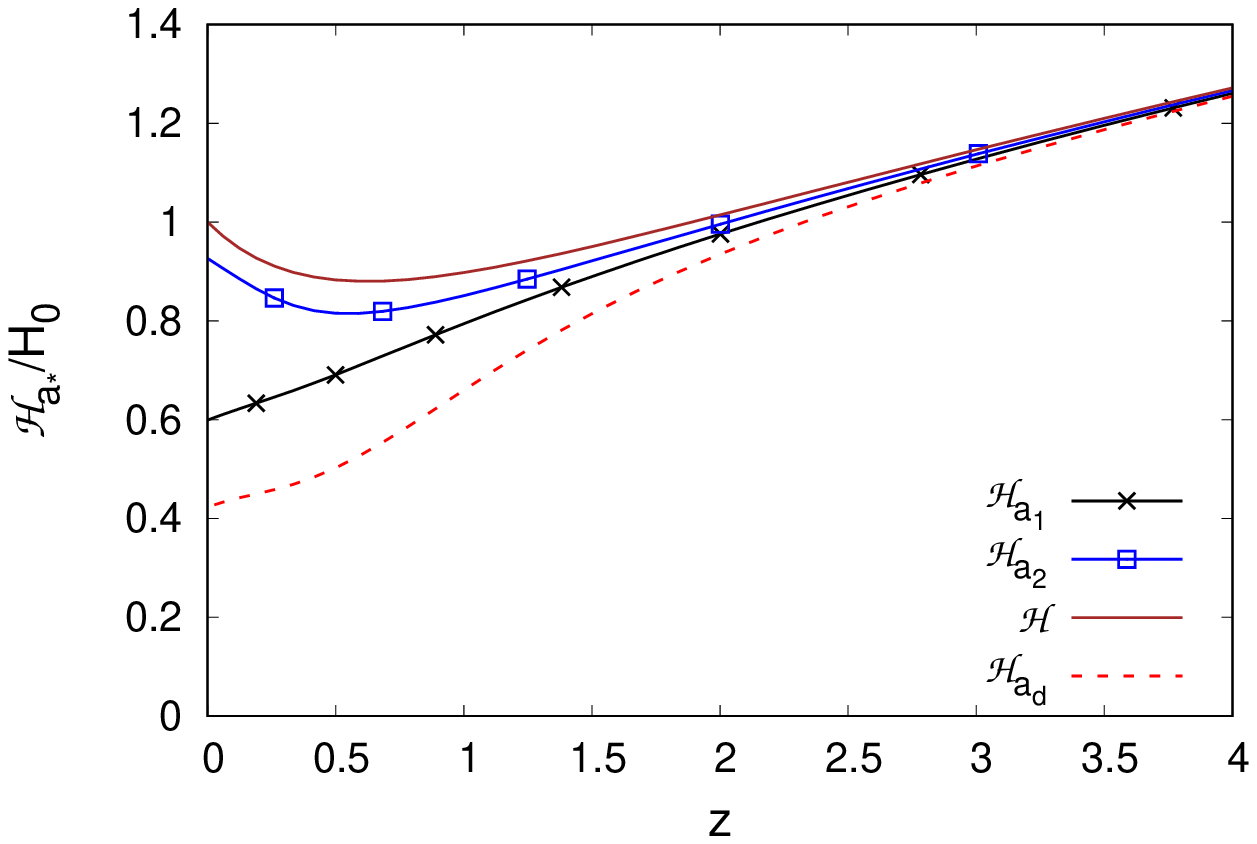}}
\epsfxsize=8.5 cm \epsfysize=5.2 cm {\epsfbox{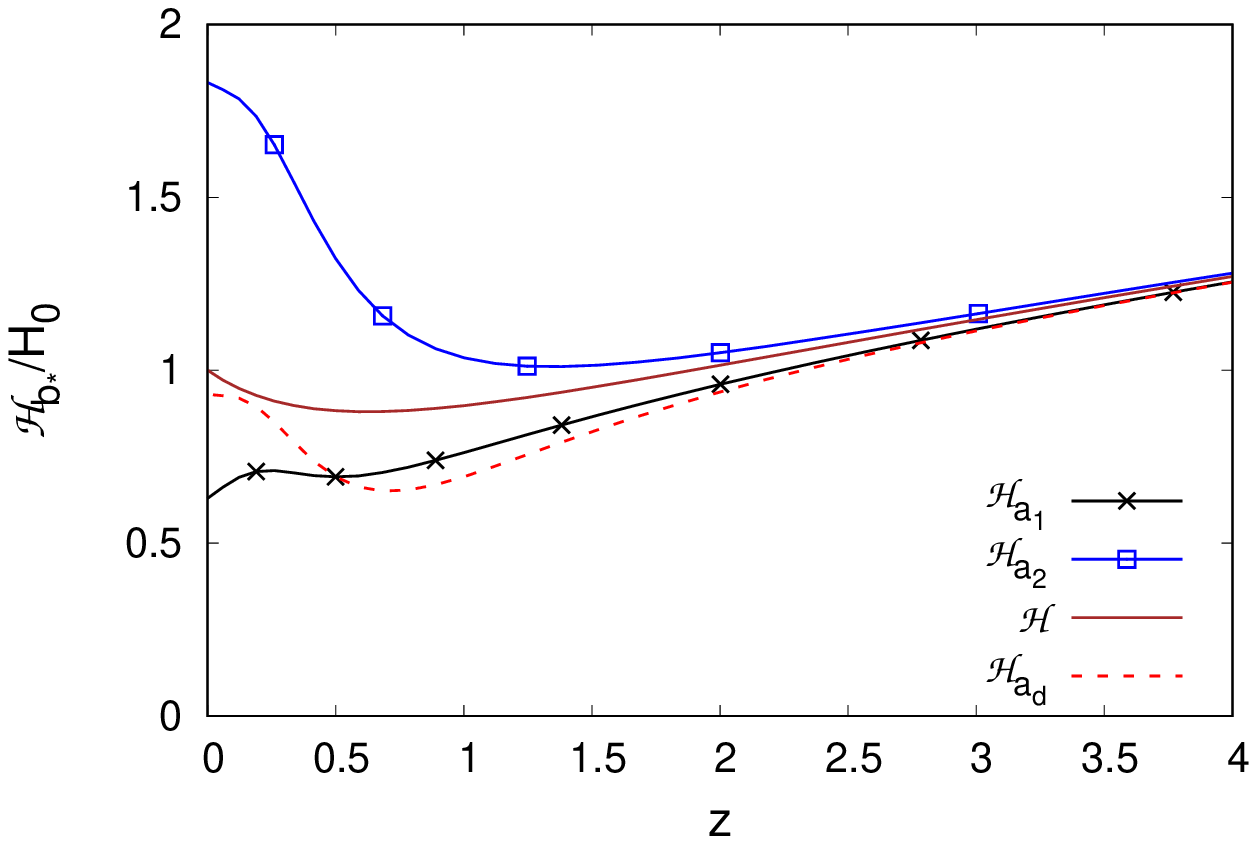}}\\
\epsfxsize=8.5 cm \epsfysize=5.2 cm {\epsfbox{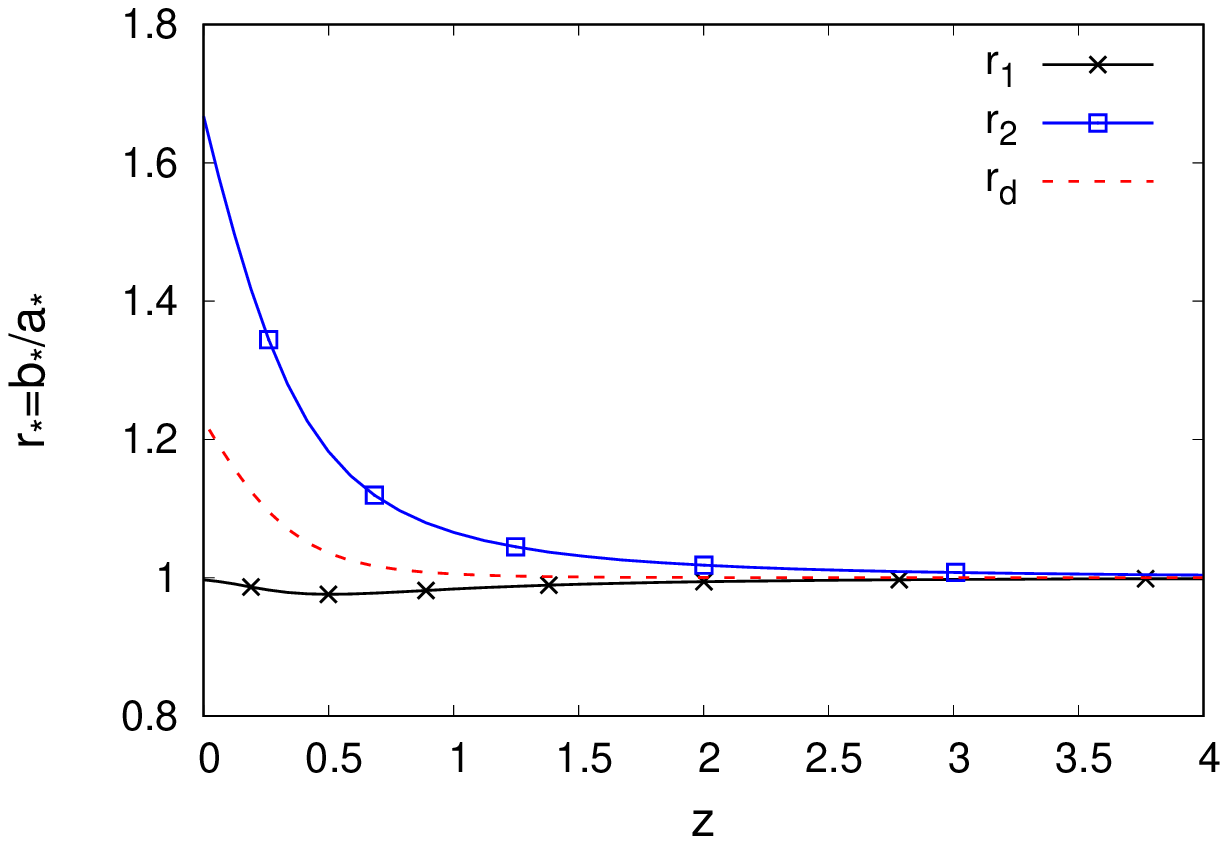}}
\epsfxsize=8.5 cm \epsfysize=5.2 cm {\epsfbox{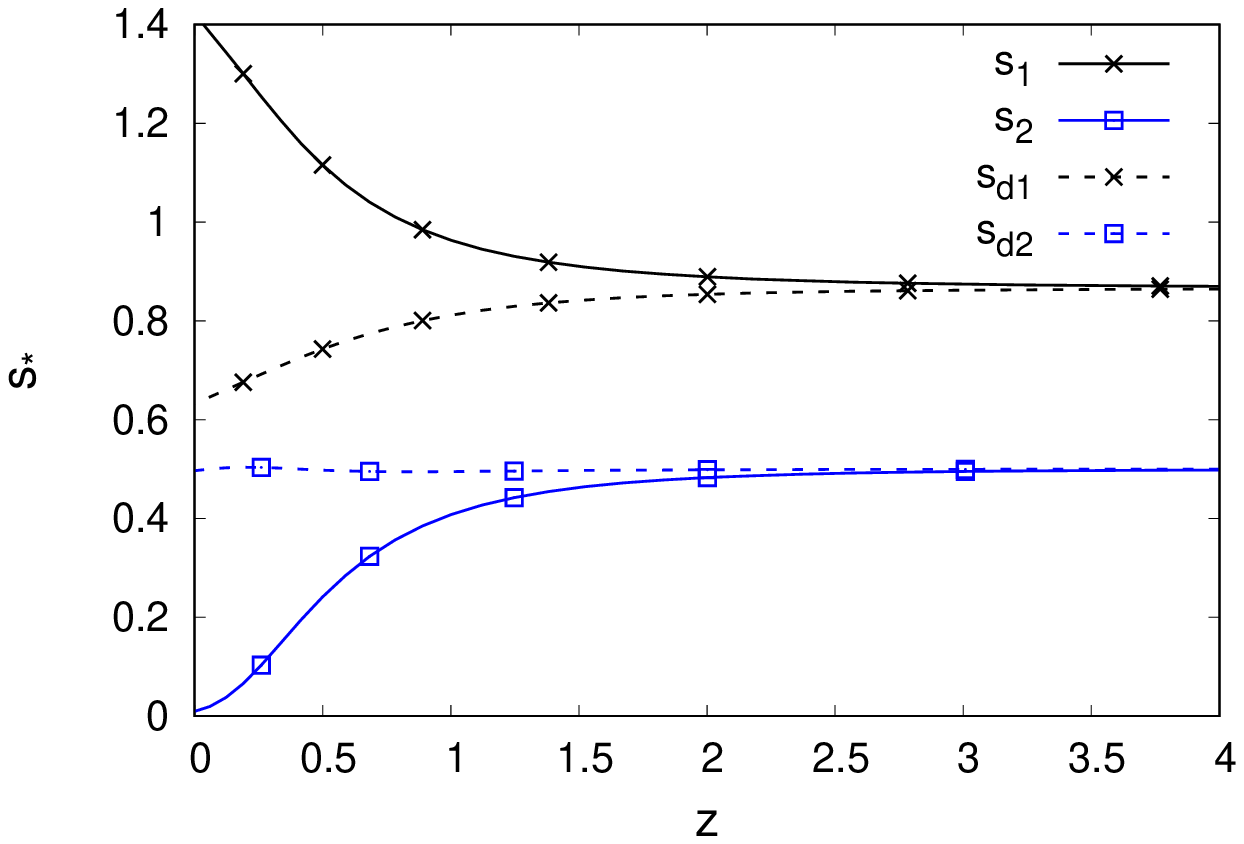}}
\end{center}
\caption{Background quantities for a solution of the form (\ref{r1-1-r2}),
where the different metrics have different conformal times (i.e., are not proportional),
but $r_1 \to 1$ at $z=0$.}
\label{fig_r_a1a2}
\end{figure*}

We now turn to even more general solutions, which still follow the $\Lambda$-CDM expansion
rate for the baryonic metric, but where
\beq
a_{\ell} \neq b_{\ell} , \;\;\; a_{\rm d} \neq b_{\rm d} .
\label{non-symm-ab}
\eeq
Then, the conformal times $\tau_*$ of the various metrics are different.
As the metrics are not proportional, already at the background level this scenario is
different from models where the baryonic and the dark matter metrics are given by
different conformal rescalings of a single Einstein-frame metric $\tilde{g}_{\mu\nu}$.

Defining the scale-factor ratios
\beq
r_{\ell}(a) = \frac{b_{\ell}}{a_{\ell}} , \;\;
r_{\rm d}(a) = \frac{b_{\rm d}}{a_{\rm d}} ,
\label{ri-def}
\eeq
the two constraints in the first line of Eq.(\ref{a-ad-a1-a2}) read as
\beq
a = s_1 a_1 + s_2 a_2 , \;\; a = s_1 r_1 a_1 + s_2 r_2 a_2 ,
\label{a-b-r}
\eeq
These two linear equations provide $\{s_{\ell}\}$ as a function of
$\{a,a_{\ell}\}$,
\beq
s_1 = \frac{a (1-r_2)}{a_1 (r_1-r_2)} , \;\;
s_2 = \frac{a (1-r_1)}{a_2 (r_2-r_1)} ,
\label{s1s2-r1r2}
\eeq
when we are given the arbitrary free functions $r_{\ell}(a)$.
As in the previous cases, $\{a_{\ell},\omega_{\ell},s_{{\rm d} \ell}\}$ are obtained
from Eqs.(\ref{omega-i-def}) and (\ref{Friedmann-00})-(\ref{Friedmann-ii}),
while $a_{\rm d}$ and $b_{\rm d}$ are obtained from the second line
of Eq.(\ref{a-ad-a1-a2}).

We require $a_*>0, b_*>0, s_{\ell}>0, s_{{\rm d}\ell}>0$, to avoid singularities.
This implies $s_1 s_2 >0$ and Eq.(\ref{s1s2-r1r2}) leads to
\beq
(1-r_1) (1-r_2) < 0 ,
\eeq
and we can choose for instance
\beq
r_1 < 1 < r_ 2 .
\label{r1-1-r2}
\eeq

The recent detections of gravitational waves from a binary neutron star merger
by the LIGO-VIRGO collaboration (GW170817) \cite{Abbott2017},
with electromagnetic counterparts in gamma-ray burst \cite{Abbott2017a}
and in UV, optical and NIR bands \cite{Cowperthwaite2017},
place very stringent limits on the speed
of gravitational waves, $|c_g-1| \leq 3\times 10^{-15}$ \cite{Abbott2017a}.
For the bimetric action (\ref{Sgrav-def}), we have two gravitons associated
with the two Einstein-Hilbert terms $R_{\ell}$.
We can obtain their equations of motion from the nonlinear Einstein equations
(\ref{Einstein-e1}), starting at the level of the vierbeins.
In the case of a constant scalar field $\varphi$ we recover the results obtained
from the quadratic action at the level of the metrics in
\cite{Comelli:2015pua,Gumrukcuoglu:2015nua,Brax:2016ssf}.
This gives for the first graviton $h_{1ij}$
\beqa
&& M_{\rm Pl}^2 \frac{a_1^2}{b_1} \left[ h''_{1ij} + (3 {\cal H}_{a_1}
- {\cal H}_{b_1} ) h'_{1ij} - \frac{b_1^2}{a_1^2} \nabla^2 h_{1ij} \right]
\nonumber \\
&& - a_2 ( \bar p s_1 s_2 a^2
+ \bar p_{\varphi} s_{\rm d1} s_{\rm d2} b_{\rm d} a_{\rm d} )
(h_{1ij} - h_{2ij} ) = 0 , \hspace{0.5cm}
\label{graviton-1}
\eeqa
and the equation of motion of the second graviton $h_{2ij}$ is given by the
permutation $1 \leftrightarrow 2$.
Here, we note $\bar p$ the total pressure of the baryonic sector fluids.
In the radiation and matter eras, this is simply the radiation pressure,
$\bar p = \bar p_{\gamma} = \bar\rho_{\gamma}/3$, while during the inflationary era,
it is the pressure $\bar p_{\chi} = - \bar\rho_{\chi}$ of the inflaton $\chi$.
We can see that the speed of the two gravitons is given by
$c_{g\ell} = b_{\ell}/a_{\ell}$, which differs from the speed of light when
$r_{\ell} \neq 1$.

To explain the multimessenger event GW170817, at least one of these two gravitons
must propagate at the speed of light (up to an accuracy of $10^{-15}$)
in the local and recent Universe,
$d \lesssim 40$ Mpc and $z \lesssim 0.01$.
In principle, a nonlinear screening mechanism might change the laws of gravity
and ensure convergence to General Relativity in the local environment. However,
it is unlikely that it would apply over 40 Mpc. Moreover,
in most parts of the trajectory, between the host galaxy and the Milky Way,
the local density is below or of the order of the cosmological background density.
Besides, it would require a fine-tuned cancellation to make the average speed
$c_g = 1$ over the full trajectory, inside the two galaxies and the low-density
intergalactic medium.
Then, at least one of the lapse factors $r_{\ell}$ must converge to unity
at low $z$. If both coefficients $r_{\ell}$ go to unity, we converge to the
solutions studied in section~\ref{sec:backg-common-conformal}.
For illustration, we consider in Fig.~\ref{fig_r_a1a2} the case where only one
of the coefficients $r_{\ell}$ goes to unity at low redshift, for instance $r_1$
[with again the same initial conditions $\{s_1^{(0)},s_2^{(0)}\}$ and scalar-field energy
density $\xi(a)$ as in Fig.~\ref{fig_sym_a1a2}].
In this limit, the system effectively reduces again to a single metric for the baryonic
sector. Indeed, Eq.(\ref{s1s2-r1r2}) implies that $s_2 \to 0$ if $r_1 \to 1$
(and $s_1 \to 0$ if $r_2 \to 1$). Then, the baryonic metric $g_{\mu\nu}$ becomes
proportional to the metric $g_{1\mu\nu}$. However, the dark matter metric remains
sensitive to both gravitational metrics $g_1$ and $g_2$, as $s_{\rm d 2}$ remains
nonzero, so that the baryon+dark matter system remains different from the common
conformal time scenarios of section~\ref{sec:backg-common-conformal}.
In particular, the baryonic and dark matter metrics are not proportional, so that
this scenario remains different from models where the baryonic and the dark matter
metrics are given by different conformal rescalings of a single Einstein-frame metric
$\tilde{g}_{\mu\nu}$.

We can see that in this scenario the scale factors $a_*$ remain similar to those
obtained in Fig.~\ref{fig_sym_a1a2} for the symmetric solution (\ref{symm-s1-s2}).
However, we can now distinguish the difference between the two expansion
rates ${\cal H}_{a_1}$ and ${\cal H}_{a_2}$ at low $z$.
The main difference with respect to the previous solutions is the behavior of the
lapse functions $b_{\ell}$. Thanks to the additional degrees of freedom
$r_{\ell}$, the lapses $b_{\ell}$ can behave in a significantly different way
than the scale factors $a_{\ell}$.
In the example shown in Fig.~\ref{fig_r_a1a2}, the two lapses even evolve
in different directions and cross each other at $z \simeq 0.1$. This leads to
rates that are significantly different with ${\cal H}_{b_2} > {\cal H}$.
As explained in Section~\ref{sec:backg-symmetric}, because of the lack of cosmological
constant, the gravitational expansion rates ${\cal H}_{a_{\ell}}$ are typically smaller
than the $\Lambda$-CDM expansion rate ${\cal H}$.
This remains true for the more general solution shown in Fig.~\ref{fig_r_a1a2}.
But the lapse functions are not so strongly constrained and it is possible to have
one of them growing faster than $a$. For the choice (\ref{r1-1-r2}) this corresponds
to $b_2$, with ${\cal H}_{b_2} > {\cal H}$.
This requires a ratio $r_2$ that significantly departs from unity at low $z$,
as seen in the lower left panel.

The coefficients $s_{\ell}$ and $s_{{\rm d}\ell}$
follow similar behaviors to those obtained in Figs.~\ref{fig_sym_a1a2}
and \ref{fig_lambda_a1a2}, with opposite deviations at low redshift for the baryonic and
dark sector coefficients.
Because of the constraint $| r_1 -1 | < 3\times 10^{-15}$ at $z=0$, the coefficient $s_2$
almost goes to zero, with $s_2 \lesssim 10^{-15}$ at $z=0$.

\section{Cosmological perturbations}
\label{sec:linear}

We have seen that it is possible to build several families of solutions that
follow a $\Lambda$-CDM expansion history for the baryonic metric.
In the case of metrics that are not proportional, the multimessenger neutron star
merger GW170817 also implies that at least one of the two gravitational metrics,
$g_1$ and $g_2$, becomes proportional to the baryonic metric (i.e., $r_{\ell}=1$)
at low redshift.

We show below that these models are actually severely constrained by the behavior
of perturbations.
Here we focus on the scalar perturbations in the quasistatic approximation,
which applies to the formation of large-scale structures.
Then, the relevant metric perturbations are set by the four gravitational potentials
$\{\phi_{\ell},\psi_{\ell}\}$ as in the usual Newtonian gauge.

\subsection{Scalar-field perturbations}
\label{sec:pert-scalar}

On small scales in the quasistatic approximation, the Klein-Gordon equation
(\ref{KG-1}) becomes
\beq
\frac{1}{a_{\rm d}^2} \nabla^2 \delta\varphi = m^2 \delta\varphi
+ \frac{\beta_{\rm dm}}{M_{\rm Pl}} \delta\rho_{\rm dm}
+ \frac{\beta}{M_{\rm Pl}} \delta\rho ,
\label{KG-QS}
\eeq
with $\delta\varphi = \varphi-\bar\varphi$,
$\delta\rho_{\rm dm} = \rho_{\rm dm}-\bar\rho_{\rm dm}$, and
$\delta\rho = \rho-\bar\rho$.
Here, we assumed nonrelativistic matter components, $p_{\rm dm}=p=0$, and
we neglected radiation fluctuations.
As $\delta\rho_{\varphi} = \delta p_{\varphi} = b_{\rm d}^{-2} \frac{d\bar\varphi}{d\tau}
\frac{\partial\delta\varphi}{\partial\tau}$, we also neglected the linear fluctuations
of the scalar field density and pressure in the quasistatic limit.
The scalar-field mass around the cosmological background is
\beqa
&& m^2 = ( \bar\rho_{\rm dm}+\bar\rho_{\varphi} ) \sum_{\ell}
\frac{d^2s_{{\rm d}\ell}}{d\varphi^2} \frac{b_{\ell}}{b_{\rm d}}
- 3 \bar\rho_{\varphi} \sum_{\ell} \frac{d^2s_{{\rm d}\ell}}{d\varphi^2}
\frac{a_{\ell}}{a_{\rm d}} \nonumber \\
&& + ( \bar\rho+\bar\rho_{\gamma} ) \sum_{\ell} \frac{d^2s_{\ell}}{d\varphi^2}
\frac{a^3 b_{\ell}}{a_{\rm d}^3 b_{\rm d}}
- \bar\rho_{\gamma} \sum_{\ell} \frac{d^2s_{\ell}}{d\varphi^2}
\frac{a^3 a_{\ell}}{a_{\rm d}^3 b_{\rm d}} .
\label{m2-def}
\eeqa
Using the relation (\ref{ds-dsd}), it is possible to express the dark sector derivatives
$d^2 s_{{\rm d}\ell}/d\varphi^2$ and $d s_{{\rm d}\ell}/d\varphi$
in terms of $d^2 s_{\ell}/d\varphi^2$ and $d s_{\ell}/d\varphi$.
It is then  possible to remove the second derivatives $d^2 s_{\ell}/d\varphi^2$
thanks to the symmetry in $\ell=1,2$, using the relations obtained by taking derivatives
with respect to $\ln a$ of the constraints $a=s_1 b_1+s_2 b_2$ and $a=s_1 a_1+s_2 a_2$.
The couplings to matter are
\beq
\beta = M_{\rm Pl} \sum_{\ell} \frac{ds_{\ell}}{d\varphi}
\frac{a^3 b_{\ell}}{a_{\rm d}^3 b_{\rm d}} , \;\;\;
\beta_{\rm dm} = M_{\rm Pl} \sum_{\ell} \frac{ds_{{\rm d}\ell}}{d\varphi}
\frac{b_{\ell}}{b_{\rm d}} .
\label{betad-beta-def}
\eeq
In Fourier space, this yields
\beq
\frac{\delta\varphi}{M_{\rm Pl}} = - \frac{3 H_0^2}{m^2+k^2/a_{\rm d}^2}
\left[ \frac{\Omega_{\rm dm 0} \beta_{\rm dm} \delta_{\rm dm}}{a_{\rm d}^3} +
\frac{\Omega_{\rm b 0} \beta \delta}{a^3} \right] ,
\label{KG-k}
\eeq
where $\delta_{\rm dm}=\delta\rho_{\rm dm}/\bar\rho_{\rm dm}$,
$\delta=\delta\rho/\bar\rho$.

\begin{figure}
\begin{center}
\epsfxsize=8.5 cm \epsfysize=5.2 cm {\epsfbox{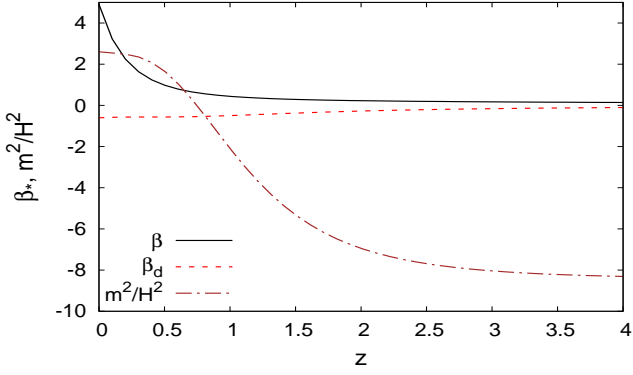}}
\end{center}
\caption{Scalar-field mass and couplings for the symmetric model of
Fig.~\ref{fig_sym_a1a2}.}
\label{fig_sym_beta_m2}
\end{figure}

It is interesting to consider the scaling in $\xi$ of the scalar-field mass and
couplings.
From Eq.(\ref{scaling-s-phi}) we have the scalings
\beqa
&& \frac{ds_*}{d\ln a} \sim 1 , \;\;\;
\frac{ds_*}{d\varphi} \sim \frac{1}{M_{\rm Pl} \sqrt{\xi}} , \;\;\;
\frac{d^2s_*}{d\varphi^2} \sim \frac{1}{M_{\rm Pl}^2 \xi} . \;\;\;
\label{scaling-ds}
\eeqa
Then, from Eq.(\ref{m2-def}) it seems that $m^2 \sim H_0^2/\xi$.
However, using the relationship (\ref{ds-dsd}) one finds that the terms of order
$1/\xi$ cancel out and we obtain
\beq
m^2 \sim H_0^2 ( \Omega_{\gamma 0} + \xi ) .
\label{scaling-m2}
\eeq
On the other hand, the couplings scale as
\beq
\beta \sim \frac{1}{\sqrt{\xi}} , \;\;\; \beta_{\rm dm} \sim \frac{1}{\sqrt{\xi}} .
\label{scaling-beta}
\eeq
Therefore, very small values of the scalar-field energy density $\xi$ yield
a very large fifth force.
This implies that we cannot take $\xi$ too small, which is why we choose
$\xi \sim \Omega_{\rm dm 0}/10$ at $z=0$ in the models that we consider in this paper.
This feature comes from the fact that we require effects of order one
from the scalar field onto the background at low redshift, $ds_*/d\ln a \sim 1$,
to generate the apparent acceleration of the baryonic metric.
This implies $ds_*/d\varphi \propto 1/\bar\varphi' \propto 1/\sqrt{\xi}$.

We show in Fig.~\ref{fig_sym_beta_m2} the scalar-field mass and couplings
for the symmetric model of Fig.~\ref{fig_sym_a1a2}.
As expected from the expression (\ref{m2-def}), the squared mass evolves
as $\bar\rho/M_{\rm Pl}^2 \sim H^2$ and it is of order $H^2$.
This means that it is negligible on scales much below the horizon, where the
quasistatic approximation (\ref{KG-QS}) applies, and does not lead to small-scale
instabilities, even when it is negative.
The couplings $\beta$ and $\beta_{\rm d}$ are of order unity and decrease at high
$z$, because $ds_*/d\varphi \to 0$. This is because we choose the
high-$z$ decay of the scalar-field energy density, determined by Eq.(\ref{xi-u-def}),
to be slow enough so that $ds_*/d\varphi \to 0$ at early times.
The baryonic and dark matter couplings have opposite signs, with
$\beta > 0 > \beta_{\rm dm}$, because we typically have $ds_{\ell}/d\ln a>0$
and $ds_{{\rm d}\ell}/d\ln a<0$, as explained in
Section~\ref{sec:backg-common-conformal} and in agreement with
Eq.(\ref{s-sd-low-z}).

The other solutions considered in Sections~\ref{sec:backg-common-conformal}
and \ref{sec:backg-different-conformal} give results similar to those found in
Fig.~\ref{fig_sym_beta_m2}.

\subsection{Einstein equations}
\label{sec:pert-Einstein}

\subsubsection{Gravitational potentials $\phi_{\ell}$ and $\psi_{\ell}$}

We study in details the behavior of linear perturbations in section~\ref{sec:freedom-linear}
below, and we provide in Appendix~\ref{sec:linear-general} explicit expressions of
the Einstein equations in the case $a_{\ell}=b_{\ell}$.
The extra two scalars added to the four Newtonian potentials that cannot be eliminated
by gauge freedom (because of the loss of the nondiagonal diffeomorphism invariance)
are not dynamical \cite{Comelli:2015pua} and there is no scalar instability.
In this section, we focus on small subhorizon scales, $k \gg {\cal H}$, in the quasistatic
approximation, where we only keep the higher-order spatial gradients.
Then, as in General Relativity, only the four gravitational potentials
$\{\phi_{\ell},\psi_{\ell}\}$ remain.
The perturbed metrics take the usual form
\beq
g_{*00} = -b_*^2 (1+2\phi_*), \;\;  g_{*ii} = a_*^2 (1-2\psi_*) ,
\eeq
while the vierbeins are diagonal with
\beq
e^0_{*0} = b_* (1+\phi_*), \;\; e^i_{*i} = a_* (1-\psi_*) .
\label{vierbein-phi}
\eeq

For nonrelativistic matter components, the (0,0) component of the Einstein equations
(\ref{Einstein-e1}) gives for the metric $g_{1\mu\nu}$
\beqa
&& \frac{2 a_1}{3 H_0^2} \nabla^2 \psi_1 =
s_1 \Omega_{\rm b 0} \delta
+ s_{\rm d 1} \Omega_{\rm dm 0} \delta_{\rm dm} + \Upsilon_{\!\psi_1}
\frac{d\ln a}{d\bar\varphi} \delta\varphi , \nonumber \\
&&
\label{Poisson-psi-1}
\eeqa
with
\beqa
&&\Upsilon_{\!\psi_1} = \left( \Omega_{\rm b 0} + \frac{\Omega_{\gamma 0}}{a} \right)
\left[ \left( 1+s_1\frac{b_1}{a} \right) \frac{ds_1}{d\ln a}
+ s_1\frac{b_2}{a}\frac{ds_2}{d\ln a} \right]  \nonumber \\
&& + \left( \Omega_{\rm dm 0} + \xi \right) \left[ \left(
1+s_{\rm d1}\frac{b_1}{b_{\rm d}} \right) \frac{ds_{\rm d1}}{d\ln a}
+ s_{\rm d1}\frac{b_2}{b_{\rm d}}\frac{ds_{\rm d2}}{d\ln a} \right] . \hspace{1cm}
\label{Upsilon-psi}
\eeqa
The $(i,j)$ components of the Einstein equations give
\beqa
&& \frac{b_1}{H_0^2} \left[ - \partial_i\partial_j (\phi_1-\psi_1)
+ \delta_{ij} \nabla^2 (\phi_1-\psi_1) \right] =  \Upsilon_{\!\phi_1}
\frac{d\ln a}{d\bar\varphi} \delta\varphi , \nonumber \\
&&
\label{Poisson-phi-1}
\eeqa
with
\beqa
&&\Upsilon_{\!\phi_1} = \frac{\Omega_{\gamma 0}}{a}
\left[ \left( 1+s_1\frac{a_1}{a} \right) \frac{ds_1}{d\ln a}
+ s_1\frac{a_2}{a}\frac{ds_2}{d\ln a} \right]  \nonumber \\
&& + 3 \xi r_{\rm d} \left[ \left(
1+s_{\rm d1}\frac{a_1}{a_{\rm d}} \right) \frac{ds_{\rm d1}}{d\ln a}
+ s_{\rm d1}\frac{a_2}{a_{\rm d}}\frac{ds_{\rm d2}}{d\ln a} \right] . \hspace{1cm}
\label{Upsilon-phi}
\eeqa
We can use the Klein-Gordon equation (\ref{KG-k}) satisfied by the scalar field
to eliminate $\delta\varphi$.
In Fourier space, this gives
\beqa
- \frac{2}{3} a_{\ell} \frac{k^2}{H_0^2} \psi_{\ell} & = & \left( 1+\gamma^{\psi_{\ell}} \right) s_{\ell} \Omega_{\rm b 0} \delta \nonumber \\
&& + \left( 1+\gamma^{\psi_{\ell}}_{\rm dm} \right) s_{{\rm d}\ell} \Omega_{\rm dm 0}
\delta_{\rm dm}
\label{Poisson-psi-2}
\eeqa
and
\beqa
&& - a_{\ell} \frac{k_i k_j-\delta_{ij}k^2}{H_0^2} (\phi_{\ell}-\psi_{\ell}) =
\gamma^{\phi_{\ell}} s_{\ell} \Omega_{\rm b 0} \delta \nonumber \\
&& \hspace{4cm} +  \gamma^{\phi_{\ell}}_{\rm dm} s_{{\rm d}\ell} \Omega_{\rm dm 0}
\delta_{\rm dm} .
\label{Poisson-phi-2}
\eeqa
The coefficients $\gamma^*_*$ arise from the fluctuations of the scalar field
$\varphi$, which generate fluctuations $\delta s_*$ of the vierbein coefficients
$s_*$ that relate the matter and gravitational metrics.
They are given by
\beqa
\gamma^{\psi_{\ell}} & = & - \frac{{\cal H}}{H_0 r_{\rm d} a^3 s_{\ell}}
\sqrt{\frac{3 a_{\rm d}}{2\xi}} \frac{\beta H_0^2}{m^2+k^2/a_{\rm d}^2}
\Upsilon_{\psi_{\ell}} , \nonumber \\
\gamma^{\psi_{\ell}}_{\rm dm} & = & - \frac{{\cal H}}
{H_0 r_{\rm d} a_{\rm d}^3 s_{{\rm d}\ell}} \sqrt{\frac{3 a_{\rm d}}{2\xi}}
\frac{\beta_{\rm dm} H_0^2}{m^2+k^2/a_{\rm d}^2}
\Upsilon_{\psi_{\ell}} , \nonumber \\
\gamma^{\phi_{\ell}} & = & \frac{{\cal H}}{H_0 r_{\rm d} a^3 r_{\ell} s_{\ell}}
\sqrt{\frac{3 a_{\rm d}}{2\xi}} \frac{\beta H_0^2}{m^2+k^2/a_{\rm d}^2}
\Upsilon_{\phi_{\ell}} , \nonumber \\
\gamma^{\phi_{\ell}}_{\rm dm} & = & \frac{{\cal H}}
{H_0 r_{\rm d} a_{\rm d}^3 r_{\ell} s_{{\rm d}\ell}} \sqrt{\frac{3 a_{\rm d}}{2\xi}}
\frac{\beta_{\rm dm} H_0^2}{m^2+k^2/a_{\rm d}^2}
\Upsilon_{\phi_{\ell}} ,
\label{gamma-psi1-b}
\eeqa
where the factors $\Upsilon_{\psi_{\ell}}$ and $\Upsilon_{\phi_{\ell}}$ are given
in Eqs.(\ref{Upsilon-psi}) and (\ref{Upsilon-phi}).
The contribution from the fifth force to the gravitational potentials
$\psi_{\ell}$ and $\phi_{\ell}$ is negligible if the coefficients $\gamma^*_*$
are much smaller than unity.
Then, we recover Einstein equations for these gravitational potentials that are close
to their standard form,
\beq
\left| \gamma^*_*\right|  \ll 1 : \;\;\;
\bea{l} \phi_{\ell} \simeq \psi_{\ell} \\ \\
- \frac{2}{3} a_{\ell} \frac{k^2}{H_0^2} \psi_{\ell} \simeq s_{\ell} \Omega_{\rm b 0}
\delta + s_{{\rm d}\ell} \Omega_{\rm dm 0} \delta_{\rm dm} \ea
\label{Poisson-ell}
\eeq

\begin{figure}
\begin{center}
\epsfxsize=8.5 cm \epsfysize=5.2 cm {\epsfbox{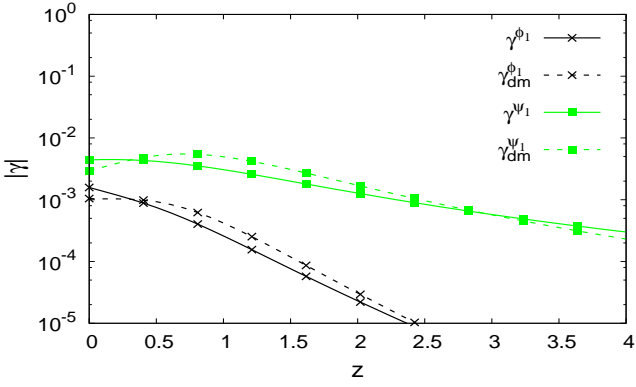}}
\end{center}
\caption{Absolute value of the coefficients $\gamma^*_*$ for the symmetric model of
Fig.~\ref{fig_sym_a1a2}, at comoving wave number $k(z)=10 {\cal H}(z)$.}
\label{fig_sym_gamma}
\end{figure}

We show in Fig.~\ref{fig_sym_gamma} the coefficients $\gamma^*_*$
for the symmetric solution of Fig.~\ref{fig_sym_a1a2}, at comoving wave number
$k(z)=10 {\cal H}(z)$.
At $z=0$, we expect from Eqs.(\ref{gamma-psi1-b})
that $|\gamma^*_*| \simeq (H_0/k)^2$ on small scales.
Indeed, we can see in the figure that for $k=10 {\cal H}$ we have
$|\gamma^*_*| \lesssim 10^{-2}$. Moreover, the amplitude shows a fast decrease
at higher $z$. Therefore, on subhorizon scales the coefficients $\gamma^*_*$
are much smaller than unity at all redshifts and we can always use the
approximations (\ref{Poisson-ell}).

The other solutions considered in Sections~\ref{sec:backg-common-conformal}
and \ref{sec:backg-different-conformal} give results similar to those found in
Fig.~\ref{fig_sym_gamma}.

\subsubsection{Baryonic gravitational potentials $\phi$ and $\psi$}

In the following, we assume that the properties (\ref{Poisson-ell}) are satisfied.
However, this is not sufficient to remove the fifth force because the dynamics of
dark matter and baryons are set by their own metric potentials $\phi_{\rm d}$ and
$\phi$. Their relationship with the potentials $\phi_{\ell}$ involves the scalar field
and will give rise to a fifth force.
Indeed, from the vierbeins (\ref{vierbein-phi}) and their relations
(\ref{vierbein-db-def}), we obtain at linear order
\beqa
a \phi & = & \sum_{\ell} b_{\ell} \left( s_{\ell} \phi_{\ell} + \delta s_{\ell} \right) ,
\nonumber \\
a \psi = & = & \sum_{\ell} a_{\ell} \left( s_{\ell} \psi_{\ell}
- \delta s_{\ell} \right) .
\label{psi-psil}
\eeqa
As for the gravitational potentials $\phi_{\ell}$ and $\psi_{\ell}$,
the fluctuations of the coefficients $s_{\ell}$ and $s_{{\rm d}\ell}$, due to the
perturbations of the scalar field $\delta\varphi$, give rise to nonstandard terms.
Using Eq.(\ref{Poisson-ell}), we obtain
\beqa
- \frac{2}{3} a \frac{k^2}{H_0^2} \phi & = & \mu^{\phi} \Omega_{\rm b 0}
\delta + \mu^{\phi}_{\rm dm} \Omega_{\rm dm 0} \delta_{\rm dm} ,
\nonumber \\
- \frac{2}{3} a \frac{k^2}{H_0^2} \psi & = & \mu^{\psi} \Omega_{\rm b 0}
\delta + \mu^{\psi}_{\rm dm} \Omega_{\rm dm 0} \delta_{\rm dm} ,
\label{Poisson-psib-1}
\eeqa
with
\beqa
&& \hspace{-0.5cm} \mu^{\phi} = \sum_{\ell} \left[ s_{\ell}^2 r_{\ell} +
\frac{{\cal H}a_{\rm d}^2}{H_0 r_{\rm d} a^3}
\sqrt{\frac{2 a_{\rm d}}{3\xi}} \frac{\beta k^2}{k^2+a_{\rm d}^2 m^2}
\frac{ds_{\ell}}{d\ln a} b_{\ell}  \right]
\nonumber \\
&& \hspace{-0.5cm} \mu^{\phi}_{\rm dm} = \sum_{\ell} \left[ s_{\ell} s_{{\rm d}\ell} r_{\ell} +
\frac{{\cal H}}{H_0 r_{\rm d} a_{\rm d}}
\sqrt{\frac{2 a_{\rm d}}{3\xi}} \frac{\beta_{\rm dm} k^2}{k^2+a_{\rm d}^2 m^2}
\frac{ds_{\ell}}{d\ln a} b_{\ell}  \right]
\nonumber \\
&& \hspace{-0.5cm} \mu^{\psi} = \sum_{\ell} \left[ s_{\ell}^2 -
\frac{{\cal H}a_{\rm d}^2}{H_0 r_{\rm d} a^3}
\sqrt{\frac{2 a_{\rm d}}{3\xi}} \frac{\beta k^2}{k^2+a_{\rm d}^2 m^2}
\frac{ds_{\ell}}{d\ln a} a_{\ell}  \right]
\nonumber \\
&& \hspace{-0.5cm} \mu^{\psi}_{\rm dm} = \sum_{\ell} \left[ s_{\ell} s_{{\rm d}\ell}  -
\frac{{\cal H}}{H_0 r_{\rm d} a_{\rm d}}
\sqrt{\frac{2 a_{\rm d}}{3\xi}} \frac{\beta_{\rm dm} k^2}{k^2+a_{\rm d}^2 m^2}
\frac{ds_{\ell}}{d\ln a} a_{\ell}  \right] .
\nonumber \\
&& \label{mu-psi-dm}
\eeqa
We recover the standard Poisson equations if $\mu^*_*=1$.

We can split the coefficients $\mu^*_*$ into two parts. The first term, of the form
$s_{\ell}^2 r_{\ell}$, is similar to a scale-independent renormalized Newton's constant
and arises from the coefficients $s_{\ell}$ that relate the various metric potentials.
The second part, of the form $ds_{\ell}/d\ln a$, arises from the fluctuations of the
scalar field through $\delta s_{\ell}$, and corresponds to a fifth force.
It is scale dependent.
Thus, we may define the renormalized Newton's constants (in units of the
natural Newton's constant, ${\cal G}_{\rm N}=1/8\pi M_{\rm Pl}^2$),
\beqa
&& {\cal G}^{\phi} = \sum_{\ell} s_{\ell}^2 r_{\ell} , \;\;\;
{\cal G}^{\phi}_{\rm dm} = \sum_{\ell} s_{\ell} s_{{\rm d}\ell} r_{\ell} , \;\;\;
{\cal G}^{\psi} = \sum_{\ell} s_{\ell}^2 , \nonumber \\
&& {\cal G}^{\psi}_{\rm dm} = \sum_{\ell} s_{\ell} s_{{\rm d}\ell} ,
\label{G-b}
\eeqa
which are all positive.

The two baryonic metric potentials $\phi$ and $\psi$ are generically different.
First, if $r_{\ell} \neq 1$ the associated effective Newton's constants
${\cal G}^{\phi}$ and ${\cal G}^{\psi}$ are different.
Second, the fifth-force contributions that enter $\phi$ and $\psi$ have the same
amplitude but opposite signs.

\begin{figure}
\begin{center}
\epsfxsize=8.5 cm \epsfysize=5.2 cm {\epsfbox{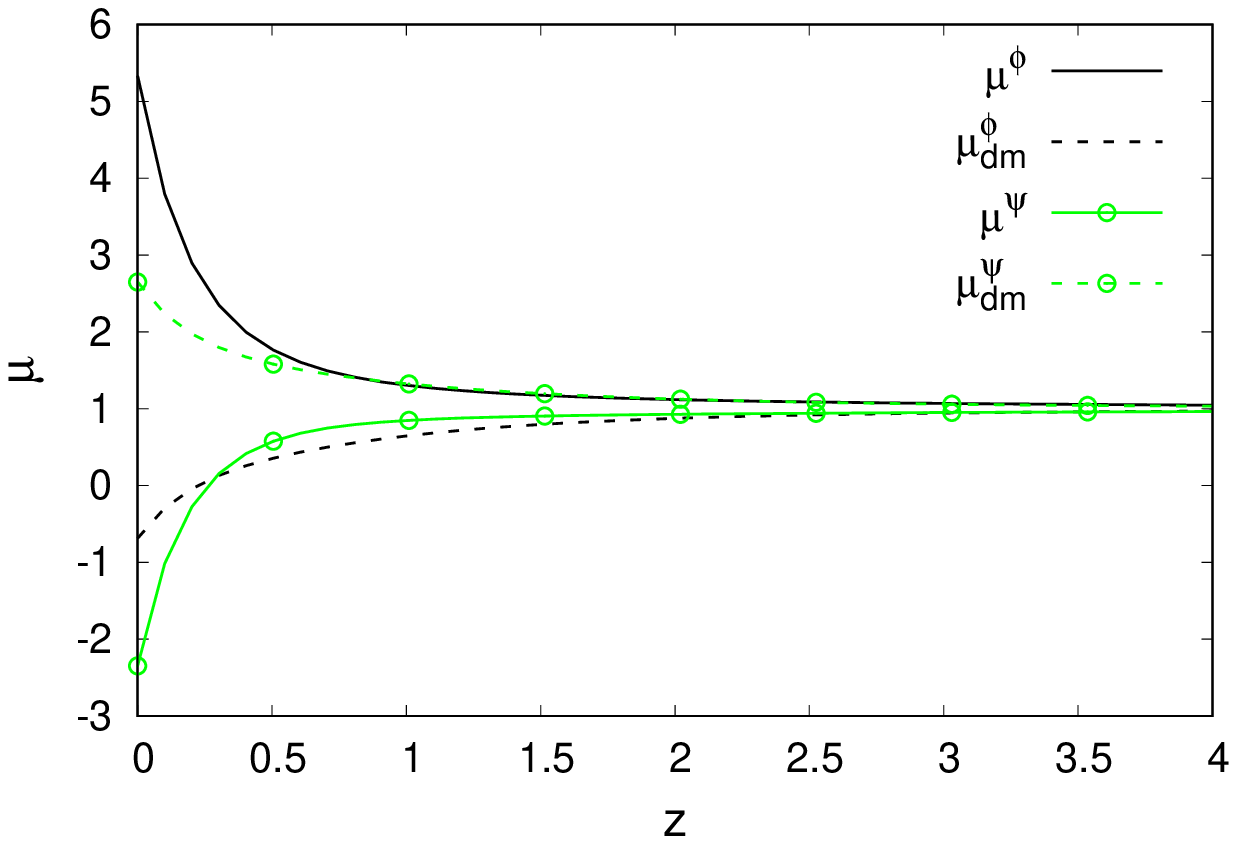}}\\
\epsfxsize=8.5 cm \epsfysize=5.2 cm {\epsfbox{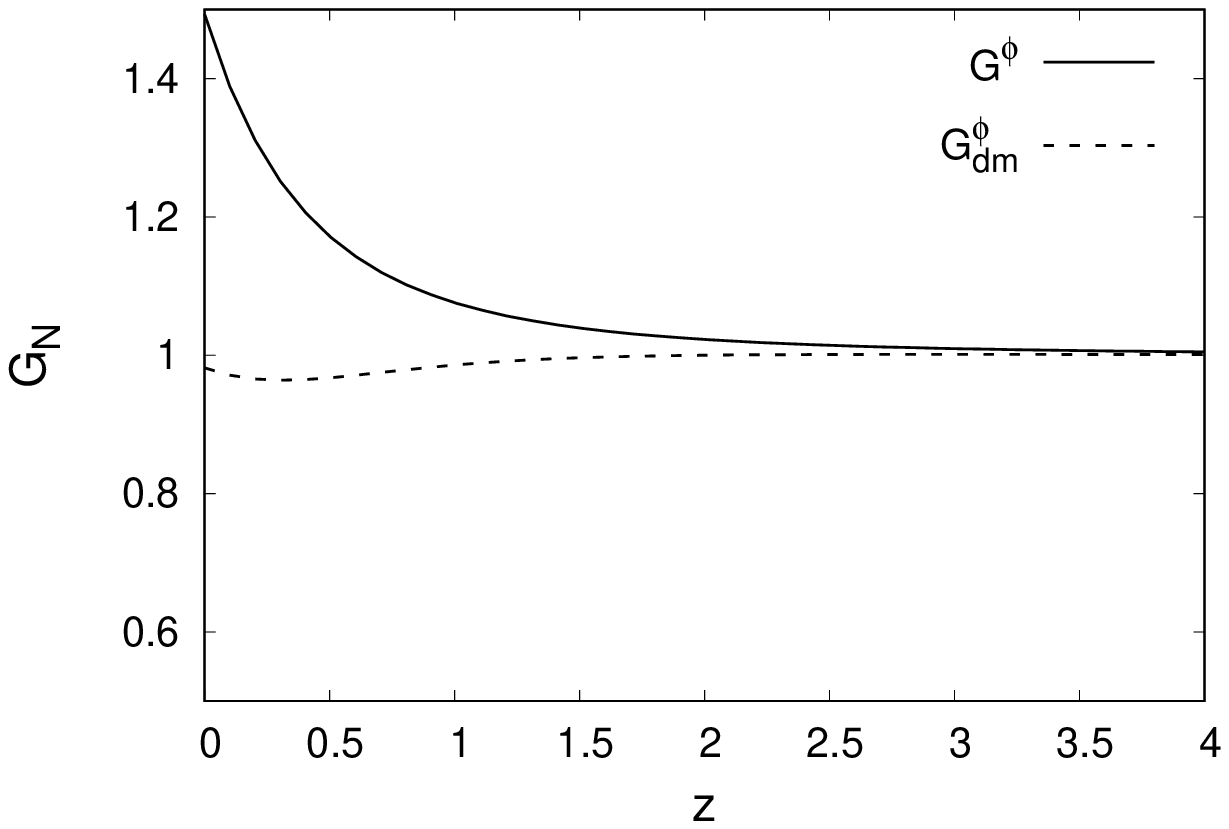}}
\end{center}
\caption{{\it Upper panel:} coefficients $\mu^{\phi}_*$ and $\mu^{\psi}_*$
for the symmetric model of Fig.~\ref{fig_sym_a1a2}, at comoving wave number
$k(z)=10 {\cal H}(z)$.
{\it Lower panel:} effective Newton constants ${\cal G}^{\phi}_*$. For this model,
${\cal G}^{\phi}_* = {\cal G}^{\psi}_*$.}
\label{fig_sym_mu}
\end{figure}

We show in Fig.~\ref{fig_sym_mu} the coefficients $\mu^{\phi}_*$ and $\mu^{\psi}_*$
for the symmetric solution of Fig.~\ref{fig_sym_a1a2}, at comoving wave number
$k(z)=10 {\cal H}(z)$, as well as the effective Newton constants.
At early times, when the scalar field has no effect and we converge to the
Einstein-de Sitter cosmology, we recover General Relativity with $\mu^*_* \to 1$
and ${\cal G}^*_* \to 1$.
At late times these coefficients show deviations of order unity.
In this regime, the comparison of the two panels shows that the coefficients
$\mu^*_*$ are dominated by the fifth-force contributions. This means that
the fifth force is greater than Newtonian gravity.
Moreover, the coefficients $\mu^{\phi}_{\rm dm}$ and $\mu^{\psi}$
become negative, which would give rise to very nonstandard behaviors.
Thus, the dark matter overdensities repel the baryonic matter at late times.

\subsubsection{Dark matter gravitational potentials $\phi_{\rm d}$ and $\psi_{\rm d}$}

In a similar fashion, the dark sector gravitational potentials $\phi_{\rm d}$ and
$\psi_{\rm d}$ obey Poisson equations of the form (\ref{Poisson-psib-1}),
\beqa
- \frac{2}{3} a_{\rm d} \frac{k^2}{H_0^2} \phi_{\rm d} & = & \mu^{\phi_{\rm d}}
\Omega_{\rm b 0} \delta + \mu^{\phi_{\rm d}}_{\rm dm} \Omega_{\rm dm 0}
\delta_{\rm dm} , \nonumber \\
- \frac{2}{3} a_{\rm d} \frac{k^2}{H_0^2} \psi_{\rm d} & = & \mu^{\psi_{\rm d}}
\Omega_{\rm b 0}  \delta + \mu^{\psi_{\rm d}}_{\rm dm} \Omega_{\rm dm 0}
\delta_{\rm dm} ,
\label{Poisson-psid-1}
\eeqa
with
\beqa
&& \hspace{-0.5cm} \mu^{\phi_{\rm d}} = \sum_{\ell} \left[ s_{{\rm d}\ell} s_{\ell} \frac{r_{\ell}}{r_{\rm d}} +
\frac{{\cal H}a_{\rm d}^2}{H_0 r_{\rm d}^2 a^3}
\sqrt{\frac{2 a_{\rm d}}{3\xi}} \frac{\beta k^2}{k^2+a_{\rm d}^2 m^2}
\frac{ds_{{\rm d}\ell}}{d\ln a} b_{\ell}  \right]
\nonumber \\
&& \hspace{-0.5cm} \mu^{\phi_{\rm d}}_{\rm dm} = \sum_{\ell} \left[ s_{{\rm d}\ell}^2 \frac{r_{\ell}}{r_{\rm d}} +
\frac{{\cal H}}{H_0 r_{\rm d}^2 a_{\rm d}}
\sqrt{\frac{2 a_{\rm d}}{3\xi}} \frac{\beta_{\rm dm} k^2}{k^2+a_{\rm d}^2 m^2}
\frac{ds_{{\rm d}\ell}}{d\ln a} b_{\ell}  \right]
\nonumber \\
&& \hspace{-0.5cm} \mu^{\psi_{\rm d}} = \sum_{\ell} \left[ s_{{\rm d}\ell} s_{\ell} -
\frac{{\cal H}a_{\rm d}^2}{H_0 r_{\rm d} a^3}
\sqrt{\frac{2 a_{\rm d}}{3\xi}} \frac{\beta k^2}{k^2+a_{\rm d}^2 m^2}
\frac{ds_{{\rm d}\ell}}{d\ln a} a_{\ell}  \right]
\nonumber \\
&& \hspace{-0.5cm} \mu^{\psi_{\rm d}}_{\rm dm} = \sum_{\ell} \left[ s_{{\rm d}\ell}^2  -
\frac{{\cal H}}{H_0 r_{\rm d} a_{\rm d}}
\sqrt{\frac{2 a_{\rm d}}{3\xi}} \frac{\beta_{\rm dm} k^2}{k^2+a_{\rm d}^2 m^2}
\frac{ds_{{\rm d}\ell}}{d\ln a} a_{\ell}  \right] \! .
\label{mu-psid-dm}
\eeqa

The renormalized Newton's constants are now
\beqa
&& {\cal G}^{\phi_{\rm d}} = \sum_{\ell} s_{{\rm d}\ell} s_{\ell} \frac{r_{\ell}}{r_{\rm d}} , \;\;\;
{\cal G}^{\phi_{\rm d}}_{\rm dm} = \sum_{\ell} s_{{\rm d}\ell}^2 \frac{r_{\ell}}{r_{\rm d}} ,
\nonumber \\
&& {\cal G}^{\psi_{\rm d}} = \sum_{\ell} s_{{\rm d}\ell} s_{\ell} , \;\;\;
{\cal G}^{\psi_{\rm d}}_{\rm dm} = \sum_{\ell} s_{{\rm d}\ell}^2 ,
\label{G-d}
\eeqa
which are again positive.
The comparison with Eq.(\ref{G-b}) shows that the cross-terms are related by
\beq
{\cal G}^{\phi}_{\rm dm} =  r_{\rm d}  {\cal G}^{\phi_{\rm d}}  , \;\;\;
{\cal G}^{\psi}_{\rm dm} = {\cal G}^{\psi_{\rm d}} .
\eeq

\begin{figure}
\begin{center}
\epsfxsize=8.5 cm \epsfysize=5.5 cm {\epsfbox{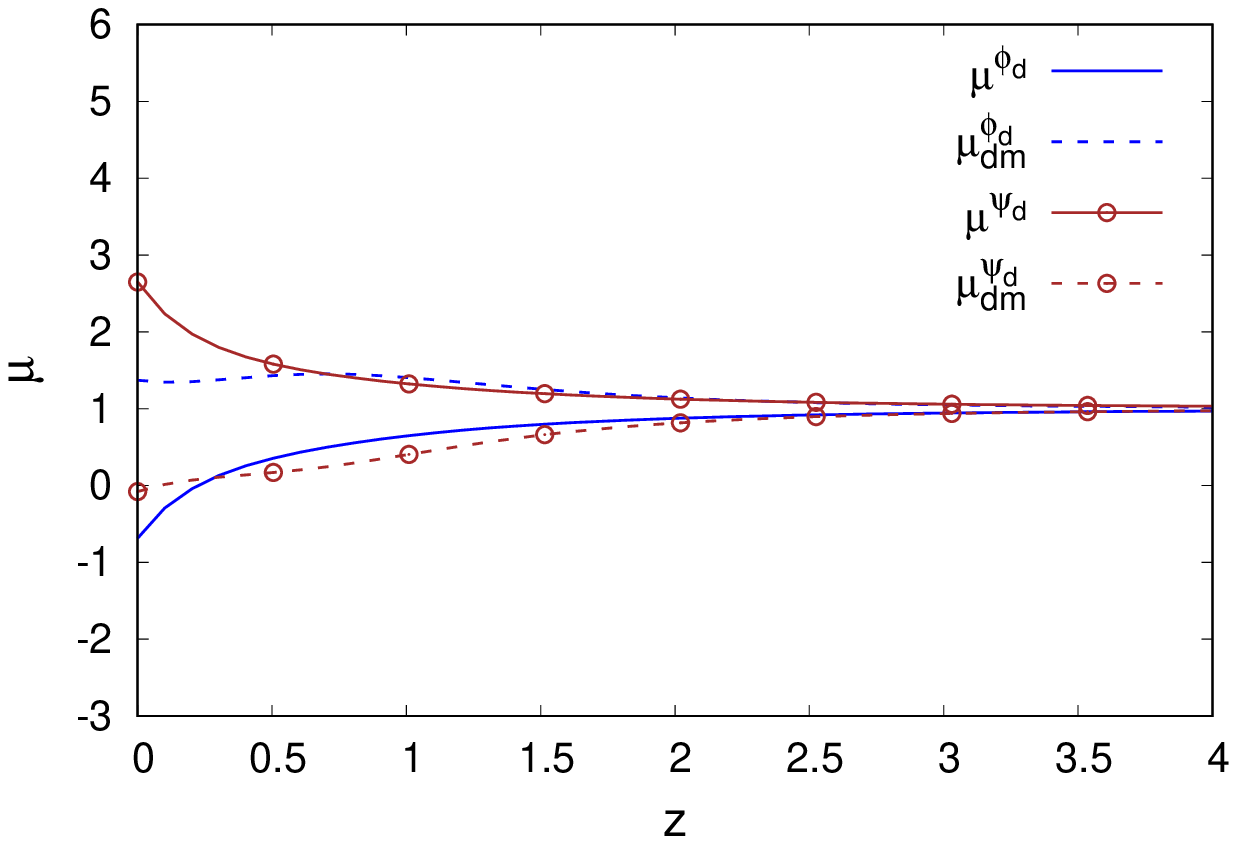}}\\
\epsfxsize=8.5 cm \epsfysize=5.5 cm {\epsfbox{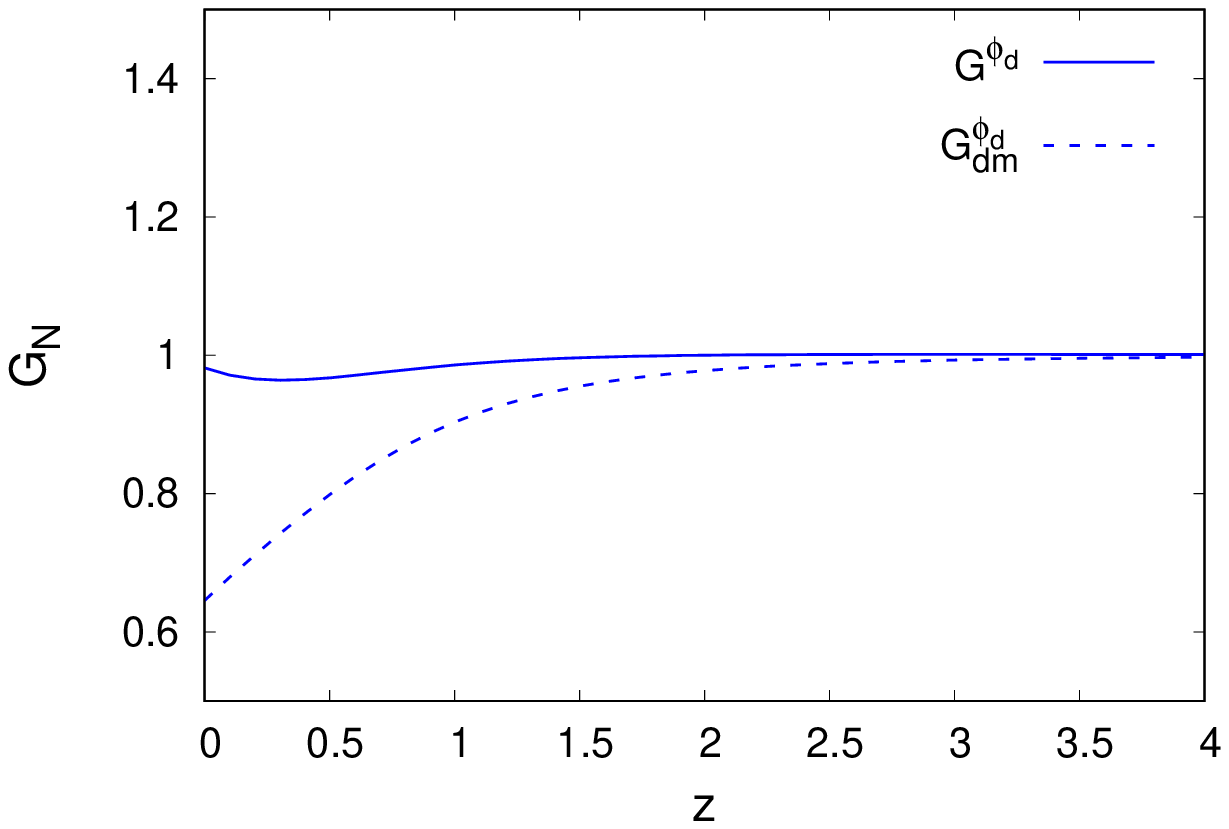}}
\end{center}
\caption{{\it Upper panel:} coefficients $\mu^{\phi_{\rm d}}_*$ and $\mu^{\psi_{\rm d}}_*$
for the symmetric model of Fig.~\ref{fig_sym_a1a2}, at comoving wave number
$k(z)=10 {\cal H}(z)$.
{\it Lower panel:} effective Newton constants ${\cal G}^{\phi_{\rm d}}_*$ and
${\cal G}^{\psi_{\rm d}}_*$.}
\label{fig_sym_mud}
\end{figure}

We show in Fig.~\ref{fig_sym_mud} the coefficients $\mu^{\phi_{\rm d}}_*$ and
$\mu^{\psi_{\rm d}}_*$ for the symmetric solution of Fig.~\ref{fig_sym_a1a2},
at comoving wave number $k(z)=10 {\cal H}(z)$, as well as the effective Newton constants.
We obtain behaviors that are similar to those found in Fig.~\ref{fig_sym_mu}
for the baryonic metric potentials.
At late times the fifth force is again greater than Newtonian gravity and
can lead to repulsive effects between baryons and dark matter.

\subsection{Density and velocity fields}
\label{sec:linear-density}

In their Jordan frame, associated with the metric $g_{\mu\nu}$, the baryons follow the
usual equation of motion $\nabla_{\mu} T^{\mu}_{\nu}=0$. This gives the standard
continuity and Euler equations
\beqa
&& \frac{\partial \rho}{\partial\tau} + \nabla \cdot (\rho \vv) + 3 {\cal H} \rho = 0 ,
\nonumber \\
&& \frac{\partial\vv}{\partial\tau} + ( \vv\cdot\nabla) \vv + {\cal H} \vv
= - \nabla \phi .
\label{Euler-b}
\eeqa
Using the Poisson equation (\ref{Poisson-psib-1}), we obtain the evolution equation
of the linear baryonic matter density contrast,
\beqa
\frac{\partial^2\delta}{(\partial \ln a)^2} + \left[ 1+\frac{d\ln{\cal H}}{d\ln a} \right]
\frac{\partial\delta}{\partial \ln a} & = & \frac{3 H_0^2}{2 a {\cal H}^2} \left[
\mu^{\phi} \Omega_{\rm b 0} \delta \right. \nonumber \\
&& \left. + \mu^{\phi}_{\rm dm} \Omega_{\rm dm 0} \delta_{\rm dm} \right] .
\hspace{1cm}
\label{delta-b}
\eeqa

The dark matter also follows its usual equation of motion,
$\nabla_{{\rm d}\mu} T^{\mu}_{\nu}=0$, where $\nabla_{{\rm d}\mu}$ is now the
covariant derivative associated with the dark sector metric $g_{{\rm d}\mu\nu}$.
This gives the continuity and Euler equations
\beqa
&& \frac{\partial \rho_{\rm dm}}{\partial\tau} + \nabla \cdot (\rho_{\rm dm}
\vv_{\rm dm}) + 3 {\cal H}_{a_{\rm dm}}  \rho_{\rm dm} = 0 ,
\nonumber \\
&& \frac{\partial\vv_{\rm dm}}{\partial\tau} + ( \vv_{\rm dm}\cdot\nabla) \vv_{\rm dm}
+ (2 {\cal H}_{a_{\rm d}} - {\cal H}_{b_{\rm d}} ) \vv_{\rm dm} = - r_{\rm d}^2
\nabla \phi_{\rm d} , \nonumber \\
&&
\label{Euler-dm}
\eeqa
where $\tau$ is still the conformal time of the baryonic metric.
Using the Poisson equation, the evolution equation
of the linear dark matter density contrast reads as
\beqa
&& \frac{\partial^2\delta_{\rm dm}}{(\partial \ln a)^2} + \left[ \frac{2 {\cal H}_{a_{\rm d}}
- {\cal H}_{b_{\rm d}}}{\cal H} +\frac{d\ln{\cal H}}{d\ln a} \right]
\frac{\partial\delta_{\rm dm}}{\partial \ln a} = \frac{3 r_{\rm d}^2 H_0^2}{2 a_{\rm d} {\cal H}^2}
\nonumber \\
&& \times \left[
\mu^{\phi_{\rm d}} \Omega_{\rm b 0} \delta + \mu^{\phi_{\rm d}}_{\rm dm}
\Omega_{\rm dm 0} \delta_{\rm dm} \right] .
\label{delta-dm}
\eeqa

\begin{figure}
\begin{center}
\epsfxsize=8.5 cm \epsfysize=5.5 cm {\epsfbox{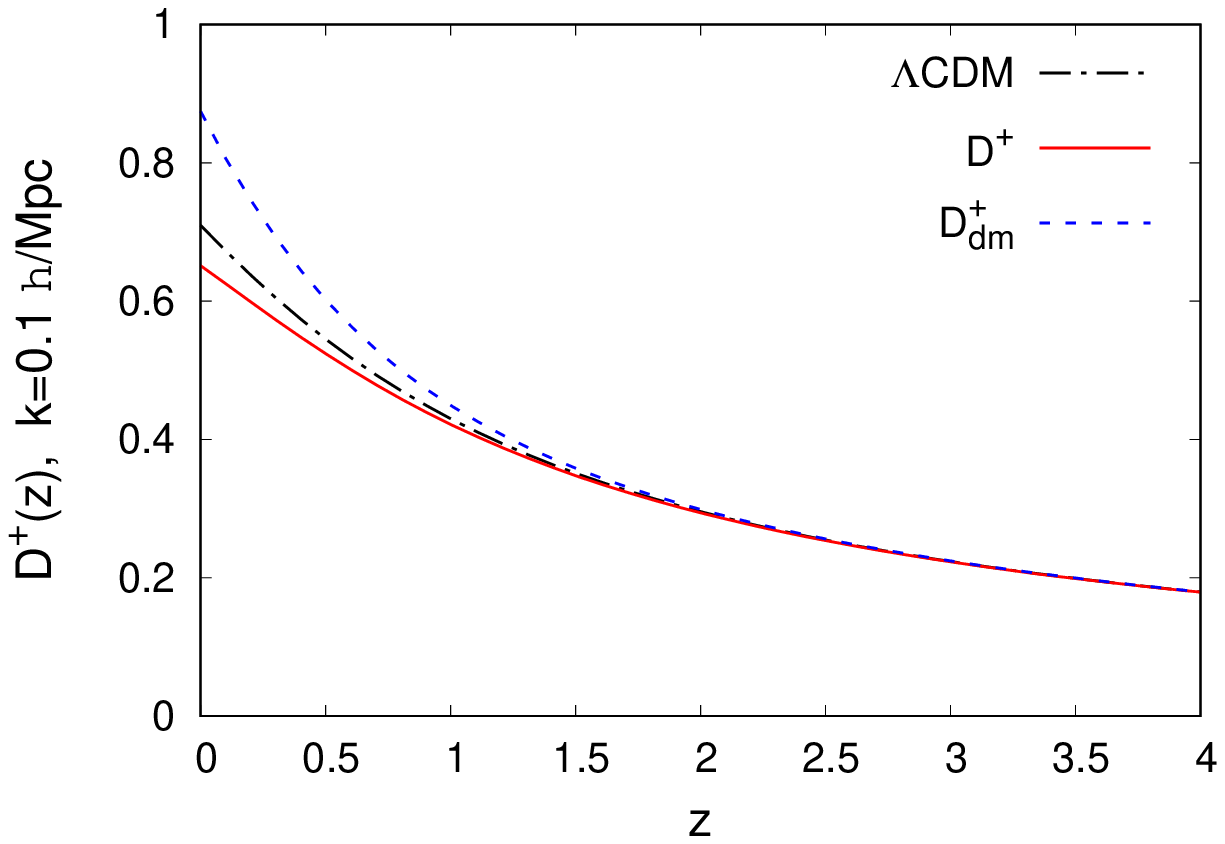}}\\
\epsfxsize=8.5 cm \epsfysize=5.5 cm {\epsfbox{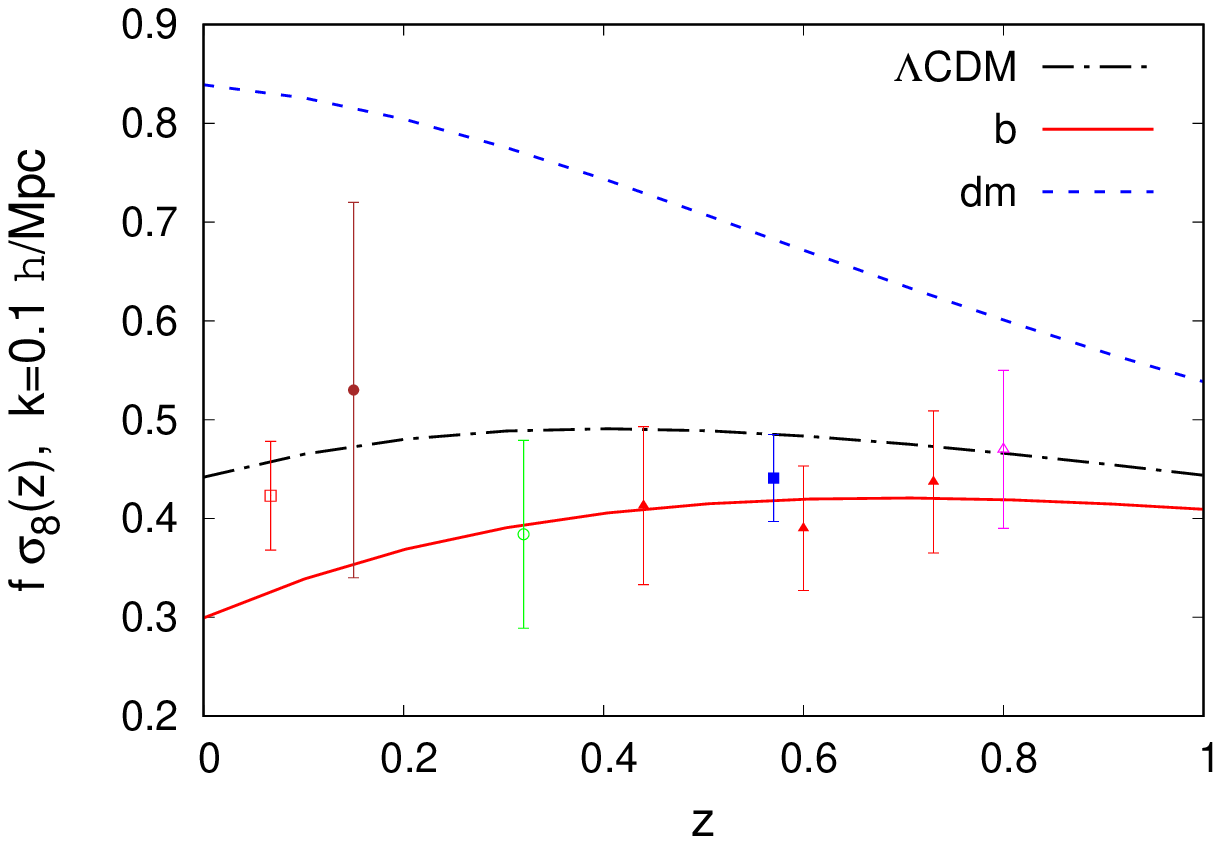}}
\end{center}
\caption{{\it Upper panel:} linear growing modes $D^+(k,a)$ and
$D^+_{\rm dm}(k,a)$, for the symmetric model of Fig.~\ref{fig_sym_a1a2},
at comoving wave number $k=0.1 h/{\rm Mpc}$.
{\it Lower panel:} growth factors $f \sigma_8$ and $f_{\rm dm} \sigma_{\rm dm 8}$.}
\label{fig_sym_D}
\end{figure}

\begin{figure}
\begin{center}
\epsfxsize=8.5 cm \epsfysize=5.5 cm {\epsfbox{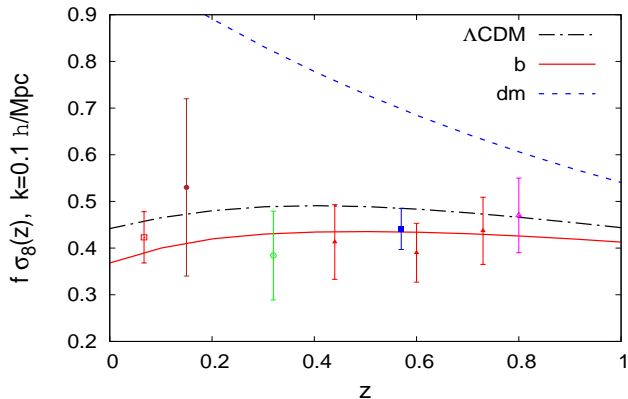}}
\end{center}
\caption{Growth factors $f \sigma_8$ and $f_{\rm dm} \sigma_{\rm dm 8}$, for the model
of Fig.~\ref{fig_r_a1a2} with different conformal times.}
\label{fig_sig8_r}
\end{figure}

The baryonic and dark matter linear growing modes are coupled and given by the
system of equations (\ref{delta-b}) and (\ref{delta-dm}).
We show in Fig.~\ref{fig_sym_D} their behavior as a function of redshift for the
comoving wave number $k=0.1 h/{\rm Mpc}$.
At high redshift they follow the $\Lambda$-CDM
reference, but at low redshift the dark matter perturbations grow faster than
in the $\Lambda$-CDM cosmology whereas the baryonic perturbations grow
more slowly.
This is more clearly seen in the lower panel, as the growth rate $f_*=d\ln D^+_*/d\ln a$
amplifies the deviations from the $\Lambda$-CDM cosmology because of the
time derivative.

The data points in Fig.~\ref{fig_sym_D} are only given to compare the magnitude
of the deviation of the growth factor with observational error bars, but do not provide
a meaningful test. Indeed, the Newton constant obtained in this scenario is amplified
at $z=0$, as seen in Fig.\ref{fig_sym_mu}. This means that to compare with data
we would need to run this model again by normalizing Newton's constant to its value
at $z=0$ instead of $z\to\infty$, as we have done so far.
We do not go further in this direction in this paper, because this model is already ruled
out by the large time derivative $d\ln {\cal G}/{dt} \sim 0.7 H_0$ at $z=0$, as we discuss
in the next section.

Nevertheless, it is interesting to note that this model leads to a slower growth for the baryonic
density perturbations than in the $\Lambda$-CDM cosmology. This is due to the decrease
of the gravitational attraction of dark matter onto baryonic matter, shown by the
coefficient $\mu^{\phi}_{\rm dm}$ in Fig.~\ref{fig_sym_mu}, which even turns negative
at $z \lesssim 0.1$ (i.e., the fifth force between dark matter and baryons becomes repulsive).
This is a distinctive feature of this model, as most modified-gravity scenarios amplify
the growth of large-scale structures.

We show in Fig.~\ref{fig_sig8_r} the growth factors obtained for the case
(\ref{r1-1-r2}) of Fig.~\ref{fig_r_a1a2}, where the different metrics have different conformal
times. This actually gives similar results for the linear growth of large-scale structures.

\subsection{Gravitational slip}
\label{sec:Gravitational slip}

\begin{figure}
\begin{center}
\epsfxsize=8.5 cm \epsfysize=5.2 cm {\epsfbox{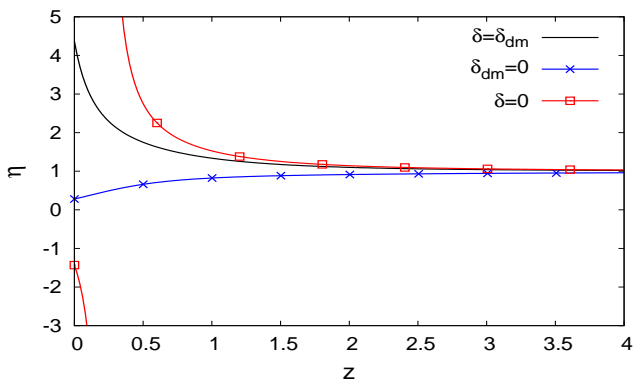}}
\end{center}
\caption{Gravitational slip $\eta$ of Eq.(\ref{eta-def}) for several values of the baryon to dark matter
ratio $\delta/\delta_{\rm dm}$.}
\label{fig_eta_mu}
\end{figure}

Because the fifth force enters with opposite signs in the $\phi$ and $\psi$
gravitational potentials, see Eq.(\ref{mu-psi-dm}), the lensing potential
$\phi_{\rm lens}=(\phi+\psi)/2$, which deflects light rays, and the dynamical potential $\phi$,
which determines the trajectory of massive bodies, are different.
This means that the lensing mass of clusters of galaxies (deduced from lensing observations)
and the dynamical mass (deduced from the galaxy velocity dispersion or the pressure profile of
the hot gas in hydrostatic equilibrium) are also different.
This is measured by the ratio $\eta$, which we define as
\beq
\eta = \frac{\phi+\psi}{2\phi} = \frac{1}{2}
+ \frac{\mu^{\psi} \Omega_{\rm b0}\delta+\mu^{\psi}_{\rm dm}\Omega_{\rm dm0} \delta_{\rm dm}}
{2 [ \mu^{\phi} \Omega_{\rm b0}\delta+\mu^{\phi}_{\rm dm}\Omega_{\rm dm0} \delta_{\rm dm} ]} .
\label{eta-def}
\eeq

We show in Fig.~\ref{fig_eta_mu} the gravitational slip $\eta$ on subhorizon scales,
for $\delta=\delta_{\rm dm}$, $\delta_{\rm dm}=0$ (which corresponds to cases where
$\rho\gg \rho_{\rm dm}$), and $\delta=0$ (for $\rho_{\rm dm} \gg \rho$).
In agreement with Fig.~\ref{fig_sym_mu}, the three curves converge
to the General Relativity value $\eta=1$ at high redshift and show deviations of order
unity at low $z$. Because the couplings to baryons and dark matter are different,
the gravitational slip $\eta$ depends on the relative amount of baryons and dark matter
in the lens. On cosmological scales down to clusters of galaxies, which are the largest\
collapsed structures, we expect $\delta\simeq \delta_{\rm dm}$. This gives $\eta > 1$
at low $z$, hence the lensing mass would be greater than the dynamical mass.
This ratio can reach a factor three at $z < 0.1$, but in practice, most cosmological lenses
are at redshifts $z \gtrsim 0.5$, as the lensing efficiency goes to zero as the source
redshift vanishes. This gives $1<\eta \lesssim 1.7$.
On the other hand, on subgalactic scales where baryons dominate, the gravitational slip is
smaller than unity so that the lensing mass is smaller than the dynamical mass by a factor
three at $z=0$.
In the case where dark matter dominates, $\eta$ goes to infinity at $z\sim 0.3$ and becomes
negative at lower redshift. This is because $\phi$ goes through zero and changes sign.
This follows from $\mu^{\phi}_{\rm dm}<0$, as seen in Fig.~\ref{fig_sym_mu}.
This implies a repulsive fifth force from dark matter onto baryons, which dominates
when the lens is mostly made of dark matter.
This regime should not be reached in practice, as we have $\delta_{\rm dm} \sim \delta$
on large scales, where the separation of baryons from dark matter due to the fifth force
has not yet had time to be efficient, as seen by the small impact on the linear growing modes
in Fig.~\ref{fig_sym_D}, whereas we typically have $\rho \gg \rho_{\rm dm}$ on subgalactic
scales because radiative cooling processes make baryons collapse further and eventually form\
stars.

\section{Dynamical degrees of freedom and linear perturbations}
\label{sec:freedom-linear}

In this section, we study the behavior of linear perturbations around the cosmological
background for the tensor, vector, and scalar sectors,
without using the quasistatic approximation.
This allows us to count the number of dynamical degrees of freedom, beyond the simple
counting of components described in section~\ref{sec:degrees-of-freedom} above.
The number of perturbative degrees of freedom in bigravity theories has been
discussed in \cite{Comelli:2015pua,Gumrukcuoglu:2015nua,Brax:2016ssf}.
They obtained the behavior of scalar, vector and tensor modes by expanding
the action up to quadratic order over the fluctuations.
We present an alternative derivation,
starting directly from the vierbeins as for our derivation of the nonlinear Einstein
equations (\ref{Einstein-e1}).
This also allows us to  implement explicitly the discussion of
section~\ref{sec:degrees-of-freedom} and to show how the 32 components of
the vierbeins can be reduced to the expected 16 components by successive
gauge choices, associated with the diagonal Lorentz and diffeomorphism invariances
and with the symmetry constraint (\ref{Ymunu-sym}). Then, constraint equations
further reduce the number of dynamical degrees of freedom.
We find that there are no ghosts at the level of the quadratic action around
the cosmological background.

In Minkowski space-time, i.e. in vacuum, the bimetric action (\ref{Sgrav-def})
reduces to two independent copies of General Relativity. Therefore,
it shows $2\times 2=4$ dynamical degrees of freedom (associated with the
two massless gravitons of the tensor sector), without ghosts nor dangerous instabilities.
In the following, we focus on perturbations around the cosmological background,
with nonzero mean density and pressure and with cosmological expansion.

\subsection{Vierbein and metric perturbations, quadratic action}
\label{sec:Quadratic-action}

Starting from the vierbeins $\delta e^{a}_{\ell \mu}$, the metric perturbations
$\delta g_{\ell \mu\nu}$ are given from the definition (\ref{vierbein-12-def}) by
\beq
\delta g_{\ell \mu\nu} = ( \delta e^{a}_{\ell \mu} e^b_{\ell \nu}
+ e^{a}_{\ell \mu} \delta e^b_{\ell \nu} ) \eta_{ab} .
\eeq
For the diagonal background (\ref{e-diag-def}) this simplifies to
\beqa
&& \delta g_{\ell 00} = -2 b_{\ell} \delta e^0_{\ell 0} , \;\;\;
\delta g_{\ell 0i} = \delta g_{\ell i0} = a_{\ell} \delta e^i_{\ell 0} - b_{\ell} \delta e^0_{\ell i} ,
\nonumber \\
&& \delta g_{\ell ij} = a_{\ell} ( \delta e^i_{\ell j} + \delta e^j_{\ell i} ) .
\eeqa
The perturbations of the matrices $X^{\mu}_{\ell\nu}$ defined in Eq.(\ref{Xmunu-def})
also simplify as
\beqa
&& \delta X^0_{20} = - \frac{\delta e^0_{20} b_1}{b_2^2} + \frac{\delta e^0_{10}}{b_2} , \;\;\;
\delta X^i_{20} = - \frac{\delta e^i_{20} b_1}{a_2 b_2} + \frac{\delta e^i_{10}}{a_2} ,
\nonumber \\
&& \delta X^0_{2i} = - \frac{\delta e^0_{2i} a_1}{b_2 a_2} + \frac{\delta e^0_{1i}}{b_2} , \;\;\;
\delta X^i_{2j} = - \frac{\delta e^i_{2j} a_1}{a_2^2} + \frac{\delta e^i_{1j}}{a_2} .
\nonumber \\
&&
\eeqa
The permutation $1 \leftrightarrow 2$ provides $\delta X^{\mu}_{1\nu}$.

As in General Relativity, we can split the gravitational perturbations in scalar, vector and
tensor modes. As in \cite{Brax:2016ssf}, we can do so at the level of the vierbeins
and we can write
\beqa
&& \delta e^0 _{\ell 0} = b_{\ell} \phi_{\ell} \, , \;\;\;
\delta e^{0}_{\ell i} = a_{\ell} [ - \partial_i V_{\ell} + C_{\ell i} ] \, , \nonumber \\
&& \delta e^{i}_{\ell 0} = b_{\ell} [ - \partial^i W_{\ell} + D^i_{\ell} ] \, , \nonumber \\
&& \delta e^{i}_{\ell j} = a_{\ell} [ - \psi_{\ell} \delta^i_j + \partial^i \partial_j U_{\ell}
+ \partial_j V^i_{\ell} + \partial^i W_{\ell j} + h^i_{\ell j} ] \, , \hspace{1cm}
\label{vierbeins-SVT}
\eeqa
where the spatial indices are raised and lowered with $\delta^{ij}$ and $\delta_{ij}$,
so that $\partial^i=\partial_i$ and $V^i_{\ell}=V_{\ell i}$.
The transversality conditions are
\[
\partial^i C_{\ell i} = \partial_i D^i_\ell = \partial_i V^i_{\ell} = \partial^i W_{\ell i} = 0,\ \
\partial_i h^i_{\ell j} = 0 \, ,
\]
and tracelessness corresponds to
\[
h^{i}_{\ell i} = 0 .
\]
This provides the perturbations of the gravitational metrics as
\beqa
&& \hspace{-0.4cm} \delta g_{\ell 00} = -2 b_{\ell}^2  \phi_{\ell} , \;\;\;
\delta g_{\ell 0i} = a_{\ell} b_{\ell} [ \partial_i ( V_{\ell}-W_{\ell} ) + D_{\ell i}-C_{\ell i} ] ,
\nonumber \\
&&  \hspace{-0.4cm} \delta g_{\ell ij} = a_{\ell}^2 [ -2 \psi_{\ell} \delta_{ij}
+ 2 \partial_i\partial_j U_{\ell} + \partial_i (V_{\ell j}+W_{\ell j} ) \nonumber \\
&& \hspace{1.2cm} + \partial_j (V_{\ell i}+W_{\ell i} ) + h_{\ell ij} + h_{\ell ji} ] .
\label{dgmunu}
\eeqa
The baryonic and dark vierbeins and metrics obey the same decompositions,
obtained from the combinations (\ref{vierbein-db-def}).

This gives 32 components for the two gravitational metrics: 10 scalars
$\{\phi_{\ell}, V_{\ell}, W_{\ell}, \psi_{\ell}, U_{\ell}\}$, 8 vectors
$\{C_{\ell i}, D^i_{\ell}, V^i_{\ell}, W_{\ell i} \}$, and 2 nonsymmetric tensors
$h^i_{\ell j}$.
As explained in section~\ref{sec:degrees-of-freedom}, this can be reduced to 16
components when we use the invariance under the diagonal Lorentz transformations
and diffeomorphisms, and the symmetry constraints (\ref{Ymunu-sym}).
It is convenient to handle the Lorentz invariance and the symmetry constraints
(\ref{Ymunu-sym}) through the variables $\delta Z_{\ell\mu\nu}$ introduced
in Eq.(\ref{deltaZ-def}).This suppresses  the Lorentz degeneracies associated with the
vierbeins by the condition $\delta Z_{\ell\mu\nu}=\delta Z_{\ell\nu\mu}$, which implies
\beq
h_{\ell ij} = h_{\ell ji} , \;\; D_{\ell i} = - C_{\ell i} , \;\; V_{\ell i} = W_{\ell i}, \;\;
W_{\ell} = - V_{\ell}
\eeq
and  removes $2\times 6$ components. The diagonal diffeomorphism invariance
still remains.

We study first the dynamics of tensor, vector and scalar perturbations,
in the early-time regime where the scalar field is constant and the background
follows the simple solution (\ref{EdS-0}), i.e., all metrics show the same
Hubble expansion rate.
Then, dark and baryonic matter can be unified in the same matter sector
as $s_{{\rm d}\ell}=s_{\ell}$.
In all three cases, the explicit computation of the Einstein equations (\ref{Einstein-e1})
at linear order shows that the perturbations separate in two decoupled sectors,
$S_+$ and $S_-$.
The sector $S_+$ involves the matter perturbations, which act as source terms in
the Einstein equations, and the matter metric defined from Eq.(\ref{vierbein-db-def}),
which gives
\beq
h_{\mu\nu} = s_1^2 h_{1\mu\nu} + s_2^2 h_{2\mu\nu} ,
\label{h+_def}
\eeq
where $h_{\ell\mu\nu}$ are the linear metric perturbations of the two gravitational metrics, defined by $g_{\ell\mu\nu}=a^2(\eta_{\mu\nu}+2h_{\ell\mu\nu})$.
We find that the Einstein equations of this sector are identical to General Relativity.
Therefore, in this regime there is no deviation from General Relativity in the sector
probed by matter and by observations.
The ``hidden'' sector $S_-$ has no matter source terms and only involves the
hidden metric components $h_{-\mu\nu}$, defined by
\beq
h_{-\mu\nu} = h_{1\mu\nu} - h_{2\mu\nu} .
\label{h-_def}
\eeq
Its equations of motion differ from those of General Relativity by mass terms.
[The components $h_{-\mu\nu}$ do not directly define a metric, because if
we define the vierbeins $e^a_{-\mu}=s_1^{-1} e^a_{1\mu} - s_2^{-1} e^a_{2\mu}$,
which would imply (\ref{h-_def}), we find that the background vierbeins
$\bar{e}^a_{-\mu}$ vanish.]

We can check that the equations of motion of the hidden sector $S_-$ can be derived
from the quadratic action defined by the standard expression
\beq
\delta^2 S = \int d^4x \; \frac{1}{2} \delta\left( \frac{\delta S}{\delta Z_{\mu\nu}} \right)
\delta Z_{\mu\nu} ,
\label{d2S-Z}
\eeq
but where again we work at the level of the vierbeins and use the variables
$\delta Z_{\mu\nu}$ of Eq.(\ref{deltaZ-def}).
For instance, using Eq.(\ref{de-dZ}) we obtain for the quadratic part that arises from the
first gravitational action $S_1= \int d^4x (M_{\rm Pl}^2/2) \sqrt{-g_1} R_1$
the expression
\beqa
\delta^2 S_1 & = & - \frac{M_{\rm Pl}^2}{2} \int d^4x \; \delta\left[ \sqrt{-g_1}
\left( G_1^{\mu\sigma} X^{\nu}_{2\sigma} + G_1^{\nu\sigma} X^{\mu}_{2\sigma} \right)
\right] \nonumber \\
&& \times \Theta_{\mu\nu} \, \delta Z_{1\mu\nu} ,
\label{d2S-S1}
\eeqa
where $\Theta_{\mu\nu}=1$ or $1/2$ was introduced in (\ref{de-dZ}).
We recognize the structure of the left-hand side of the Einstein equations
(\ref{Einstein-e1}).
The contribution $\delta^2 S_2$ from the second gravitational action $S_2$
can be obtained from Eq.(\ref{d2S-S1}) by the permutation $1\leftrightarrow 2$.
There are also similar contributions from the matter action.
For the matter sector $S_+$ this procedure is more complex because of the coupling
to matter. This involves term linear and quadratic in matter perturbations,
which enforce the coupling between fluid and metric perturbations and the equations
of motion of the fluid. These terms do not contribute to the hidden sector $S_-$,
as can be seen from a direct computation of the Einstein equations from
Eq.(\ref{Einstein-e1}).

\subsection{Tensor modes}
\label{sec:tensors}

In the tensor sector, we consider the evolution of metric perturbations over the background.
Because we do not consider matter sources and there are no tensor gauge transformations,
the computation from the Einstein equations coincide with the one from the quadratic
action (\ref{d2S-Z}) where we do not include matter perturbations.
Then, the quadratic action separates as
\beq
\delta^2 S =  \delta^2 S_+ +  s_1^2 s_2^2 \; \delta^2 S_-
\label{d2S-split}
\eeq
with
\beq
\delta^2 S_+ =  \int d^4x \, a^2 M_{\rm Pl}^2
\left[ h_{ij}^{' \, 2} - (\nabla h_{ij})^2 \right]
\label{d2S-h}
\eeq
and
\beq
\delta^2 S_- =  \int d^4x \left\{ a^2 M_{\rm Pl}^2 \left[ h_{- \, ij}^{' \, 2}
- (\nabla h_{- \, ij})^2 \right] + a^4 \bar{p} \; h_{- \, ij}^2 \right\}
\label{d2S-h-}
\eeq
where the sum is only over the independent components.
Thus, at the quadratic order, the action $\delta^2 S_+$ of the matter sector
is identical to that of General Relativity, while there exists a second decoupled
sector that differs from General Relativity by a new mass term.
This leads to $2\times 2=4$ dynamical degrees of freedom in the tensor sector.

We recover the results obtained in
\cite{Comelli:2015pua,Gumrukcuoglu:2015nua,Brax:2016ssf}.
Omitting the indices $ij$, the two uncoupled gravitons obey the equations of motion
\beqa
&& h'' + 2 {\cal H} h'  - \nabla^2 h = 0 ,
\label{h-matter-evol} \\
&& h''_- + 2 {\cal H} h'_-  - \nabla^2 h_-
- \frac{a^2 {\bar p}}{M_{\rm Pl}^2} h_- = 0 .
\label{h-_evol}
\eeqa
The massless graviton $h$ of the baryonic and dark matter metric
evolves as in General Relativity. On subhorizon scales it propagates with the
speed of light. On scales greater than the horizon it contains a constant mode and
a decaying mode that evolves as $h' \propto a^{-2}$.
This physical mode (in the sense that it is the one seen by the matter metric) is governed
by Eq.(\ref{h-matter-evol}) throughout all cosmological eras and does not mix with the
hidden graviton $h_-$.

The second hidden graviton $h_-$ has a negative squared mass in the radiation era,
as $\bar p=\bar p_{\gamma} >0$, which becomes negligible in the matter era.
In the radiation era, we have ${\cal H}=1/\tau$ and
$a=\sqrt{\Omega_{\gamma 0}} H_0 \tau$.
The hidden massive graviton $h_-$ obeys the equation of motion
$h''_- + \frac{2}{\tau} h'_-  - \nabla^2 h_-  - \frac{1}{\tau^2} h_- = 0 .$
It oscillates on subhorizon scales.
On superhorizon scales it contains both a decaying mode
and a growing mode
\beq
k \ll {\cal H} : \;\;\;  h^-_- \propto a^{-(1+\sqrt{5})/2} , \;\;\;
h^+_- \propto a^{(\sqrt{5}-1)/2} ,
\label{tensor-power-law}
\eeq
associated with the tachyonic instability.
In the matter era, we have ${\cal H}=2/\tau$ and $a \propto \tau^2$.
The mass of the second graviton $h_-$ becomes negligible and it behaves like
the massless graviton, with a constant mode and a decaying mode $\propto a^{-3/2}$.

Although $h_-$ is not seen by the matter, it should remain small at all epochs so that
the perturbative approach applies. This implies that the initial tensor fluctuations
at the onset of the radiation era must be sufficiently small.
This is easily satisfied as the squared mass
turns positive during the inflation era and the graviton decays \cite{Comelli:2015pua}.
During the inflationary stage, the tensor evolution equation is still given by
Eq.(\ref{h-_evol}), where $\bar p$ is now the pressure
$\bar p_{\chi}=-\bar\rho_{\chi}$ of the inflaton $\chi$.
Because we now have $\bar p_{\chi}<0$ the squared mass becomes positive
and there is no tachyonic instability, and on superhorizon scales there are only two
decaying modes
\beqa
&& k \ll {\cal H} : \;\;\;  h^{c}_- \propto a^{-3/2} \cos\left( \frac{\sqrt{3}}{2} \ln a \right) ,
\nonumber \\
&& \hspace{1.5cm} h^s_- \propto a^{-3/2} \sin\left( \frac{\sqrt{3}}{2} \ln a \right) .
\label{h-inflation-superhorizon}
\eeqa
Let us consider a mode $k$ that remains above the horizon until the end of the
radiation era, $k \leq a_{\rm eq} H_{\rm eq}$. It crosses the horizon during the
inflationary stage at the time $a_k = k/H_I$, where $H_I$ is the constant Hubble
expansion rate of the inflationary de Sitter era.
Then, the amplitude of the tensor mode $h_-$ at the end of the radiation era reads as
\beq
h_-(a_{\rm eq}) = h_-(a_k) \left( \frac{a_f}{a_k} \right)^{-3/2} \left( \frac{a_{\rm eq}}{a_f} \right)^{(\sqrt{5}-1)/2} ,
\eeq
where $a_f$ is the scale factor at the end of the inflationary era.
For $H_I \sim 10^{-5} M_{\rm Pl} \sim 10^{13} {\rm GeV}$,
$a_f \sim 10^{-28}$, $a_{\rm eq} \sim 10^{-3}$, we find that all modes with
$k \leq a_{\rm eq} H_{\rm eq}$ remain in the perturbative regime,
$h_-(a_{\rm eq}) \ll 1$, provided $h_-(a_k) \ll 10^{24}$.
As we expect $h_-(a_k) \sim H_I/M_{\rm Pl} \sim 10^{-5}$, if the tensor fluctuations
are generated by the quantum fluctuations, all modes remain far in the perturbative
regime until the end of the radiation era.
This is due to their decay during the inflationary stage on superhorizon scales,
and to their small initial values associated with quantum fluctuations.

Therefore, the main constraint from the tensor sector is the measurement
of the speed of gravitational waves from the binary neutron star merger
GW170817 \cite{Abbott2017}, which implies that at least one of the lapse
factors $r_{\ell}$ is unity at $z=0$, as discussed in
section~\ref{sec:backg-different-conformal}.

\subsection{Vector modes}
\label{sec:vectors}

In the vector sector, the perturbations of the energy-momentum tensor are
\beqa
&& \delta T^{00} = 0 , \; \delta T^{0i} = a^{-2} \left[ \bar\rho v^i -2 \bar{p} C^i \right] ,
\nonumber \\
&& \delta T^{ij} = - 2 a^{-2} \bar{p} [ \partial^i V^j + \partial^j V^i ] .
\hspace{1cm}
\eeqa
As in General Relativity, the equations of motion of matter,
$\nabla_{\mu} T^{\mu\nu}=0$,
decouple from the Einstein equations and read as
\beq
\frac{\partial}{\partial\tau} [ (\bar\rho+\bar p) U^i ]
+ 4 {\cal H} [ (\bar\rho+\bar p) U^i ] = 0 ,
\eeq
where we introduced the usual gauge invariant velocity,
\beq
U^i = v^i - 2 C^i = v^i - 2 (s_1^2 C^i_1 + s_2^2 C^i_2) .
\label{U-def}
\eeq
The equations of motion follow from the Einstein equations (\ref{Einstein-e1}).
One can check that they also follow from a quadratic action
that separates as in (\ref{d2S-split}), with
\beq
\delta^2 S_+ = \int d^4x \; a^2 M_{\rm Pl}^2 \left[ \nabla( V'_j+C_j) \cdot \nabla( V'_j+C_j) \right] ,
\label{S+_vector}
\eeq
and
\beqa
\delta^2 S_- & = & \int d^4x \; \biggl\lbrace a^2  M_{\rm Pl}^2
\left[ \nabla( V'_{- j}+C_{- j}) \cdot \nabla( V'_{- j}+C_{- j}) \right]  \nonumber \\
&& + a^4 \left[ \frac{3\bar\rho+\bar{p}}{2} C_{- j}^2
+ \bar{p} (\nabla V_{- j})^2 \right] \biggl \rbrace  .
\label{S-_vector}
\eeqa
Here, for the matter sector $S_+$ we focused on the solution $U^i=0$ of the matter
conservation equation (\ref{U-def}).
As for the tensors, the action $\delta^2 S_+$ of the matter metric is identical to General
Relativity, while the second decoupled sector $\delta^2 S_-$ is modified by a new mass
term that vanishes in the Minkowski space-time.

Therefore, as in General Relativity, there are no vector dynamical degrees of freedom
left in $\delta^2 S_+$, if we set $U^i=0$.
One can see that $C_i$ is not dynamical. Its ``equation of motion''
reads as $C_i=-V'_i$. Substituting into the action gives $\delta^2 S_+=0$,
so that $V_i$ is arbitrary. This is due to the diffeomorphism invariance of
General Relativity.

In the action $\delta^2 S_-$, $C_{- i}$ is again nondynamical. Its equation of motion
reads in Fourier space
\beq
C_{- i}(\vk) = - \frac{2 M_{\rm Pl}^2 k^2}{2 M_{\rm Pl}^2 k^2 + a^2 (3\bar\rho+{\bar p})}
V'_{- i}(\vk) ,
\label{C-_V'-}
\eeq
and substituting into the action gives
\beqa
\delta^2 S_- & = & (2\pi)^3 \int d\vk d\tau \; a^4 k^2 \biggl\lbrace
\frac{M_{\rm Pl}^2 (3\bar\rho + {\bar p})}{2 M_{\rm Pl}^2 k^2 + a^2 (3\bar\rho+{\bar p})}
\nonumber \\
&& \times V'_{- j}(\vk) V'_{- j}(-\vk) + {\bar p} \, V_{- j}(\vk) V_{- j}(-\vk) \biggl\rbrace .
\hspace{1cm}
\eeqa
The vector $V_{- i}$ is now dynamical when $3\bar\rho+{\bar p} \neq 0$.
Therefore, we have two dynamical degrees of freedom in the vector sector,
associated with the hidden vector $V_{- i}$.

Its equation of motion reads
\beq
- \frac{\partial}{\partial\tau} \left[ \frac{a^4 M_{\rm Pl}^2 (3\bar\rho + {\bar p})}
{2 M_{\rm Pl}^2 k^2 + a^2 (3\bar\rho+{\bar p})} V'_{- j}(\vk) \right]
+ a^4 {\bar p} V_{- j}(\vk) = 0 .
\label{V-_vector-eq}
\eeq
Thus, the mode $V_{- i}$ shows a gradient instability on subhorizon scales
in the radiation and matter eras, where $\bar p = \bar p_{\gamma} >0$, and
we recover the results obtained in
\cite{Comelli:2015pua,Gumrukcuoglu:2015nua,Brax:2016ssf}.

Let us consider in turns the inflationary, radiation and matter eras.
In the inflationary era, Eq.(\ref{V-_vector-eq}) gives on subhorizon scales
$V''_-+4{\cal H} V'_- + k^2 V_- =0$ (where we omit the index $i$), so that the
vector mode $V_-$ oscillates with frequency $\omega=\pm k$.
On superhorizon scales, we obtain
$V''_- - \frac{2}{\tau} V'_- + \frac{3}{\tau^2} V_- =0$.
This is the same evolution equation as for the tensor modes, and we obtain the
same two decaying solutions as in Eq.(\ref{h-inflation-superhorizon}).

In the radiation era, on subhorizon scales we obtain
$V''_- -\frac{k^2}{5} V_- =0$. This gradient instability leads to the two
exponential modes
\beq
k \gg {\cal H} : \;\;\; V_-^{\pm} \propto e^{\pm k\tau/\sqrt{5}} .
\label{vector-gradient-radiation-era}
\eeq
On superhorizon scales we again recover the same behavior as for tensors,
$V''_- + \frac{2}{\tau} V'_- - \frac{1}{\tau^2} V_- = 0$,
with the power-law growing and decaying modes (\ref{tensor-power-law}).

In the matter era, we obtain on subhorizon scales
$V''_- + \frac{2}{\tau} V'_- - \frac{8 k^2}{9 {\cal H}_{\rm eq}^2 \tau^2} V_-
= 0$, which gives the power-law growing and decaying modes
\beq
k \gg {\cal H} : \;\;\; V_-^{\pm} \propto \tau^{\lambda_\pm} \;\;\;
\mbox{with} \;\;\;
\lambda_\pm = - \frac{1}{2} \pm \frac{1}{2} \sqrt{1 + \frac{32 k^2}{9 {\cal H}_{\rm eq}^2}} .
\label{vector-gradient-matter-era}
\eeq
On superhorizon scales we have
$V''_- + \frac{4}{\tau} V'_- - \frac{4\tau_{\rm eq}^2}{\tau^4} V_- = 0$.
Long after the radiation-matter equality, $\tau \gg \tau_{\rm eq}$,
this gives a constant mode and a decaying mode $V_- \propto \tau^{-3}$.

Let us estimate the magnitude of the unstable vector mode $V_-$ at $z=0$,
for a wave number $k$ that goes beyond the horizon at $a_k$ during the inflationary
stage and goes below the horizon at $a'_k$ during the radiation era.
Collecting the results above, we obtain
\beqa
&& k V_- \sim k V_-(a_k) \left( \frac{H_I}{H_{\rm eq}} \right)^{-3/2}
\left( \frac{a_{\rm eq}}{a_f} \right)^{(\sqrt{5}+2)/2} \nonumber \\
&& \times \left( \frac{k}{a_{\rm eq} H_{\rm eq}}
\right)^{(4-\sqrt{5})/2} e^{\frac{k}{a_{\rm eq} H_{\rm eq}} ( \frac{1}{\sqrt{5}} - \frac{\sqrt{2}}{3} \ln a_{\rm eq} )} .
\label{vector-estimate-1}
\eeqa
After horizon exit during the inflationary era, this mode first decays as $a^{-3/2}$
until the end of the inflationary era at $a_f$.
Next, it grows as $a^{(\sqrt{5}-1)/2}$ during the radiation era, until it enters the horizon.
Then, its subhorizon behavior deviates from the one of the tensor mode $h_-$
as it shows the exponential growth (\ref{vector-gradient-radiation-era})
until the matter era starts where it shows the power-law growth
(\ref{vector-gradient-matter-era}).
These last two stages give the exponential factor in Eq.(\ref{vector-estimate-1}),
which is actually dominated by the matter era growth factor.
If we assume that at horizon exit during the inflationary stage we have
$C_- \sim V_-' \sim k V_- \sim H_I/M_{\rm Pl}$,
we obtain for $H_I \sim 10^{-5} M_{\rm Pl}$ that $k V_- \ll 1$ at $z=0$
for $k \ll 0.3 h {\rm Mpc}^{-1}$.
Therefore, on weakly nonlinear scales and below, the growth of the ``hidden'' vector
jeopardizes the perturbativity of the model, and the gravitational metrics
$g_1$ and $g_2$ become non-linear in this regime.
This implies that the initial vector seeds at horizon exit during the inflationary era
should be suppressed or that the scenario must be supplemented by additional
mechanisms that damp the growth of this vector mode on small scales
at high redshift.

\subsection{Scalar modes}
\label{sec:scalars}

The same decoupling as for tensors and vectors occurs for scalars.
The matter sector $S_+$ is again identical to General Relativity.
The hidden sector $S_-$ does not couple to matter and differs from
General Relativity by mass terms.
Its quadratic action reads as
\beqa
&& \delta^2 S_- = \int d^4x \; a^2 M_{\rm Pl}^2 \biggl\lbrace \phi_- \left[ -3 {\cal H}^2 \phi_-
- 9 {\cal H}^2 \psi_- \right. \nonumber \\
&& \left. + 3 {\cal H}^2 \nabla^2 U_- - 6 {\cal H} \psi'_-
- 4 {\cal H} \nabla^2 V_- + 2 {\cal H} \nabla^2 U'_- + 2 \nabla^2 \psi_- \right]
\nonumber \\
&& + \psi_- \left[ 3 ( {\cal H}' - {\cal H}^2 ) \psi_- - 2 ( 2 {\cal H}' + {\cal H}^2 ) \nabla^2 U_-
+ 8 {\cal H} \nabla^2 V_- \right. \nonumber \\
&& \left. - 4 {\cal H} \nabla^2 U'_- + 3 \psi''_- - \nabla^2 \psi_- + 4 \nabla^2 V'_-
- 2 \nabla^2 U''_- \right] \nonumber \\
&& + ( 4 {\cal H}^2 - {\cal H}' ) (\nabla V_-)^2 \biggl \rbrace .
\label{S-_scalar}
\eeqa
The equations of motion in this scalar sector can be easily carried out and result in no dynamical degree of freedom.
Indeed, as the scalar $U$ only enters linearly,
it is nondynamical and it provides a constraint equation that allows us to substitute
for $\psi''$. The scalar $V$ is also nondynamical, and its equation of motion allows
us to substitute for $\psi'$. Then, $\phi$ only enters linearly, hence
it is also nondynamical and provides another constraint equation, when $\psi$
is also nondynamical.
Thus, there are no new dynamical degrees of freedom in the scalar sector.

\subsection{Dynamical degrees of freedom in Einstein-de Sitter space-time}
\label{sec:degrees-freedom}

To summarize the results from the previous sections, in Minkowski space-time
we have two copies of General Relativity and 4 dynamical degrees of freedom,
associated with the two massless gravitons.

Around the cosmological background, chosen to be the early Universe Einstein - de Sitter solution (\ref{EdS-0}) of the equations of motion, the quadratic action separates as a part
$\delta^2 S_+$ that describes the metric seen by matter and a part $\delta^2 S_-$ that
describes a second hidden metric. The first part $\delta^2 S_+$ remains identical
to General Relativity, with 2 dynamical degrees of freedom associated with the
massless graviton.
The second part $\delta^2 S_-$ contains new mass terms.
It generates 4 degrees of freedom, associated with a massive graviton and
a transverse vector that shows a gradient instability.
At this level, there are no new scalar dynamical degrees of freedom and no ghosts.

We study the linear perturbations in the general case in appendix~\ref{sec:linear-general},
when we no longer assume $s_{\rm dl \ell}$ and $s_{\ell}$ to be equal and the
various metrics can have different Hubble expansion rates.
As the baryonic and dark matter metrics are different, the quadratic action no longer
separates in a sector $S_+$, which contains all matter and remains identical to General
Relativity, and a hidden sector $S_-$ that differs from General Relativity by mass terms
and is decoupled from matter.
However, from the Einstein equations, we find that linear perturbations behave in the same
fashion as in the simpler case presented above.
In the tensor sector, we have two massive gravitons, which at high frequency and
wave number have a negligible mass and behave as in General Relativity.
In the vector sector, we can still separate $\{C_i,V_i\}$ and $\{C_{-i},V_{-i}\}$.
Again, there are only two propagating degrees of freedom, associated with
$V_{-i}$, and they still show the gradient instability (\ref{vector-gradient-radiation-era})
in the radiation era.
In the scalar sector, no new dynamical degrees of freedom or ghost appear.

In the next section, we will analyze the existence of ghosts and the cutoff of the theory by performing a St\"uckelberg analysis.

\section{Analysis of ghosts by the St\"uckelberg method}
\label{sec:Stuckelberg}

As shown by the explicit computation of linear perturbations around the Einstein-de Sitter cosmological
background, in that case the system decouples in the two
sectors $S_+$ and $S_-$.
The sector $S_+$ contains the matter metric perturbations $\delta g_{\mu\nu}$
and the matter fluid perturbations, such as $\delta\rho$, and it coincides with General
Relativity. It is the sector relevant for observations (at this linear order).
The hidden sector $S_-$ contains the other metric components, $\delta g_{-\mu\nu}$,
and is not sourced by matter.
This shows that around the cosmological background, it is more convenient
to decompose the metric degrees of freedom in these two metrics, rather than the
two gravitational metrics $\delta g_{\ell\mu\nu}$.
In particular, it means that the two sets $\delta g_{\ell\mu\nu}$ are strongly coupled
and that one cannot study the fluctuations of $g_{1\mu\nu}$ while neglecting
its coupling to $g_{2\mu\nu}$.

In contrast, in vacuum we only have two independent Einstein-Hilbert terms,
giving rise to two independent copies of General Relativity.
Therefore, around the Minkowski background, the relevant decomposition is over the
two gravitational metrics $\delta g_{\ell\mu\nu}$.
This shows that the physics is quite different over these two backgrounds,
and different treatments are appropriate.

\subsection{Explicit quadratic action around Einstein-de Sitter background}
\label{sec:quadratic-action}

We now check with the St\"uckelberg method that there is no Boulware-Deser ghost
at the linear order of perturbations around the cosmological background.
In massive gravity or bigravity theories, a Boulware-Deser ghost
\cite{Boulware1972} can appear
in the scalar sector because of the new degrees of freedom, associated with the
additional metric or the loss of gauge invariance.
In General Relativity, there are no scalar dynamical degrees of freedom around
Minkowski or Einstein-de Sitter backgrounds because the gauge invariance removes
two scalar degrees of freedom (among the four scalar components, two are nondynamical
fields or Lagrange multipliers, and the other two are pure gauges).
In a bimetric theory like the one we consider in this paper, we have two metrics,
but only the diagonal gauge invariance is left. Therefore, as compared to
two independent copies of General Relativity, we have additional degrees of freedom,
as  one gauge invariance is missing in order to remove a few of them. Then, some of these
new degrees of freedom may turn out to be ghosts.

The sector $S_+$ being identical to General Relativity it is healthy and
it makes full use of the diagonal gauge invariance.
We will try to restore full diffeomorphism invariance by performing a St\"uckelberg analysis on the decoupled sector.
Because $S_-$ is decoupled (at linear order), we can study the quadratic action
(\ref{S-_scalar}) alone.
Around the cosmological background, a change of coordinates
$x^{\mu} \to x^{\mu} + \xi^{\mu}$ corresponds at linear order to a change of the metric
\beq
\delta g_{\mu\nu} \to \delta g_{\mu\nu} -
\frac{\partial\bar{g}_{\mu\nu}}{\partial x^{\sigma}} \xi^{\sigma}
- \bar{g}_{\sigma\nu} \frac{\partial\xi^{\sigma}}{\partial x^{\mu}}
- \bar{g}_{\mu\sigma} \frac{\partial\xi^{\sigma}}{\partial x^{\nu}}  .
\label{gmunu-gauge-inv}
\eeq
Because we have lost gauge invariance, the action $\delta^2S_-$ is not invariant
when $\delta g_{-\mu\nu}$ transforms as in (\ref{gmunu-gauge-inv}).
Following the St\"uckelberg formalism, we can introduce an additional field $\zeta^{\mu}$
to restore the gauge invariance, by writing \cite{Rubakov2008}
\beq
\delta g_{-\mu\nu} = \hat{\delta g}_{-\mu\nu} +
\frac{\partial\bar{g}_{-\mu\nu}}{\partial x^{\sigma}} \zeta^{\sigma}
+ \bar{g}_{-\sigma\nu} \frac{\partial\zeta^{\sigma}}{\partial x^{\mu}}
+ \bar{g}_{-\mu\sigma} \frac{\partial\zeta^{\sigma}}{\partial x^{\nu}}  .
\label{gmunu-zeta-def}
\eeq
Then, the action $\delta^2 S_- (\hat{\delta g}_{-\mu\nu},\zeta^{\mu})$
is invariant under the combined gauge transformation
where $\hat{\delta g}_{-\mu\nu}$ transforms as in (\ref{gmunu-gauge-inv})
while $\zeta$ transforms as $\zeta^{\mu} \to \zeta^{\mu} + \xi^{\mu}$.
By choosing another gauge condition than $\zeta^{\mu}=0$, one can often read on the
Lagrangian terms involving $\zeta^{\mu}$ the behavior of dangerous modes.
Focusing on the scalar sector, with $\zeta^{\mu}=\eta^{\mu\nu} \partial_{\nu}\pi$,
this gives for the scalar perturbations of the hidden metric $\delta g_{-\mu\nu}$,
\beqa
&& \phi_- = \hat{\phi}_- - {\cal H} \pi' - \pi'' , \;\;\;
 \psi_- = \hat{\psi}_-  + {\cal H} \pi' \nonumber \\
&& V_- = \hat{V}_-  + \pi' , \;\;\;\;
U_- = \hat{U}_- + \pi .
\label{pi-def-Stuckelberg}
\eeqa
Substituting into the quadratic action (\ref{S-_scalar}), one finds that the
St\"uckelberg field $\pi$ does not cancel out because the action $\delta^2S_-$
is not gauge invariant.
We could expect quadratic terms with up to four derivatives from
(\ref{pi-def-Stuckelberg}), which would be the usual signature of the
Boulware-Deser ghost.
However, the explicit computation from Eq.(\ref{S-_scalar}) shows that all third and
fourth-order time derivatives cancel out and the action can be written in terms of
first-order time derivatives. This means that there is no Ostrogradsky ghost, associated with
higher derivative terms in the Lagrangian, at linear order around the cosmological
background.

For completeness, the explicit expression of the action is given by
$\delta^2 S_- = \delta^2S^{(0)}_- + \delta^2S^{(1)}_- + \delta^2S^{(2)}_-$,
where $\delta^2S^{(0)}_-$ is given by Eq.(\ref{S-_scalar}) where we add a hat to the
metric variables, $\delta^2S^{(1)}_-$ is the linear part over $\pi$ and reads as
\beqa
&& \hspace{-0.5cm}  \delta^2S^{(1)}_- = \int d^4x \; a^2 M_{\rm Pl}^2
\biggl\lbrace \hat\phi_- 3 {\cal H} [ {\cal H} \nabla^2\pi - ({\cal H}^2 + 2 {\cal H}' ) \pi' ]
\nonumber \\
&& \hspace{-0.5cm}   + \hat\psi_- [ ({\cal H}^2 + 2 {\cal H}' ) (3 \pi'' - 2 \nabla^2\pi)
+ 3 ( {\cal H}^3+4{\cal H} {\cal H}'+2{\cal H}'' ) \pi' ]  \nonumber \\
&& \hspace{-0.5cm}   - 2 (2{\cal H}^2 + {\cal H}' ) (\nabla^2 \hat{V}_-) \pi'
- ( \nabla^2 \hat{U}_-)  [ ({\cal H}^2 + 2 {\cal H}' ) \pi'' \nonumber \\
&& \hspace{-0.5cm}   + ( {\cal H}^3+4{\cal H} {\cal H}'+2{\cal H}'') \pi' ] \biggl\rbrace ,
\label{d2S-pi1}
\eeqa
which only involves first-order time derivatives if we integrate $\pi''$ by parts,
and $\delta^2S^{(2)}_-$ is the quadratic part over $\pi$ and reads as
\beqa
&& \delta^2S^{(2)}_- = \int d^4x \; \frac{a^2 M_{\rm Pl}^2}{2} \big \lbrace
- 3 ({\cal H}^2 + 2 {\cal H}' ) {\cal H}' \pi'^2  \nonumber \\
&& + ( 7 {\cal H}^2 {\cal H}' + 2 {\cal H}'^2 + 2 {\cal H} ( {\cal H}^3 + {\cal H}'' ) )
(\nabla\pi)^2 \big \rbrace , \;\;\;
\label{d2S-pi2}
\eeqa
which only contains first-order time derivatives.

As was the case for the original action (\ref{S-_scalar}), we can check from the
action $\delta^2 S_- = \delta^2S^{(0)}_- + \delta^2S^{(1)}_- + \delta^2S^{(2)}_-$
that there are no propagating modes and $\pi$ is not dynamical.
This is not apparent from the quadratic part (\ref{d2S-pi2}), but $\pi$ is coupled
to the other metric components through (\ref{d2S-pi1}).
Then, for instance, $U_-$ again enters linearly into the action and provides a constraint
that removes another degree of freedom.
After successive simplifications, one finds that there are no physical dynamical modes
left.

We obtain the same result in Appendix~\ref{sec:linear-general} for the more general case
where the different metrics follow different Hubble expansion rates.

\subsection{Goldstone bosons}
\label{sec:Goldstone}

We now study how ghosts may appear beyond the linear perturbation theory
investigated in the previous section and beyond the Einstein-de Sitter case, when
the baryonic and dark matter metrics are different.
We again follow the St\"uckelberg formalism and we first show that we do not need
to explicitly compute the action to recover the previous results at linear order,
in the regime of short time and length scales as compared to the horizon and
the age of the Universe.
Next, we discuss the nonlinear terms. 
Notice that our analysis remains perturbative around FLRW backgrounds throughout
and that a full investigation of the presence of ghosts should be carried out nonperturbatively. Here, we restrict ourselves to a perturbative
analysis which provides an upper value for the cutoff scale of the theory.

As noticed above, in the absence of matter our bimetric theory reduces to two copies
of General Relativity and it is therefore ghost free.
This corresponds to the Minkowski background, and one would like to extend this result
to the case of FLRW spaces, where the coupling of the two metrics is present
through the matter actions and might reintroduce a Boulware-Deser ghost.
As in (\ref{gmunu-zeta-def}), this can be investigated by introducing four Goldstone fields
$\zeta^\mu$ whose role is to restore the full diffeomorphism invariance of the theory,
which is broken by the presence of the matter actions.
The order parameter of the breaking of the two copies of diffeomorphism invariance
to the diagonal subgroup is the Hubble parameter of the Universe.
We will see that it plays the same role as the mass term for gravitons in massive bigravity
\cite{deRham:2014naa}.

In the following, we consider the case where $a_{\ell}=b_{\ell}$ (i.e., all metrics
have a common conformal time), so that the background vierbeins are diagonal with
\be
\bar{e}^a_{\ell \mu}=  a_{\ell} \delta^a_\mu ,
\label{e-bar-ell}
\ee
and we focus on short times compared to the age of the Universe and
short distances compared to the horizon,
\be
\partial \ln h_{\mu\nu} \gg {\cal H} .
\label{appro}
\ee
Here, $h_{\mu\nu}$ stands for the metric perturbations, and ${\cal H}$ stands for the
conformal Hubble expansion rates, which we take to be of the same order for the
different metrics.
In contrast with section~\ref{sec:quadratic-action}, we do not restrict to the early-time
regime (\ref{EdS-0}). Hence the baryonic and dark matters follow different metrics
$g_{\mu\nu}$ and $g_{\rm d \mu\nu}$ with different expansion rates and
$s_{\rm d \ell}$ are different from $s_{\ell}$.

The matter actions break the two copies of diffeomorphism invariance associated
with the two Einstein-Hilbert actions. However, in the approximation
(\ref{appro}) we can reintroduce the broken symmetry invariance
by introducing St\"uckelberg fields $\phi^{\mu}_{\ell}$ and defining the composite object
\be
g_{\ell \mu\nu} = \hat{g}_{\alpha \beta}  \frac{\partial \phi^\alpha_\ell}{\partial x^\mu}
\frac {\partial \phi^\beta_\ell}{\partial x^\nu} .
\label{g-ell-phi}
\ee
The metric $g_{\ell\mu\nu}$ is now invariant under the combined transformations
\beq
\hat{g}_{\ell \mu\nu} \to \frac{\partial x^{\lambda}}{\partial x'^{\mu}}
\frac{\partial x^{\rho}}{\partial x'^{\nu}} \hat{g}_{\ell \lambda\rho} , \;\;\;
\frac{\partial \phi^{\alpha}_\ell}{\partial x^{\mu}} \to
\frac{\partial x'^{\alpha}}{\partial x^{\lambda}}
\frac{\partial \phi^{\lambda}_\ell}{\partial x'^{\mu}} .
\eeq
We recover the initial action by the gauge choice $\phi^{\mu}_{\ell}=x^{\mu}$.
This is the nonlinear extension of (\ref{gmunu-zeta-def}),
with $\phi^\mu=x^\mu + \zeta^\mu$, where we neglect derivatives of the cosmological
background $\partial \bar{g}_{\mu\nu}$ thanks to the approximation (\ref{appro}).
The definition (\ref{g-ell-phi}) can also be written at the level of the vierbeins as
\be
e^a_{\ell\mu} =  \hat{e}^a_{\ell \nu} \frac{ \partial \phi^\nu_\ell}{\partial x^\mu} .
\label{e-ell-phi}
\ee

It is convenient to separate the diffeomorphisms into the diagonal ones, which are not
broken by the presence of matter, and the broken ones in the complementary directions
which belong to the group quotient $({\rm diff}_1\times {\rm diff}_2)/ {\rm diff}_{\rm diag}$
\beq
\phi^{\mu}_{\ell} = x^{\mu} + \xi^{\mu} + \gamma_{\ell} \; \pi^{\mu} ,
\;\;\;  \gamma_1 \neq \gamma_2 .
\label{gamma1-gamma2-def}
\eeq
Here $\xi^{\mu}$ is the diagonal diffeomorphism while $\pi^{\mu}$ is an arbitrary
complementary direction, set by the constant coefficients $\gamma_{\ell}$.
Then, the vierbeins (\ref{e-ell-phi}) read
\beq
e^a_{\ell\mu} =  \hat{e}^a_{\ell \mu} + \hat{e}^a_{\ell \nu} \left(
\frac{\partial \xi^{\nu}}{\partial x^{\mu}} + \gamma_{\ell}
\frac{\partial \pi^{\nu}}{\partial x^{\mu}} \right) .
\eeq
The total action becomes $S(e^a_{\ell\mu}) \to S(\hat{e}^a_{\ell\mu},\xi^{\mu},\pi^{\mu})$,
which is independent of $\xi^{\mu}$ as the diagonal diffeomorphism invariance is not broken.
The field $\pi^\mu$  cannot be gauged away, as if annulled in $g_{2\mu\nu}$
by a diagonal change of coordinates it would reappear in the $g_{1\mu\nu}$ metric
and vice versa. Hence, the $\pi^\mu$ fields parametrize orthogonal directions
to diagonal gauge transformations.

To investigate the Boulware-Deser ghosts we can focus on the fields
$\pi^{\mu}$, which are the Goldstone bosons of the broken symmetry, and consider
the scalar mode
\be
\pi^\mu = \partial^\mu \pi ,
\ee
for a scalar $\pi$. Because of the approximation (\ref{appro}), it does not matter
whether $\partial^\mu$ is defined with respect to $\eta^{\mu\nu}$ or any of the
metrics $\bar g_{\ell}^{\mu\nu}$.

From the definition (\ref{vierbein-db-def}), the baryonic matter and dark matter vierbeins
read as
\be
e^a_{\star \mu} = s_{\star 1} \hat{e}^a_{1\mu} + s_{\star 2} \hat{e}^a_{2\mu}
+ \left( s_{\star 1} \gamma_1 \hat{e}^a_{1\nu}
+ s_{\star 2} \gamma_2 \hat{e}^a_{2\nu} \right) \partial^{\nu} \partial_{\mu} \pi ,
\label{sttar}
\ee
where the subscript $\star$ stands for ${\rm b}$ or ${\rm d}$ (i.e. baryonic or dark matter).
The field $\pi$ could only be removed from the matter action by a change of
coordinate if $\gamma_1=\gamma_2$, associated with a diagonal diffeomorphism.
On the other hand, if we choose $\gamma_1=1/s_{\star 1}^2$ and
$\gamma_2 = -1/s_{\star 2}^2$, the field $\pi$ disappears from
Eq.(\ref{sttar}) at the linear level.
This corresponds to the choice used in section~\ref{sec:quadratic-action},
where the scalar $\pi$ in (\ref{pi-def-Stuckelberg}) lived in the sector $S_-$ and
was not coupled to matter. Indeed, we can check that with this choice of coefficients
$\gamma_{\ell}$, $\pi$ contributes to the sector $S_-$ as defined by
Eq.(\ref{h-_def}) and does not contribute to the sector $S_+$ defined by
Eq.(\ref{h+_def}). This is only possible in the early-Universe regime,
where the baryonic and dark matter metrics are identical, with $s_{\ell}=s_{\rm d\ell}$.
In this section, we go beyond this regime and we do not assume $s_{\ell}=s_{\rm d\ell}$.
Then, it is not possible to find coefficients $\gamma_{\ell}$ that remove the
field $\pi$ from both the baryonic and dark matter actions.

Let us now focus on the scalar $\pi$ alone, setting the other metric modes to zero,
that is, $\hat{e}^a_{\ell\mu} = \bar{e}^a_{\ell\mu}$.
The matter vierbeins (\ref{sttar}) contain second derivatives $\partial^2\pi$.
Therefore, the equations of motion for $\pi$ coming from the matter actions may contain
up to four derivatives and may lead to the propagation of extra ghostlike modes
\cite{Rubakov2008}.
Specifically, the Euler-Lagrange terms in the equations of motion for $\pi$
coming from the matter action take the form
\beq
E_{\star1} \propto \partial^{\nu} \partial_{\mu} \left( \frac{\delta S_\star}{\delta e^a_{1\mu}} \bar{e}^a_{1\nu} \right) \propto \partial^{\nu} \partial_{\mu} \left( \sqrt{-g_\star}
T_\star^{\mu\sigma} e_{\star \nu\sigma} \right)
\label{E-star-pi-1}
\eeq
where we used the approximation (\ref{appro}) to neglect background derivatives.
Using the equation of motion of the matter, $\nabla_{\star\mu} T^{\mu\nu}_\star = 0$,
which gives
\beq
\partial_{\mu} \left( \sqrt{-g_\star} T^{\mu\nu}_\star \right) = - \sqrt{-g_\star} \Gamma^{\nu}_{\star\mu\lambda} T^{\mu\lambda}_\star ,
\eeq
and the property
\beq
\partial_\mu e_{\star a\lambda} = e_{\star a\nu} \Gamma^\nu_{\star\lambda\mu}
- e_{\star b \lambda} \omega^b_{\star a \mu} ,
\eeq
where $\omega^{\star ab}_{\mu}$ is the spin connection defined by
\beqa
\omega^{ab}_{\star \mu} & = & \frac{1}{2} e^{a \nu}_\star ( \partial_\mu e^b_{\star \nu}
- \partial_\nu e^b_{\star \mu} ) - \frac{1}{2} e^{b \nu}_{\star }
( \partial_\mu e^a_{\star \nu} - \partial_\nu e^a_{\star \mu} ) \nonumber \\
&& - \frac{1}{2} e^{a \rho}_\star e^{b \sigma}_\star ( \partial_\rho e_{\star c\sigma}
- \partial_\sigma e_{\star c \rho} ) e^c_{\star\mu} ,
\label{omega-spin-def}
\eeqa
we can write Eq.(\ref{E-star-pi-1}) as
\beq
E_{\star1} \propto \partial^{\nu} \left( \sqrt{-g_\star} T_\star^{\mu\sigma}
e_{\star b \sigma } \omega^b_{\star a\mu} \right) .
\label{E-star-pi-2}
\eeq
The matter vierbeins (\ref{sttar}) take the form
$e^a_{\star\mu} = \bar{e}^a_{\star\mu} + \bar{A}^{a\nu}_\star \partial_\nu\partial_\mu \pi$, for a given matrix $\bar{A}^{a\nu}_\star$,
and substituting into the definition (\ref {omega-spin-def}) we find
$\omega^{ab}_{\star \mu} = 0$, within the approximation (\ref{appro}).
As a result, well inside the horizon and on timescales much shorter than the age of the Universe, we find that the contributions to the equations of motion for $\pi$ coming
from the matter terms do not involve higher-order derivatives and therefore do not
give rise to ghosts.
This is similar to what happens in massive bigravity \cite{deRham:2014naa}.

This result can be understood in a simpler way that also applies to the two
Einstein-Hilbert terms. Within the approximation (\ref{appro}),
the matter vierbeins (\ref{sttar}) take the form
\beq
e^a_{\star\mu} = \bar{e}^a_{\star\mu} + \bar{e}^a_{\star\nu} \partial^{\nu}
\partial_{\mu} \left( \frac{s_{\star 1}\gamma_1 a_1 + s_{\star 2} \gamma_2 a_2}
{a_\star} \pi \right) ,
\label{e-star-diffeo}
\eeq
where we used Eq.(\ref{e-bar-ell}) for the background vierbeins.
This corresponds to a diffeomorphism
$x^\mu \to x^\mu + \partial^\mu [ (s_{\star 1}\gamma_1 a_1 + s_{\star 2} \gamma_2 a_2)
\pi /a_\star ]$,
so that the matter action reads as
$\sqrt{-g_\star} {\cal L}_\star(g_{\star\mu\nu}) = \sqrt{-g_\star}
{\cal L}_\star( \bar{g}_{\star\mu\nu})$.
The gravitational vierbeins $e^a_{\ell\mu}$ also take the form
(\ref{e-star-diffeo}), where the fraction is replaced by a simple factor $\gamma_\ell$.
Again, the invariance of the Ricci scalar under changes of coordinates implies
that the Einstein-Hilbert terms read as
$\sqrt{-g_\ell} R(g_{\ell\mu\nu}) = \sqrt{-g_\ell}
R( \bar{g}_{\ell\mu\nu})$.
Therefore, the scalar $\pi$ only appears in the two Einstein-Hilbert actions
and the two matter actions through the determinants $\sqrt{-g}$.
This gives factors of the form
\beq
\sqrt{-g} = a^4 \, \det \left( \frac{\partial \phi^\mu}{\partial x^\nu} \right) ,
\ee
with $\phi^\mu = x^\mu + \bar{A} \partial_{\mu} \pi$.
Thus, the action is a sum of four terms of the form
\beq
S \propto \int d^4x \, \frac{\bar S}{4!}
\epsilon_{\mu_1\mu_2\mu_3\mu_4} \epsilon^{\nu_1\nu_2\nu_3\nu_4}
\frac{\partial\phi^{\mu_1}}{\partial x^{\nu_1}} \frac{\partial\phi^{\mu_2}}{\partial x^{\nu_2}}
\frac{\partial\phi^{\mu_3}}{\partial x^{\nu_3}} \frac{\partial\phi^{\mu_4}}{\partial x^{\nu_4}}
\eeq
for coefficients $\bar S$ related to the Ricci scalars of the two metrics and the matter contents in baryons and CDM,
which vanish thanks to the antisymmetry of the Levi-Civita tensor
\cite{Hassan2011}.

Thus, we have found that, at leading order in the approximation (\ref{appro}),
and setting the other metric modes $h_{\ell\mu\nu}$ to zero, the action
does not contain higher-order derivatives such as $(\partial^2\pi)^2$.
This agrees with the explicit expression (\ref{d2S-pi2}) for the quadratic action,
obtained without the approximation (\ref{appro}).
There, we can see that the leading terms $M_{\rm Pl}^2 {\cal H}^2 (\partial^2\pi)^2$
cancel out and the action only includes the subleading contributions
$M_{\rm Pl}^2 {\cal H}^4  (\partial\pi)^2$, with an extra factor ${\cal H}^2$ and
two fewer derivatives on $\pi$.
Thus, there is no Boulware-Deser ghost around the cosmological background,
at all orders over $\pi$ but in the small-scale and short-time approximation
(\ref{appro}) when we neglect the other metric modes $h_{\ell\mu\nu}$.

We analyze the terms $h\partial^2\pi$ in the
appendix~\ref{sec:Goldstone-metrics-coupling}.
We find that, even when the baryonic and dark matter metrics are different,
the St\"uckelberg field $\pi$ only couples to the metric combination $h_-$
as defined in Eq.(\ref{h-_def}), as in the case $s_{\rm d\ell}=s_\ell$ that was
explicitly considered in section~\ref{sec:quadratic-action}.
Besides, such terms $h\partial^2\pi$ can be written in terms of first-order
time derivatives, after integrating by parts over $\pi''$, hence they do not give
rise to ghosts.

\subsection{Cutoff scale}
\label{sec:cutoff}

We now investigate at which scale the terms we have neglected above
may introduce a ghost.
As can be seen from the explicit action (\ref{d2S-pi2}) and the terms in
$M_{\rm Pl}^2 {\cal H}^4  (\partial\pi)^2$, the canonically normalized
St\"uckelberg field $\tilde\pi$ is given by
\beq
\tilde\pi = \Lambda_3^3 \, \pi \;\;\; \mbox{with} \;\;\;
\Lambda_3 = \left( M_{\rm Pl} H^2 \right)^{1/3} ,
\label{pi-cano}
\eeq
up to a numerical factor of order unity.
Introducing the canonically normalized gravitons
$\tilde{h}_{\mu\nu} = M_{\rm Pl} h_{\mu\nu}$,
the terms that we have neglected above correspond to couplings between
$\tilde\pi$ and $\tilde{h}$ and derivatives of the background. They take the form
\beq
M_{\rm Pl}^2 H^2 \frac{\tilde{h}^n H^{2m-p} \partial^{p} \tilde\pi^{m}}
{M_{\rm Pl}^n \Lambda_3^{3m}} =
\frac{\tilde{h}^n \partial^p \tilde\pi^m}{\Lambda^{n+p+m-4}} ,
\eeq
and they are suppressed by a scale $\Lambda$ with
\beq
\Lambda = \Lambda_3 \left( \frac{\Lambda_3}{H} \right)^{(2n+2m-p-2)/(n+m+p-4)} .
\eeq
We have $n \geq 0$, $2m-p \geq 0$, and $\Lambda_3 \gg H$.
Therefore, $\Lambda \geq \Lambda_3$, except in the case $n=0$ and
$2m-p=1$. This corresponds to the combination $H \partial^{2m-1} \tilde\pi^m$,
where one partial derivative on $\tilde\pi$ is replaced by a background derivative $H$.
We have already found that there is no ghost in the quadratic action; therefore,
such a term can only give rise to ghosts if $m \geq 3$. This yields
for the lowest cutoff scale
\beq
\Lambda_{\rm cut} = \Lambda_3 \left( \frac{H}{\Lambda_3} \right)^{1/4}
= (M_{\rm Pl} H^3)^{1/4} ,
\eeq
which corresponds to $\Lambda_{\rm cut} \sim 1 {\rm a.u.} \sim 10^{-6} {\rm pc}$.
Therefore, at energies below $\Lambda_{\rm cut}$ there is no ghost in the model, but
the theory cannot be trusted on scales smaller than one astronomical unit,
and new contributions must be added to the action to ensure that there are
no ghosts. On the other hand, it can be used as an effective theory on all larger scales,
which are relevant for cosmology. 
The cutoff scale that we have deduced may be modified by
nonperturbative effects which are not investigated here.

The fact that the cutoff scale is of order 1 a.u. prevents our analysis from being applicable in most parts of the Solar System. However,
close to compact objects, or in the Solar System, on scales greater than 1 a.u.
and in the weak gravitational field regime, we can use the quadratic theory
described in section~\ref{sec:quadratic-action}  if we can neglect the dark
matter.  We can then separate the action in the sectors $S_+$ and $S_-$, with the dangerous
mode $\pi$ living in the sector $S_-$ at this order. Therefore, the field $\pi$
does not couple to matter and never enters the nonlinear regime due to matter
overdensities. At the classical level, $\pi=0$ is a solution of the equations of motion
(with all $h_{-\mu\nu}=0$), even when there are baryonic matter fluctuations.
Then, there is no need for a Vainshtein mechanism, down to the scale
$\Lambda_{\rm cut}^{-1}$.

\section{Links with doubly coupled bigravity}
\label{sec:dou}

The models that we have constructed have similarities with doubly coupled bigravity
\cite{Comelli:2015pua,Gumrukcuoglu:2015nua,Brax:2016ssf}.
In doubly coupled bigravity, there is no scalar field, and hence the Jordan-frame vierbein
couplings $s_*$ are constant, such as
\be
s_{\ell} = s_{{\rm d}\ell} = s_\ell^{(0)} ,
\ee
with a universal coupling to all types of matter, i.e. baryons, CDM and radiation.
In both the matter and radiation eras, the scale factors are in the symmetric case
\be
a_\ell= b_\ell , \;\;\; {\cal H}_{\ell} = {\cal H} ,
\label{nat}
\ee
implying that the two metrics are proportional. The late-time acceleration of the
expansion of the Universe is obtained by adding a potential term
\be
S_V= \Lambda^4 \int d^4 x \sum_{ijk\ell} m^{ijk\ell} \epsilon^{\mu\nu\rho\sigma}
\epsilon_{abcd} e^{a}_{i\mu} e^{b}_{j\nu} e^{c}_{k\rho} e^{d}_{\ell\sigma}
\ee
comprising one scale and a completely symmetric tensor $m^{ijk\ell}$ which,
up to rescaling, is associated to four coupling constants.
This term is responsible for the late-time acceleration where $\Lambda^4$ plays the role
of the vacuum energy. Moreover, the potential term gives rise
to a mass matrix for the gravitons whose order of magnitude corresponds to
$\Lambda^4 /M_{\rm Pl}^2 \sim H_0^2$, i.e very light gravitons.

At the background level, and as long as the scalar field is negligible, the bimetric models
considered here coincide with the bigravity theories. They differ when it comes to the
phase of acceleration. In bigravity, this is simply realized as $\Lambda^4$ plays the role of
dark energy. In scalar-bimetric models, there is no vacuum energy and
the acceleration is simply due to the rapid variation of the scalar factors
$s_*(\varphi)$, which imply that the baryonic and dark matter metrics do not mimic the
ones of the Einstein-de Sitter space-time.
In the acceleration phase in doubly coupled bigravity
\be
r_2 \neq r_1 ,
\ee
that is, the two gravitational metrics do not have the same conformal time.
For scalar-bimetric models, we have seen that natural models obey $r_1=r_2=1$
even at late times.
In a similar fashion, in bigravity the consistency of the Friedmann equations
gives a constraint equation that admits two branches of solutions
\cite{Comelli:2015pua,Gumrukcuoglu:2015nua,Brax:2016ssf}, the interesting
one for cosmology being ${\cal H}_{a_1}/r_1 = {\cal H}_{a_2}/r_2$ as noticed in
Eq.(\ref{branch-2}). In our case, the scalar field provides an additional degree of
freedom and there is no such constraint. As in General Relativity, the Friedmann
equations and the equations of motion of the various fluids are automatically consistent.
This follows from the fact that Eq.(\ref{ds-dsd}) is no longer a constraint equation,
because of the scalar-field dynamics.
As we checked in section~\ref{sec:backg-scalar-field-eom}, the equation of motion
of the scalar field is not independent of the Friedmann equations and of the equations
of motion of the other fluids, as it can be derived from the latter.

When it comes to the scalar perturbations, bigravity in the doubly coupled case and
scalar-bimetric models differ more drastically as the cosmological perturbations of the
scalar field imply the existence of a scale, related to its effective mass, such that for large
enough wave numbers gravity is modified. This leads to a fifth force that is of order
of the Newtonian force on cosmological scales at $z=0$.
Moreover, as the scalar field evolves
in the late-time Universe, the effective Newton constants (it is not unique anymore but
depends on the species) drift with time. This has also an effect on cosmological
perturbations.

Vector and tensor perturbations in the radiation era have similar behaviors in doubly
coupled bigravity and scalar-bimetric models, with both tensor and vector instabilities.
In the matter era, the nontrivial mass matrix for the two gravitons in doubly coupled
bigravity implies that the two gravitons oscillate leading to birefringence
\cite{Brax:2017hxh}.
Moreover, in doubly coupled bigravity, the speed of the gravitational waves differs from unity
in the late-time Universe as the ratio between the two lapse functions of the two metrics
is not equal to one anymore. This is severely constrained by the LIGO/VIRGO observations.
In contrast, in scalar-bimetric models we have shown that symmetric solutions where
$a_\ell=b_\ell$ can be obtained even during the acceleration phase.
In this case, the speed of the gravitational waves is always unity.
Moreover, at the linear level, there is no mixing between the tensor and vector
instabilities that affect the ``hidden'' modes and the matter metrics.

Finally, let us note another analogy between the bimetric models presented here and doubly coupled bigravity. The breaking of the
full diffeomorphism invariance to the diagonal subgroup is parametrized by the mass of the gravity $m$ in the latter and the Hubble expansion rate $H$ in the former.
In both cases, the strong coupling scale is given by $\Lambda_3= (v^2 M_{\rm Pl})^{1/3}$ where $v=H,m$ is the order parameter of each case. At energies
larger than this scale, ghosts are present, and a completion of the models is required. Notice that in the scalar-bimetric models ghosts may actually appear at the lower scale
$(H^3 M_{\rm Pl})^{1/4}$. In both theories, around compact objects in the weak gravitational regime for distances larger than their respective cutoff scales,  the scalar Goldstone mode decouples without the need for the Vainshtein mechanism.

On the other hand, as we are now going to analyze, the time variation of the scalar field in scalar-bimetric
models poses new problems which are late-time issues, i.e. not only restricted to the radiation era contrary to what happens in bigravity \cite{Brax:2016ssf}.

\section{Recovering General Relativity on small scales ?}
\label{sec:cancel}

As shown in Fig.~\ref{fig_sym_mu}, the scenarios obtained so far are not consistent with
small-scale tests of General Relativity. First, the fifth force is too large,
being about twice stronger than Newtonian gravity at $z=0$, as measured
by the ratio $\mu^{\phi}/{\cal G}^{\phi}-1$.
Second, the time derivative of the effective Newton constant is too high at $z=0$,
with $d\ln {\cal G}/{dt} \sim 0.7 H_0$ whereas the Lunar Laser Ranging (LLR) experiment gives
the upper bound $0.02 H_0$ ($d\ln {\cal G}/{dt} < 1.3\times 10^{-12} \, {\rm yr}^{-1}$)
\cite{Williams:2004qba}. Strictly speaking,  this constraint lies beyond the realm of validity of the models
as coming from scales below 1 a.u.. On the other hand, less stringent constraints on the planetary orbits exist \cite{Uzan:2002vq}  at the $10^{-11}$ level and should be fulfilled. Hence,
we will use the LLR bound as a template for any UV completion of scalar-bimetric models.
Third, the change of the Newton constant from its large-redshift value to its current
value is too large.
Indeed, we obtain an increase of ${\cal G}$ of about $50\%$ from its high-$z$ asymptote
to its value at $z=0$.
Here, we normalized the Planck mass at $z=\infty$ to its measured value in the Solar System
today, and defined the cosmological parameters in terms of the same Planck mass in
Eq.(\ref{Omega-def}). Instead, we should normalize both the Newton constant at $z=0$
and the cosmological parameters (i.e., the matter densities) to the measured value of
${\cal G}_{\rm N0}$.
However, we would face the same problem. Because we have no dark energy, to recover the
$\Lambda$-CDM expansion at high $z$ with the same background densities, we need the effective
Newton constant at high $z$ to be the same as in the $\Lambda$-CDM scenario,
which is also the measured value today. Thus, we need ${\cal G}_{\rm N}$ at $z=0$ to be
equal to ${\cal G}_{\rm N}$ at $z \gg 1$, unless we modify the dark matter and radiation
densities by a similar amount (with respect to the $\Lambda$-CDM reference).
However, it is not possible to change the background densities by $50\%$ while keeping
a good agreement with the CMB and BBN constraints.

These three problems are not necessarily connected. In modified-gravity models,
the fifth force is assumed to be damped in the local environment
by nonlinear screening mechanisms
(which use the fact that the Solar System length scale is much smaller than
cosmological distances and/or the local density is much higher than the cosmological
background densities).
However, it is usually assumed that the time dependence of the Newton constant,
and often its value, remain set by the cosmological background, which acts as
a boundary condition.
In particular, derivative screening such as the Vainshtein screening,
where the nonlinear terms are invariant under $\varphi \to \varphi + \alpha t$
with arbitrary $\alpha$, does not seem to prevent a slow drift of Newton constant.
Then, unless the local Newton constant can be significantly decoupled from the
cosmological background solution (e.g., through a more efficient screening that remains
to be devised), we need to modify the background solution itself to decrease
both $d\ln {\cal G}/{dt}(z=0)$ and $\Delta{\cal G}={\cal G}(z=0)-{\cal G}(z=\infty)$.

\subsection{Reducing $d\ln {\cal G}/{dt}$}
\label{sec:dlnGNdt}

\subsubsection{Constant ${\cal G}$ ?}
\label{sec:G-constant}

The most elegant way to reduce $d\ln {\cal G}/{dt}$ below the Hubble timescale
would be to keep it (almost) constant, so that one would not  need any tuning to decrease the
time derivative precisely at $z=0$. Moreover, this would ensure that ${\cal G}$ would be  about the same
at $z=0$ and $z\gg 1$.

\paragraph{Scenarios with common conformal time}

\begin{figure}
\begin{center}
\epsfxsize=8.5 cm \epsfysize=5.2 cm {\epsfbox{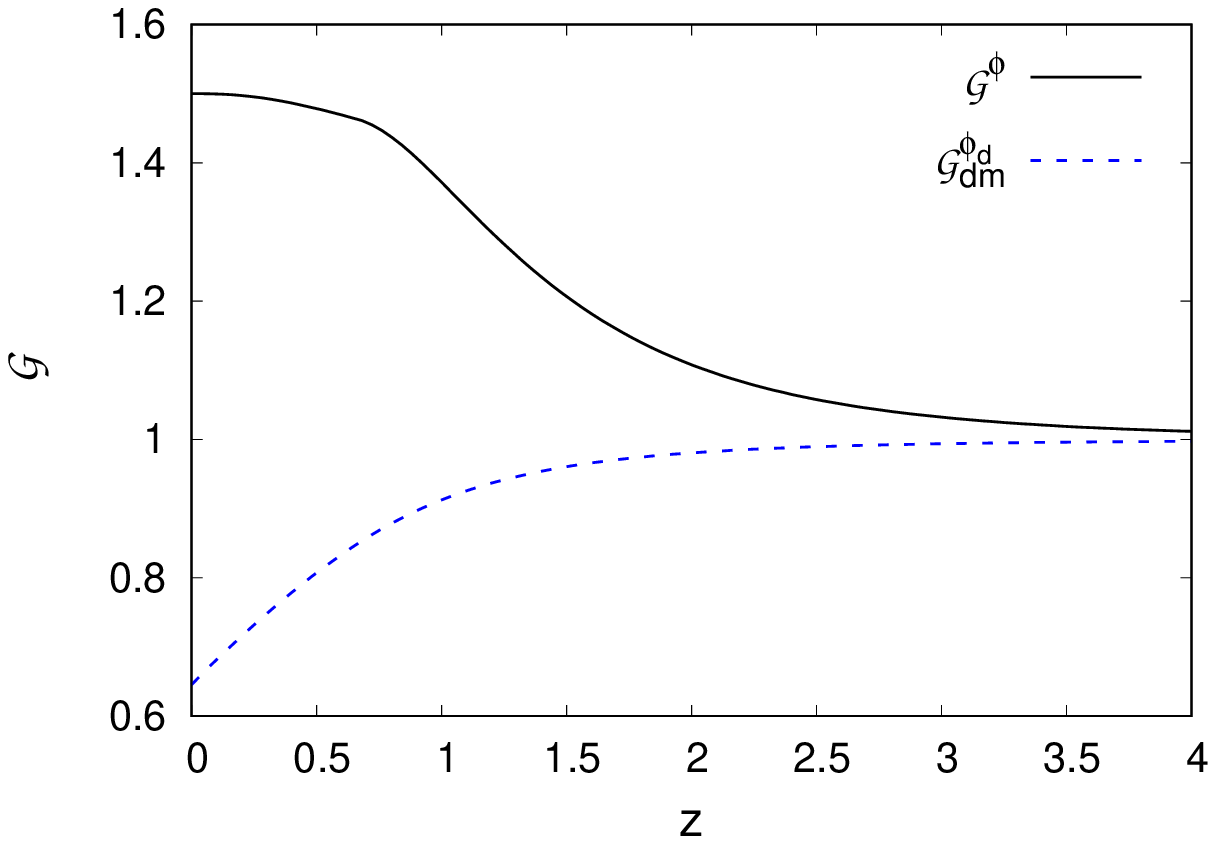}}
\epsfxsize=8.5 cm \epsfysize=5.2 cm {\epsfbox{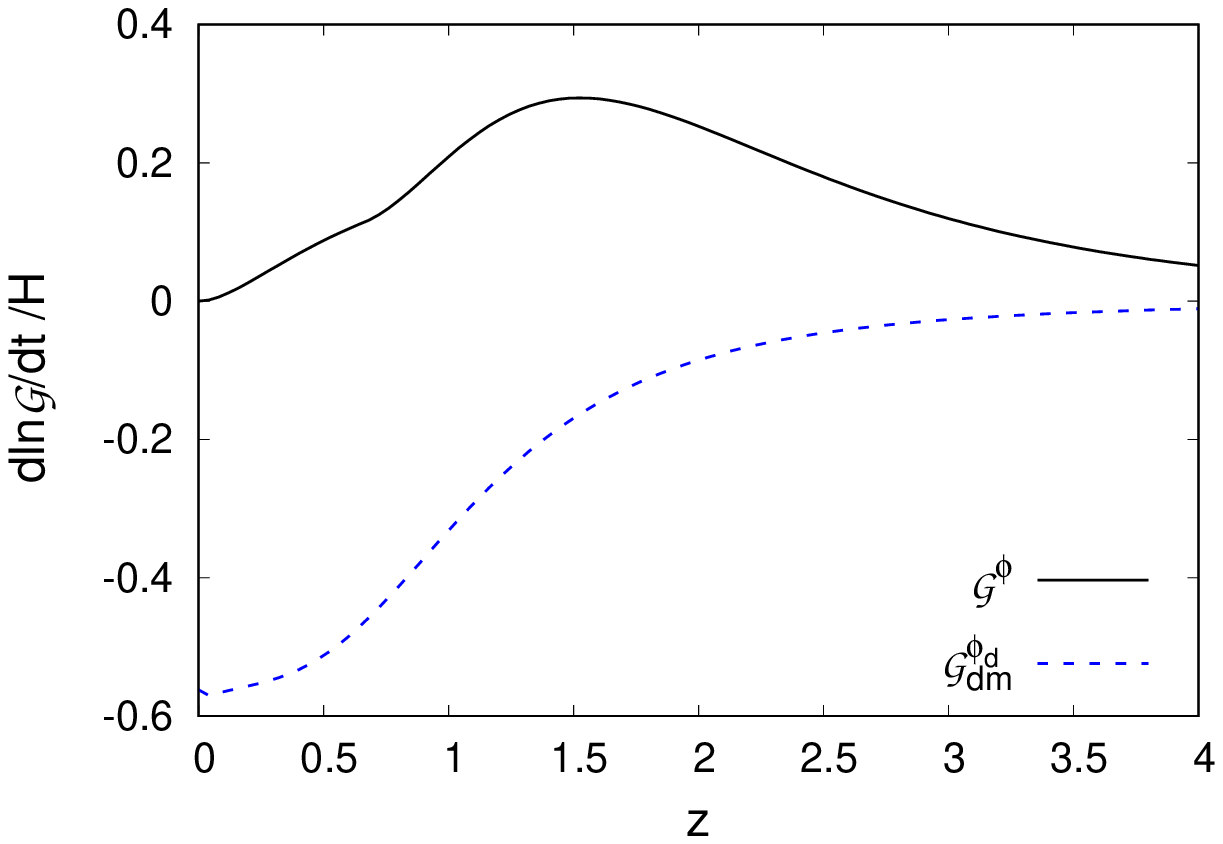}}
\end{center}
\caption{{\it Upper panel:} baryonic sector and dark sector Newton constants, normalized
to ${\cal G}_{\rm N}$.
{\it Middel panel:} time derivatives $d\ln {\cal G}/{dt}$ normalized to $H$.
{\it Lower panel:} growth factors $f \sigma_8$ and $f_{\rm dm} \sigma_{\rm dm 8}$.}
\label{fig_GN_lambda}
\end{figure}

Let us first consider the case of the scenarios with $r_{\ell}=1$, described in
section~\ref{sec:backg-common-conformal}.
Then, from Eq.(\ref{G-b}) a constant ${\cal G}$ corresponds to a constant $\lambda$
in Eq.(\ref{lambda-def}). Unfortunately, the solution (\ref{s2-lambda})
does not exist for any $\lambda(a)$, as the argument of the square root needs to remain
positive. Numerically, we found that it is not possible to keep a constant $\lambda(a)=1$,
at all times.
This can be understood from the behavior of the scales factors $a_{\ell}$.
As noticed in Fig.~\ref{fig_lambda_a1a2} and explained below Eq.(\ref{Friedmann-00-low-z}),
the behavior of the scale factors $a_*$ and Hubble expansion rates ${\cal H}_*$ are almost
independent of the evolution of the coefficients $s_{\ell}$, because we impose
a $\Lambda$-CDM-like expansion for the baryonic metric. This implies that the ratios
$a_{\ell}/a$ decrease with time, as the gravitational metrics $g_{\ell \mu\nu}$ follow
an expansion close to the Einstein-de Sitter prediction (because we do not put any cosmological
constant or dark-energy component that would play the same role).
Then, to keep the square root real in Eq.(\ref{s2-lambda}),
$\lambda(a)$ must typically increase with time. In any case, its value at $z=0$ must be greater
than unity. From the values of $a_{\ell}/a$ read in Fig.~\ref{fig_lambda_a1a2},
we find $\lambda(z=0) \gtrsim 1.5$. This means that Newton's constant ${\cal G}$ at $z=0$
must be about $50 \%$ greater than its value at high redshift.

We show in Fig.~\ref{fig_GN_lambda} the Newton constants for the baryonic and dark sectors
obtained in this manner, with the function $\lambda$ used for Fig.~\ref{fig_lambda_a1a2}
such that $d\lambda/da=0$ at $z=0$.
This allows us to reduce $d\ln {\cal G}^{\phi}/H{dt}$ at all redshifts below $0.3$,
and make it smaller than the Lunar Laser Ranging upper bound at $z=0$.
On the other hand, for the dark sector we still have the generic feature
$d\ln {\cal G}^{\phi_{\rm d}}_{\rm dm}/H{dt}$ of order unity at $z=0$.
Making $\lambda(a)$ almost constant at low $z$ is not so artificial, in the sense that it is
a simple constraint on the coefficients $s_{\ell}$, which are likely to be correlated in any case.
Moreover, the plateau for ${\cal G}^{\phi}$ can be reached at $z \gtrsim 1$, and does not need
to be tuned at $z=0$ precisely.
However, a few numerical tests suggest that it is difficult, or impossible, to make the transition
for ${\cal G}^{\phi}$ occur at much higher redshifts, such as $z =10$. This tends to make
$s_2$ negative at intermediate redshifts, amplifying the dip already seen in
Fig.~\ref{fig_lambda_a1a2}, and we prefer to keep the coefficients $s_{\ell}$ positive
(but this requirement may be unnecessary).

From the arguments discussed above, if the sum $s_1^2+s_2^2$ reaches a constant value
at late times, or satisfies a finite upper bound, the decrease of the ratios $a_{\ell}/a$ must
eventually stop in the future (a simple case is where each coefficient $s_{\ell}$ eventually
becomes constant). Then, as the gravitational metrics, the baryonic metric must recover an
Einstein-de Sitter expansion, unless the energy density and pressure of the scalar field
become dominant. Therefore, in this framework where the acceleration of the expansion
is not due to an additional dark-energy fluid, the self-acceleration is only a transient
phenomenon.
An alternative would be that the Newton constant resumes its growth in the future,
but this would introduce an additional tuning as the slow down of $d\ln {\cal G}^{\phi}/H{dt}$
would be a transient phenomenon that must be set to occur precisely around $z \simeq 0$.

It is interesting to note that the nonsymmetric solutions, such as (\ref{lambda-def}),
give rise to behaviors beyond those obtained in models where the baryonic and
dark matter metrics are simply given by different conformal rescalings of a single
Einstein-frame metric. There, we only have two free functions, $A(\varphi)$ and
$A_{\rm d}(\varphi)$, with $g_{\mu\nu}= A^2 \tilde{g}_{\mu\nu}$ and
$g_{{\rm d}\mu\nu}= A_{\rm d}^2 \tilde{g}_{\mu\nu}$. This would correspond for instance
to $s_1 = A$ and $s_2=0$, that is, there is no second gravitational metric.
As there is only one coupling $A$, both the baryonic scale factor $a$ and the baryonic
Newton constant ${\cal G}^{\phi}$ depend on $A(\varphi)$ and run at the same rate.
This means that it is not possible to have a self-accelerated expansion, driven by
$A(\varphi)$, while keeping ${\cal G}^{\phi}$ constant.
In the bimetric scenario, even in the common conformal time case, we can take advantage
of the two free functions $s_1(\varphi)$ and $s_2(\varphi)$ to keep a constant
Newton strength ${\cal G}^{\phi}$ while having self-acceleration.
However, as explained above, this can only happen for a finite time
(if we require $s_{\ell}>0$) and we cannot reduce the gap
$\Delta{\cal G}={\cal G}(z=0)-{\cal G}(z=\infty)$.
Therefore, this scenario is not sufficient to make the model agree with observational
constraints.
Presumably, increasing the number of metrics, hence of degrees of freedom and free functions
of the model, would make it increasingly easy to reconcile a constant Newton strength with
self-acceleration.

\paragraph{Scenarios with different conformal times}

In the case of the scenario (\ref{r1-1-r2}), with $r_{\ell}\neq 1$,
we explicitly checked that we can build solutions such that ${\cal G}^{\phi}$
remains constant at all times, by tuning the factors $r_{\ell}$.
More precisely, from Eq.(\ref{G-b}) a constant ${\cal G}^{\phi}$ corresponds to
\beq
\frac{d {\cal G}^{\phi}}{d\ln a} = 0 : \;\;\;
\sum_{\ell} 2 s_{\ell} r_{\ell} \frac{ds_{\ell}}{d\ln a} + s_{\ell}^2 \frac{dr_{\ell}}{d\ln a} = 0 .
\label{dG-def-0}
\eeq
Using the expressions (\ref{s1s2-r1r2}), we can write $\{ds_{\ell}/d\ln a\}$ in terms
of $\{dr_{\ell}/d\ln a\}$.
This determines for instance the derivative $dr_{2}/d\ln a$ while keeping $r_1$ free,
so that this family of solution is still parametrized by a free function $r_1(a)$.
However, this usually gives ${\cal G}^{\psi} \neq {\cal G}^{\phi}$, see Eq.(\ref{G-b}),
with a relative deviation of order unity.
To be consistent with Solar System data, in particular with the Shapiro time
delay that measures the travel time of light rays in gravitational potentials,
we must have $|\psi/\phi-1| \leq 5 \times 10^{-5}$ \cite{Will2006}.
On the other hand, as explained in section~\ref{sec:backg-different-conformal}
we need $r_1=1$ (or $r_2=1$) at $z=0$ to comply with the multimessenger
gravitational waves event GW170817.
This would give both $s_2=0$ and ${\cal G}^{\psi} = {\cal G}^{\phi}$ at $z=0$.
However, when we try to combine Eq.(\ref{dG-def-0}) with $r_1 \to 1$ at $z=0$
in a few numerical tests, we find singular behaviors with $b_2$ becoming negative
before $z=0$ and $a_2 \to 0$ at $z=0$. This is somewhat reminiscent of the impossibility
to achieve a constant ${\cal G}^{\phi}$ in the simpler case $r_1=r_2=1$ shown
in Fig.~\ref{fig_GN_lambda}.
Because the scenarios $r_{\ell}\neq 1$ already require some tuning, with
$| r_1-1 | < 3\times 10^{-15}$ at $z=0$, we do not investigate further this family of solutions.

\subsubsection{Constant $s_{\ell}$ at late times}
\label{sec:late-time}

\begin{figure}
\begin{center}
\epsfxsize=8.5 cm \epsfysize=5.2 cm {\epsfbox{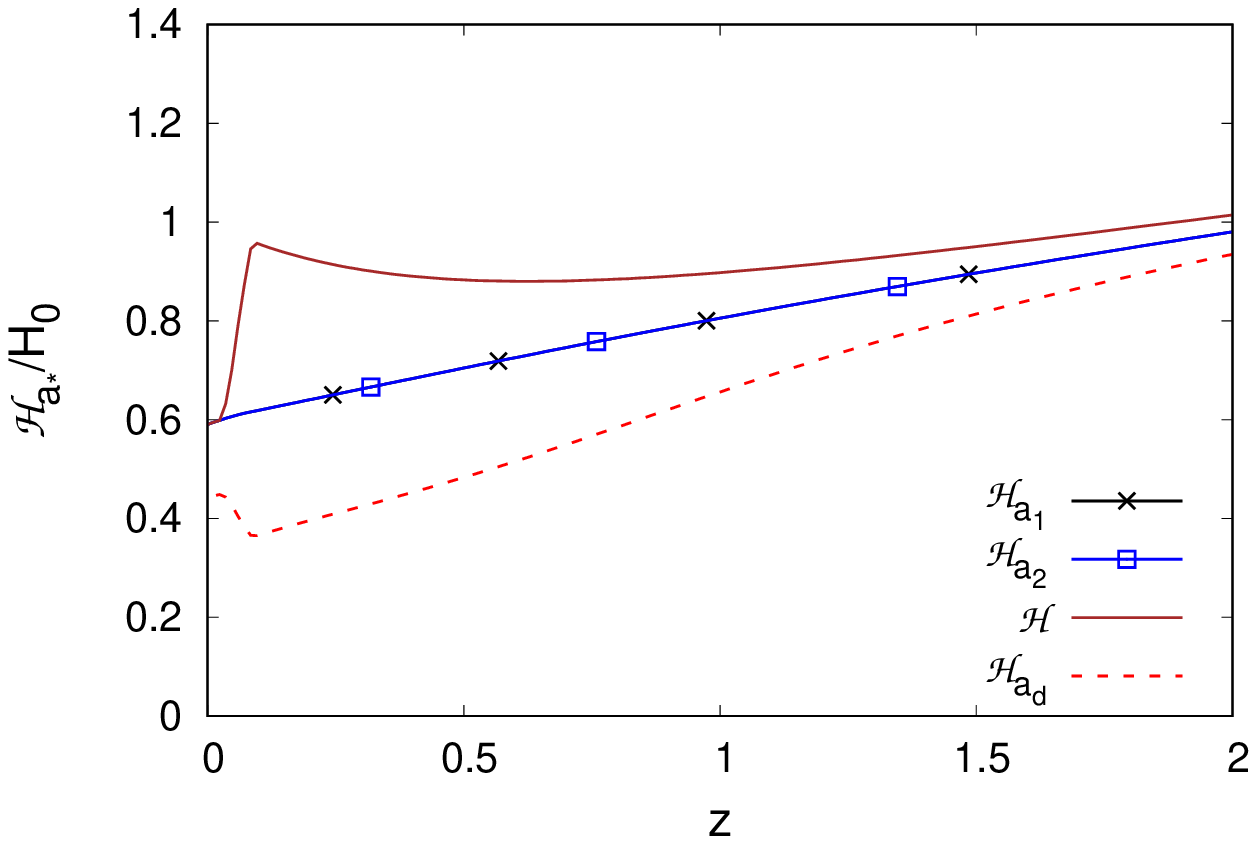}}
\epsfxsize=8.5 cm \epsfysize=5.2 cm {\epsfbox{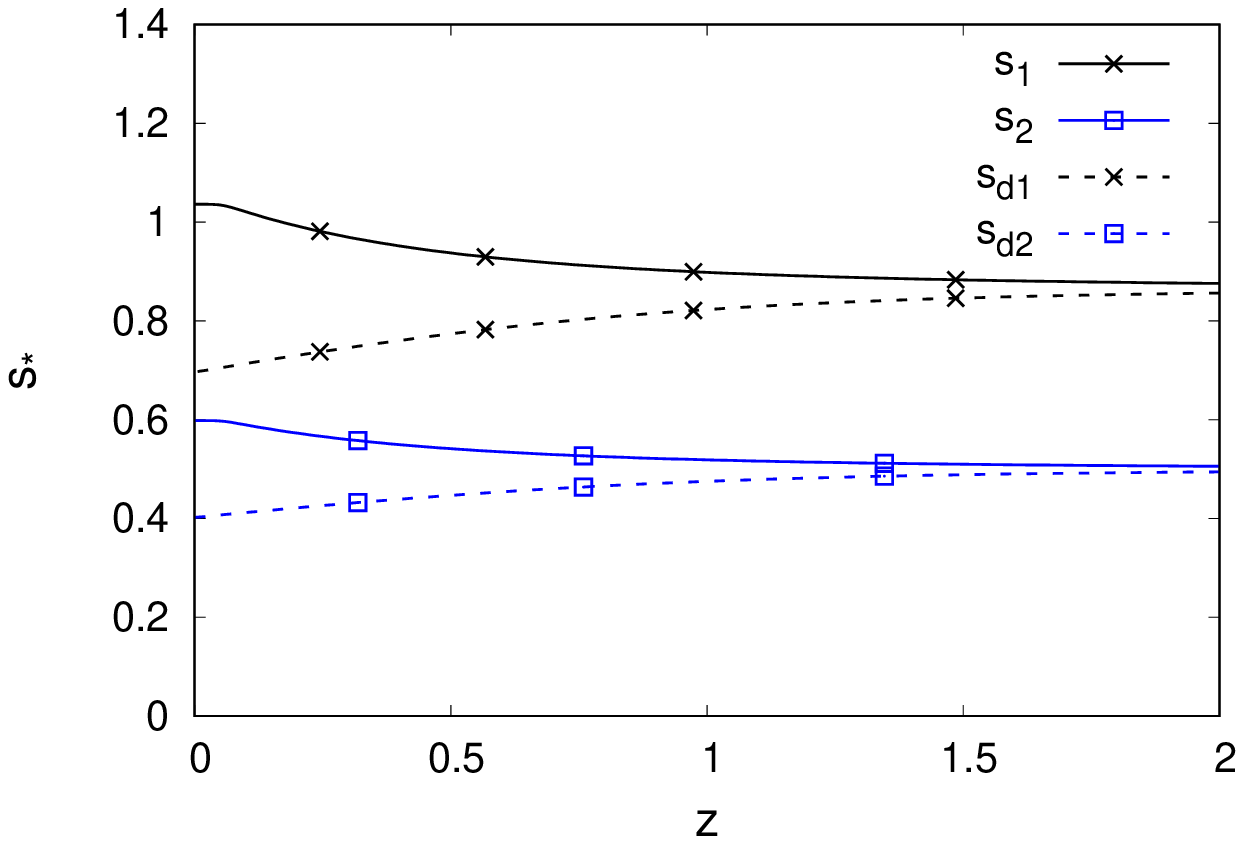}}
\end{center}
\caption{Conformal Hubble expansion rates (upper panel) and coefficients $s_*$
(lower panel) as a function of redshift, for a solution where the baryonic coefficients
$s_{\ell}$ are constant at late times.}
\label{fig_sym-s-const_a1a2}
\end{figure}

A natural solution to obtain a small $d\ln {\cal G}/{dt}$ at low $z$ is to consider
models where the coefficients $s_{\ell}$ reach a constant at late times.
This also removes any fifth force on baryons, as $\beta=0$ from
Eq.(\ref{betad-beta-def}).
However, this also makes the baryonic metric expansion rate converge again
to an Einstein-de Sitter behavior, in agreement with the simple solution of
section~\ref{sec:backg-EdS}. The deviation of $s_1^2+s_2^2$ from unity in this late-time
asymptote again corresponds to a different value for the associated Newton's constant,
as compared with the one obtained at high redshift.

We show in Fig.~\ref{fig_sym-s-const_a1a2}
our results for the symmetric solution of Fig.~\ref{fig_sym_a1a2}, which is modified
at late times so that the baryonic coefficients are constant for $a>0.9$.
In terms of these coefficients, this model is rather simple as the accelerated expansion
of the Universe is a transient phenomenon, due to the transition of the coefficients
$s_i$ between two constant asymptotes.
By requiring the Hubble expansion rate to follow the $\Lambda$-CDM history
until $z \gtrsim 0.1$, we make the transition to the final Einstein-de Sitter behavior occur
in a very small redshift interval.
This leads to a sharp decrease for the baryonic expansion rate ${\cal H}(z)$,
which suddenly drops to the expansion rate ${\cal H}_{a_1}={\cal H}_{a_2}$ of the
gravitational metrics.
This also leads to a sudden increase in the growth rate of large-scale structures, which resumes
the faster growth associated with Einstein-de Sitter cosmologies.

Even though the change of the coefficients $s_{\ell}$ is very small, as compared with the
solution of Fig.~\ref{fig_sym_a1a2}, this leads to a change for the Hubble expansion rate of
order unity. Indeed, by making the coefficients $s_i$ constant at late times, we change
their time derivative $ds_i/d\tau$ from a quantity of order $1/H_0$ to zero over a small
time $\Delta\tau$. This yields a divergent second derivative
$d^2s_{\ell}/d\tau^2 \propto 1/(\Delta\tau)$.
However, from Eq.(\ref{a-ad-a1-a2}) we can see that $d\ln {\cal H}/d\ln a$, being a second
derivative of the scale factor, contains a term such as $d^2s_{\ell}/d\tau^2$ and also
grows as $1/(\Delta\tau)$. Then, even if we let the transition time $\Delta\tau$ go to zero
the change of ${\cal H}$ remains finite and of order unity, in agreement with
Fig.~\ref{fig_sym-s-const_a1a2}.
The drop of $H(z)$ at low $z$ to about $60\%$ of the $\Lambda$-CDM extrapolation
$H_0$ implies a deviation of the distance modulus, $\mu=5 \log_{10}(d_L/10 {\rm pc})$,
of $\Delta\mu=-5 \log_{10}(0.6) \simeq 1.1$.
However, the dispersion of the distance modulus of observed type Ia supernovae in the range
$0.01 < z < 1$ is of order $0.3$, before binning \cite{Betoule2014}, and does not show
such a steep step.
Therefore, the Hubble diagram shown in Fig.~\ref{fig_sym-s-const_a1a2}
is ruled out by low-redshift supernovae.

In addition, we still have a total increase of ${\cal G}$ of about $50\%$ between
the high-$z$ and low-$z$ values of the effective Newton constant.
Therefore, this scenario would not solve this third problem in any case.

\subsection{Need for screening beyond quasistatic chameleon mechanisms}
\label{sec:screening}

We have seen in the previous section that the coefficients $s_{\ell}$ are unlikely
to have reached constant values by $z=0$, to be consistent with the low-$z$
Hubble diagram. This yields a fifth force that is of the same order as the Newtonian
force on cosmological scales. All scenarios also imply a decrease of order $50\%$ of the
effective Newton constant at higher redshifts, which makes it impossible to recover
the reference $\Lambda$-CDM expansion rate unless the matter and radiation densities
are also modified. This means that such scalar-bimetric models can only satisfy
observational constraints if gravity in the Solar System is decoupled from
its behavior on cosmological scales.

Within modified-gravity scenarios, the recovery of General Relativity on small scales
is often achieved by introducing nonlinear screening mechanisms that damp
the effect of the fifth force.
For instance, chameleon screening makes the scalar field short ranged in high-density
environments, because its effective potential and its mass depend on the matter
density. In a similar fashion, dilaton and symmetron scenarios damp the fifth force
by making its coupling vanish in high-density environments, following the
Damour-Polyakov screening.

It is interesting to note that these screening mechanisms cannot appear in the models
considered in this paper, because the scalar field always remains in the linear regime.
A first way to see this is from Eq.(\ref{KG-k}), which yields
$\delta\varphi/M_{\rm Pl} \sim v^2$ for a structure of virial velocity
$v^2 \sim {\cal G}M/r$, mass $M$ and radius $r$.
Then, in nonrelativistic environments, from clusters of galaxies to the Solar System,
where $v^2 \ll 1$, we have $\delta\varphi \ll \bar\varphi$ as we found in
Fig.~\ref{fig_sym_a1a2} that $\bar\varphi \sim M_{\rm Pl}$.
This also implies that $\delta s_{\ell} \ll \bar{s}_{\ell}$. Thus, from clusters of
galaxies to the Solar System the fluctuations of the scalar field remain small
and are not sufficient to significantly modify the coefficients $s_{\ell}$.
This means that the effective Poisson equation (i.e., the coefficients $\mu^*_*$)
keeps the same deviation from General Relativity on all these scales.

This configuration can be compared with the usual chameleon or Damour-Polyakov
screenings, shown by $f(R)$ or Dilaton and Symmetron models. There, the
Jordan-frame metric is typically related to the Einstein-frame metric
by a conformal coupling, $g_{\mu\nu} = A^2(\varphi) \tilde{g}_{\mu\nu}$.
The fifth force $c^2 \nabla \ln A$ again arises from the fluctuations
of this metric coefficient $A$, through the fluctuations of the scalar field.
However, in these models which typically include a cosmological constant,
either explicitly or as the nonzero minimum of some potential,
the conformal coupling always remains very close to unity, $|A-1| \lesssim 10^{-5}$.
This ensures that one follows the $\Lambda$-CDM background while having effects
on cosmological structures that can be of order unity, with $\delta A \sim \phi$.
The very small variation of the background value of $\bar{A}$ also means that it is
easy to introduce a screening mechanism, because the spatial perturbations
of $\delta \varphi$ and $\delta A$ can be of the same order as those of the cosmological
background over $\delta z \sim 1$, so that the nonlinear regime is easily reached
(this may be more easily understood from a tomographic point of view).
In the model considered in this paper, the difficulty arises from the fact that
we require background variations of order unity for the coefficients $s_{\ell}$,
which play a role similar to $A^2(\varphi)$ in the conformal coupling models,
whereas spatial variations should remain of order $10^{-5}$ of the same
order as the standard Newtonian potential.
This implies that spatial fluctuations of the scalar-field value are not sufficient
to reach the nonlinear regime.
This analysis agrees with the ``no-go'' theorem of Ref.\cite{Wang:2012kj},
which concludes from the same arguments that usual chameleon models
cannot provide a self-acceleration of the Universe, and must rely on a form of dark energy
(typically a hidden cosmological constant, written as the nonzero minimum of some
potential).

A way out of this difficulty is to introduce screening mechanisms that do not rely
on the scalar-field value, but on its derivatives. Then, even though $\delta\varphi$
remains small, its spatial derivatives $\partial^n\delta\varphi$ can be large
on small enough scales. This corresponds to K-mouflage and Vainshtein mechanisms.
This can be achieved by adding terms in $(\partial \varphi)^4/M^4$ or
$\Box \varphi (\partial \varphi)^2/M^3$.
In this case, these nonlinear terms dominate over the simple kinetic terms at short distance
depending on the value of $M$. As a result, the coupling
of the scalar field to the baryons (and incidentally the one to dark matter) is reduced and
local tests of gravity are satisfied.
However, this only solves the fifth-force problem, and it does not solve the problems associated
with the value of Newton constant and its time drift.
(In these screening scenarios, they are usually assumed to be set by the cosmological
background, which acts as a boundary condition.)

The analysis above implicitly assume the quasistatic approximation, where the scalar
field relaxes to its environment-dependent equilibrium and screening appears through the
spatial variations of its mass, coupling or inertia.
If the quasistatic approximation is violated, the configuration may be more complex.
In fact, from the analysis of section~\ref{sec:dlnGNdt}, we can see that we need
a local value of Newton constant that is decoupled from the one on cosmological scales.
More precisely, we need its local value to remain equal to its background value at high $z$,
before the dark-energy era.
This calls for a new screening mechanism, or a more efficient implementation of
K-mouflage or Vainshtein screening, that goes beyond the quasistatic approximation
and decouples the small-scale Newton constant from its current large-scale cosmological value.
For instance, the local Newton constant should remain equal to the one at the formation
of the Solar System.
All this requires altering the models and imposing stringent restrictions on the possible
UV completions of the models that must be introduced in the Solar System below 1 a.u..

\section{Conclusion}
\label{sec:conclusion}

We have seen in this paper that the scalar-bimetric model allows one to recover
an accelerated expansion without introducing a cosmological constant or an almost constant
dark-energy density.
This relies on the time-dependent mapping
between the gravitational metrics $g_1$ and $g_2$
and the baryonic and dark matter metrics $g$ and $g_{\rm d}$.
Because at late times the deviation between the $\Lambda$-CDM and Einstein-de Sitter
backgrounds is of order unity, the coefficients $s_{\ell}$ that define this mapping
must show variations of order unity.

When all metrics have the same conformal time, the expansion rates
of the gravitational and dark matter metrics are almost independent of the details of the model
[e.g., the shape of the functions $s_{\ell}(\varphi)$], once we require a $\Lambda$-CDM expansion
for the baryonic metric. Then, the gravitational metrics remain close to an Einstein-de Sitter
expansion (because there is no dark energy), while the dark matter metric
behaves in a way opposite to the baryonic metric, with a stronger deceleration than in the
Einstein-de Sitter case.
When the conformal times are different, the scale factors $a_*$ can show slightly different
behaviors, and even more so the lapse factors $b_*$.
This scenario is very strongly constrained by the multimessenger event GW170817,
which requires that at least one of the two gravitons propagates at the speed of light
at $z<0.01$. This implies that at least one of the ratios $b_{\ell}/a_{\ell}$ must be unity
at low $z$. This also implies that the baryonic metric becomes independent at low $z$
of the gravitational metric where $c_g \neq 1$, but the dark matter metric still remains sensitive
to both gravitational metrics.

As the coefficients $s_{\ell}$ must show variations of order unity to provide
a self-acceleration, we generically have deviations of order unity for the
effective Newton constants and for the contribution from the fifth force to the
dynamical potential seen by particles.
The dynamics of baryonic and dark matter perturbation show distinctive features,
due to the fact that they couple to different metrics and that their mappings evolve
in opposite fashions.
While the total force (Newtonian gravity and fifth force) from baryons onto baryons,
and from dark matter onto dark matter, is typically amplified at low redshift,
the cross-force between baryons and dark matter is damped and even turns negative.
This means that dark matter and baryons would tend to segregate (although this does not have
the time to happen by $z=0$ on large scales).
Then, the growth of dark matter density fluctuations is amplified (because of the stronger
self-gravity) while the growth of baryonic density fluctuations is decreased on cosmological
scales (because of the lower cross-gravity, as dark matter is dominant on large scales).
This could provide interesting features; for instance, most modified-gravity models predict
instead an amplification of baryonic density perturbations.

However, before a detailed comparison with cosmological observations, these models
present major difficulties with small-scale tests of gravity.
First, the fifth force is of the same order as Newtonian gravity.
Second, the baryonic effective Newton constant generically evolves on Hubble time scales.
Third, it is greater than its high-$z$ value by about $50\%$.
These features are related to the self-acceleration, which implies modifications
of order unity on Hubble timescales.

Thanks to the two couplings associated with the two gravitational metrics,
it is possible to keep the baryonic effective Newton strength almost constant at low $z$.
(By keeping the sum $s_1^2+s_2^2$ constant while the two coefficients vary.)
This is beyond the reach of simpler models where the baryonic metric would be given
by a conformal rescaling of a single Einstein-frame metric [which provides a single
coupling $A(\varphi)$].
However, this can only work for a finite time. Either the baryonic and dark matter metrics
eventually recover an Einstein-de Sitter expansion in the future, or the Newton coupling
resumes its growth in the future.
In this framework, it is more natural to make the self-acceleration only a transient
phenomenon, associated with the running of the couplings $s_{\ell}(\varphi)$ between
two constant asymptotes (where the fifth force and the running of Newton constants
disappear).
(The alternative scenario, where the coefficients $s_{\ell}(\varphi)$ have already reached
their constant asymptote at low $z$, is rejected by measurements of the Hubble expansion
rate, from low-$z$ supernovae or local standard candles such as cepheids.)
However, this cannot reduce the gap between the high-$z$ and low-$z$ values of
Newton's constant.

On small scales, Solar System tests of gravity imply that we must recover General Relativity.
In modified-gravity scenarios, this is often achieved by introducing nonlinear
screening mechanisms that damp the effect of the fifth force.
As in the case of single-metric and single-field models, we explain that a chameleon mechanism
cannot work. It cannot efficiently screen the fifth force in a self-accelerated model.
This leaves derivative screening mechanisms, such as K-mouflage and Vainshtein
screenings.
Therefore, the scalar-field Lagrangian must be supplemented by higher-order derivative terms,
that become dominant on small scales and provide the convergence to General Relativity.
on small scales by damping the fifth force.
However, we need to go beyond usual implementations, as we also require the local Newton
constant to be decoupled from its cosmological value and to remain equal to its high-redshift
value.
Then, the sum $s_1^2+s_2^2$ is no longer required to be almost constant at low $z$ and
this extends the family of realistic models to all solutions with common conformal time.
As the cutoff scale of the model is of order 1 a.u., the compliance with  Solar System tests
for the bimetric models would have to be analyzed thoroughly  once UV completions have been constructed. In particular, they  would have to avoid all the local
issues that we have detailed here. This is beyond the present work.

This paper only provides a first study of such bimetric models with self-acceleration.
We have shown that basic requirements already strongly constrain these scenarios.
We leave for future works a detailed study to determine whether such scenarios
can be consistent with cosmological data at the perturbative level.
However, the main challenge is to devise adequate screening mechanisms within appropriate UV completions, if they exist.
This would also have a great impact on other modified-gravity models, by providing an explicit scenario
where  gravity on cosmological scales could be decoupled from Solar System tests.
Finally, another issue concerns the  stability of  the hidden vector modes, as one would like to go beyond the linear regime and guarantee that they do not mix with the matter metrics. This is beyond the scope of the present work.

\appendix

\section{Linear perturbations in the general case $s_{\rm d\ell} \neq s_{\ell}$}
\label{sec:linear-general}

We provide in this appendix the Einstein equations for linear perturbations in the general
case, where we no longer assume $s_{\rm d\ell}$ and $s_{\ell}$ to be identical.
This allows us to go beyond the early-time Einstein-de Sitter phase
(\ref{EdS-0}). In particular, we no longer have $s_1^2+s_2^2=1$ nor
$a_{\ell} = s_{\ell} a$ and ${\cal H}_{\ell}={\cal H}$.
However, we restrict to the case $a_{\ell}=b_{\ell}$, to ensure that the graviton speeds
remain equal to the speed of light.

Because the two types of matter (baryons and dark matter) now follow different metrics,
the quadratic action can no longer be neatly split in a sector $S_+$, which contains all
matter variables and remains identical to General Relativity, and a sector $S_-$ that
is completely decoupled from matter and deviates from General Relativity (and can
include new degrees of freedom due to the loss of one diffeomorphism invariance).
Then, in this appendix we directly work at the level of the Einstein equations.
The vierbein and metric perturbations are again defined as in Eqs.(\ref{vierbeins-SVT})
and (\ref{dgmunu}).

\subsection{Tensor modes}
\label{sec:tensors-general}

For tensors, the Einstein equations (\ref{Einstein-e1}) give
\beqa
h''_{1ij} + 2 {\cal H}_1 h'_{1ij} - \nabla^2 h_{1ij}
& = & \frac{a_2 \sum_{*} s_{*1} s_{*2} a_*^2 \bar p_*}
{a_1 M_{\rm Pl}^2} \nonumber \\
&& \times (h_{1ij} - h_{2ij} )  ,
\label{graviton-1-general}
\eeqa
and a symmetric equation with respect to $1 \leftrightarrow 2$.
Here $*={\rm b, d}$ stands for the baryonic and dark matters, and we sum
over both matter sectors.
In the early-time regime (\ref{EdS-0}), these two equations can be diagonalized
as in (\ref{h-matter-evol})-(\ref{h-_evol}).
At high frequencies, $\omega \gg {\cal H}$, and high wave numbers, $k \gg {\cal H}$,
we recover the Minkowski limit of General Relativity, with two massless gravitons that propagate as in Minkowski
vacuum, $h''_{1ij} - \nabla^2 h_{1ij} = 0$.
This is not surprising, as the bimetric theory (\ref{S-def}) reduces to two copies of General
Relativity in vacuum.
In particular, we recover $2\times 2$ dynamical degrees of freedom.

\subsection{Vector modes}
\label{sec:vectors-general}

For vectors, the Einstein equations (\ref{Einstein-e1}) and the continuity
equations give
\beqa
\nabla^2 ( V'_{1i}+C_{1i} ) & = & \frac{a_2 \sum_* s_{*1} s_{*2} a_*^2 (3\bar\rho_*+\bar p_*)}
{2 a_1 M_{\rm Pl}^2} \nonumber \\
&& \times ( C_{1i} - C_{2i} ) ,
\label{vector-C-general}
\eeqa
and
\beqa
V''_{1i}+C'_{1i} + 2 {\cal H}_1 (V'_{1i}+C_{1i}) & = &
\frac{a_2 \sum_* s_{*1} s_{*2} a_*^2 \bar p_*}
{a_1 M_{\rm Pl}^2} \nonumber \\
&& \times ( V_{1i} - V_{2i} ) ,
\label{vector-V-general}
\eeqa
and the symmetric equations with respect to $1 \leftrightarrow 2$.
Again, the left-hand side corresponds to General Relativity and the right-hand side
is a new mass coupling term between the two gravitational metrics that is proportional
to the background matter content ($\bar\rho_*, \bar{p}_*$, hence to ${\cal H}^2$)
and to the products $s_{*1} s_{*2}$. It vanishes in vacuum or when one coupling $s_{*\ell}$ is zero.
In the early-time regime (\ref{EdS-0}), this system can be diagonalized as in
(\ref{S+_vector})-(\ref{S-_vector}).
By combining Eq.(\ref{vector-C-general}), multiplied by $a_1^2$, with its symmetric,
we obtain
\beq
a_1^2 ( V'_{1i}+C_{1i} ) + a_2^2 ( V'_{1i}+C_{1i} ) = 0 .
\eeq
This automatically implies that the same combination obtained from
Eq.(\ref{vector-V-general}) is also satisfied.
This ``loss'' of one equation is related to the diagonal vector gauge freedom.
Here $C_{+i}=a_1^2 C_{1i}+a_2^2 C_{2i}$, which generalizes Eq.(\ref{h+_def})
beyond the early-time regime.
Defining again $C_{-i}=C_{1i}-C_{2i}$ and $V_{-i}=V_{1i}-V_{2i}$,
we find that Eq.(\ref{C-_V'-}) generalizes to
\beq
C_{- i} = \frac{-2 a_1 a_2 M_{\rm Pl}^2 k^2 V'_{-i}}{2 a_1 a_2 M_{\rm Pl}^2 k^2 + (a_1^2+a_2^2) \sum_* s_{*1} s_{*2} a_*^2 (3\bar\rho_*+{\bar p}_*)} ,
\label{C-_V'-general}
\eeq
and at high frequencies and wave numbers we obtain the equation of motion
\beq
V''_{-i} =  \frac{\sum_* s_{*1} s_{*2} a_*^2 {\bar p}_*}
{\sum_* s_{*1} s_{*2} a_*^2 (3\bar\rho_*+{\bar p}_*)}  2 k^2 V_{-i} ,
\eeq
which generalizes Eq.(\ref{V-_vector-eq}).
In particular, we recover the same gradient instability (\ref{vector-gradient-radiation-era})
as in the Einstein-de Sitter phase, whatever the values of the coefficients $s_{*\ell}$.

In contrast with the case of tensors, the high frequency and high wave number limit
is not so straightforward and does not coincide with a naive Minkowski limit
where we put $\bar\rho_*=\bar{p}_*=0$ and ${\cal H}_{\ell}=0$
in Eqs.(\ref{vector-C-general})-(\ref{vector-V-general}). This is because the loss of the
nondiagonal gauge invariance leads to a new vector degree of freedom (here $V_{-i}$)
that cannot be ``forgotten'' and implies a different limit than the naive expectation
of two Minkowski copies.

\subsection{Scalar modes}
\label{sec:scalars-general}

For scalars, the Einstein equations (\ref{Einstein-e1}) give
\beqa
&& - 6 {\cal H}_1^2 \phi_1 - 6 {\cal H}_1 \psi'_1 + 2 \nabla^2\psi_1
- 4 {\cal H}_1 \nabla^2 V_1 + 2 {\cal H}_1 \nabla^2 U'_1 \nonumber \\
&& = \frac{\sum_* s_{*1} a_*^3 \delta\rho_*}{a_1 M_{\rm Pl}^2}
+ \frac{a_2 \sum_* s_{*1} s_{*2} a_*^2 \bar\rho_*}{a_1 M_{\rm Pl}^2}
[ 3 (\psi_1-\psi_2) \nonumber \\
&& - \nabla^2 (U_1-U_2) ] ,
\label{E00-general}
\eeqa
\beqa
&& - 4 {\cal H}_1 \phi_1 - 4 \psi'_1 + 8 ( {\cal H}'_1 - {\cal H}_1^2 ) V_1
= \frac{2 \sum_* s_{*1} a_*^3 (\bar\rho_* \!+\! \bar{p}_*) v_*}{a_1 M_{\rm Pl}^2}
\nonumber \\
&& - \frac{a_2 \sum_* s_{*1} s_{*2} a_*^2 (\bar\rho_*+3\bar{p}_*)}{a_1 M_{\rm Pl}^2}
(V_1-V_2) ,
\label{E0i-general}
\eeqa
\beqa
&& U''_1 - 2 V'_1 + 2 {\cal H}_1 ( U'_1 - 2 V_1) + \psi_1 - \phi_1 =
\nonumber \\
&& \frac{a_2 \sum_* s_{*1} s_{*2} a_*^2 \bar{p}_*}{a_1 M_{\rm Pl}^2} (U_1-U_2) ,
\label{Eij-general}
\eeqa
\beqa
&& 2 (2 {\cal H}'_1+{\cal H}^2_1) \phi_1 + 2 \psi''_1 + 4 {\cal H}_1 \psi'_1
+ 2 {\cal H}_1 \phi'_1
= \frac{\sum_* s_{*1} a_*^3 \delta p_*}{a_1 M_{\rm Pl}^2}
\nonumber \\
&& + \frac{a_2 \sum_* s_{*1} s_{*2} a_*^2 \bar{p}_*}{a_1 M_{\rm Pl}^2}
[ - (\phi_1-\phi_2) + 2 (\psi_1-\psi_2) ] ,
\label{Eii-general}
\eeqa
and the symmetric equations with respect to $1 \leftrightarrow 2$.
Again, the left-hand side and the matter source terms on the right-hand side are identical
to General Relativity. There are new mass coupling terms on the right-hand side
that are proportional to ($\bar\rho_*, \bar{p}_*$, i.e. ${\cal H}^2$) and $s_{*1} s_{*2}$.
Here we did not include the perturbations of the scalar field $\varphi$,
which corresponds to $\gamma^*_*=0$ in the quasistatic equations
(\ref{Poisson-psi-2})-(\ref{Poisson-phi-2}).

\subsection{Nonpropagation of the Goldstone mode}
\label{sec:non-propagation-Goldstone}

In the Einstein-de Sitter phase, where the quadratic action can be split over the two
sectors $S_+$ and $S_-$, we could see from the explicit action (\ref{S-_scalar})
or from the St\"uckelberg analysis in section~\ref{sec:quadratic-action}
that the scalar mode associated with the breaking of the nondiagonal diffeomorphism
does not propagate.
Here we provide an alternative check that such a mode cannot sustain decoupled
propagation at high frequencies and wave numbers, even beyond the
Einstein-de Sitter phase.

As in Eq.(\ref{pi-def-Stuckelberg}),
we introduce the St\"uckelberg scalar $\pi$ associated with the nondiagonal
diffeomorphism by writing
\beq
\phi_{\ell} = \hat\phi_{\ell} - {\cal H}_{\ell} \gamma_\ell \pi' - \gamma_{\ell} \pi'' ,
\;\;\;
\psi_{\ell} = \hat\psi_{\ell} + {\cal H}_{\ell} \gamma_{\ell} \pi' ,
\eeq
\beq
V_{\ell} = \hat{V}_{\ell} + \gamma_{\ell} \pi' ,
\;\;\;
U_{\ell} = \hat{U}_{\ell} + \gamma_{\ell} \pi ,
\eeq
where $\gamma_1 \neq \gamma_2$ are constant.
The case $\gamma_1=\gamma_2$ would be associated with the diagonal diffeomorphism.
Substituting into Eqs.(\ref{E00-general})-(\ref{Eii-general}) and only keeping the
$\pi$ terms, we obtain
\beqa
6 {\cal H}_1 ( {\cal H}'_1 - {\cal H}^2_1 ) \gamma_1 \pi' & =  &
\frac{a_2 \sum_* s_{*1} s_{*2} a_*^2 \bar\rho_*}{a_1 M_{\rm Pl}^2}
[ ( \gamma_1-\gamma_2) \nabla^2\pi   \nonumber \\
&& - 3 ({\cal H}_1 \gamma_1 - {\cal H}_2 \gamma_2) \pi' ] ,
\eeqa
\beq
4 ( {\cal H}_1^2 - {\cal H}_1' ) \gamma_1 \pi' = \frac{a_2 \sum_* s_{*1} s_{*2} a_*^2
(\bar\rho_*+3 \bar{p}_*)}{a_1 M_{\rm Pl}^2} ( \gamma_1-\gamma_2) \pi' ,
\eeq
\beq
0 = \frac{a_2 \sum_* s_{*1} s_{*2} a_*^2 \bar{p}_*}{a_1 M_{\rm Pl}^2}
( \gamma_1-\gamma_2) \pi ,
\eeq
\beqa
&& 2 ( {\cal H}_1'' - {\cal H}_1^3 - {\cal H}_1 {\cal H}_1' ) \gamma_1 \pi' =
\frac{a_2 \sum_* s_{*1} s_{*2} a_*^2 \bar{p}_*}{a_1 M_{\rm Pl}^2}
\nonumber \\
&& \times [ ( \gamma_1-\gamma_2) \pi''
+ 3 ({\cal H}_1 \gamma_1 - {\cal H}_2 \gamma_2) \pi' ] .
\label{Eii-pi-general}
\eeqa
They take the expected form involving
${\cal H}^2 \partial^2\pi = \frac{s_{*1}s_{*2} a_*^2 \bar\rho_*}{M_{\rm Pl}^2} \partial^2\pi$
(where some derivatives $\partial$ can be replaced by factors ${\cal H}$),
as these terms must disappear in the naive Minkowski limit
${\cal H} \to 0$ and $\bar\rho_* \to 0$.

In the limit of high frequencies and wave numbers, $\omega\gg {\cal H}$ and
$k \gg {\cal H}$, the last relation (\ref{Eii-pi-general}) gives $\pi '=0$ if
$\bar{p}_*=0$ or $\pi'' \sim {\cal H} \pi'$ if $a_*^2 \bar{p}_* \sim M_{\rm Pl}^2 {\cal H}^2$.
Therefore, the scalar $\pi$ cannot develop decoupled high frequency modes and does not propagate.

\section{Coupling of the Goldstone mode to the metrics}
\label{sec:Goldstone-metrics-coupling}

\subsection{General case}

In this appendix, we explore the role played by the Goldstone boson $\pi$ in the
modification of gravity.
More precisely, we derive the coupling $h\partial^2\pi$ that was neglected in
section~\ref{sec:Goldstone} and we check that it agrees with the explicit
expression (\ref{d2S-pi1}) in the early-time regime where $s_{\rm d\ell}=s_{\ell}$.
As in section~\ref{sec:Goldstone}, we go beyond this early-time regime and
we allow the baryonic and dark matter metrics to be different, but we focus
on short lengths and timescales as compared with the Hubble parameter,
using the approximation (\ref{appro}).

The St\"uckelberg fields $\phi^\mu_\ell$ are introduced as in Eq.(\ref{e-ell-phi}),
\be
e^a_{\ell\mu} =  \hat{e}^a_{\ell \nu} \frac{ \partial \phi^\nu_\ell}{\partial x^\mu} ,
\;\;\;
\hat{e}^a_{\ell \mu}  = a_{\ell} ( \delta^a_\mu + \hat h^a_{\ell\mu} ) ,
\label{e-ell-phi-appendix}
\ee
where $\hat{h}^a_{\ell\mu}$ parametrize the deviations from the FLRW background.
We again separate the diffeomorphisms into the diagonal ones, which are not broken
by the presence of matter, and the broken ones in the complementary directions which
belong to the group quotient $({\rm diff}_1\times {\rm diff}_2)/ {\rm diff}_{\rm diag}$. We
choose in the following the particular combination
\beq
\phi^\mu_\ell = x^{\mu} + \xi^{\mu} + \xi^\mu_\ell
\label{phi-xi-xi-ell}
\eeq
with
\be
\xi_1^\mu= \frac{1}{s_{1}s_{\rm d1}} \pi^{\mu} , \ \
\xi^\mu_2 = - \frac{1}{s_{2} s_{\rm d2}} \pi^{\mu} .
\label{xi-1-2-def}
\ee
This corresponds to the choice $\gamma_1=1/s_1 s_{\rm d1}$ and
$\gamma_2= -1/s_2 s_{\rm d2}$ in the main text (\ref{gamma1-gamma2-def}).
As in sections~\ref{sec:quadratic-action} and \ref{sec:Goldstone}, we focus on the
scalar mode that would be associated with a Boulware-Deser ghost, and we write
\beq
\pi^{\mu} = \partial^{\mu} \pi = \eta^{\mu\nu} \partial_{\nu} \pi .
\eeq
To simplify expressions, we always define $\partial^\mu\pi$ by the metric
$\eta^{\mu\nu}$ in the following.

The total action does not depend on the diagonal diffeomorphism $\xi^{\mu}$,
which we set to zero in the following.
We now derive the terms $\hat{h}\partial^2\pi$ that arise from the
Einstein-Hilbert Lagrangians, which we write as
\beq
L_{\rm EH}(e^a_{\ell\mu}) = \sqrt{-g_{\ell}} R(g_{\ell\mu\nu}) .
\eeq
Because of the invariance of the Ricci scalar under change of coordinates,
we have from Eq.(\ref{e-ell-phi-appendix})
\beq
L_{\rm EH}(e^a_{\ell\mu}) =  \det(\partial_{\mu}\phi^{\nu}_{\ell})
L_{\rm EH}(\hat{e}^a_{\ell \nu}) .
\label{L-EH-def}
\eeq
To obtain the terms $\hat{h}\partial^2\pi$, we only need to work at linear order over
$\hat{h}$ and $\pi$ separately.
At linear order over $\pi$, we have from Eq.(\ref{phi-xi-xi-ell})
\be
\det(\partial_\mu \phi_\ell^\nu)  = 1+\partial_\sigma \xi^{\sigma}_\ell .
\ee
On the other hand, at linear order over $\hat{h}$ we have
\beqa
L_{\rm EH}(\hat{e}^{a}_{\ell\mu}) & = & L_{\rm EH}(\bar{e}^{a}_{\ell\mu})
- \sqrt{-\bar{g}_\ell} \bar{G}^{\mu\nu}_\ell \delta\hat{g}_{\ell\mu\nu} \nonumber \\
& = & \bar{L}_{\rm EH} - \sqrt{-\bar{g}_\ell} \bar{G}^{\mu\mu}_\ell \eta_{\mu\mu}
2 a_{\ell}^2 \hat{h}^{\mu}_{\ell\mu} .
\eeqa
In the second line we used the fact that the background Einstein tensors are diagonal
and we sum over $\mu$.
Substituting into Eq.(\ref{L-EH-def}), we find that the term $\hat{h}\partial^2\pi$
that arises from the Einstein-Hilbert Lagrangians is
\beq
L_{\rm EH}(e^{a}_{\ell\mu}) \supset  - \sqrt{-\bar{g}_\ell} \bar G^{\mu\mu}_\ell
\eta_{\mu\mu} 2 a_\ell^2 \hat{h}^{\mu}_{\ell\mu} \partial_\sigma \xi^{\sigma}_\ell .
\label{coupling-h-pi-1}
\eeq
Along the diagonal, the background Einstein equations (\ref{Einstein-e1}) read
(no summation over $\mu$)
\be
M_{\rm Pl}^2 \sqrt{-{\bar g}_\ell} \bar G^{\mu\mu}_\ell a_{\ell} =
s_{\ell} \sqrt{-{\bar g}} \bar T^{\mu\mu} a + s_{\rm d\ell} \sqrt{-\bar{g}_{\rm d}}
\bar T_{\rm d}^{\mu\mu} a_{\rm d} .
\ee
Then, using Eq.(\ref{xi-1-2-def}), the term $\hat{h}\partial^2\pi$ arising from
the two Einstein-Hilbert actions is
\beqa
&& \frac{M_{\rm Pl}^2}{2} [ L_{\rm EH}(e^{a}_{1\mu})
+ L_{\rm EH}(e^{a}_{2\mu}) ] \supset
- \sqrt{- \bar g}  \bar T^{\mu\mu} \eta_{\mu\mu} a \nonumber \\
&& \times \left( \frac{a_1}{s_{\rm d1}} \hat{h}^{\mu}_{1\mu} - \frac{a_2}{s_{\rm d2}}
\hat{h}^{\mu}_{2\mu} \right) ( \partial_\sigma \partial^\sigma \pi )
- \sqrt{- \bar g_{\rm d}}  \bar T_{\rm d}^{\mu\mu} \eta_{\mu\mu} a_{\rm d} \nonumber \\
&& \times \left( \frac{a_1}{s_{1}} \hat{h}^{\mu}_{1\mu} - \frac{a_2}{s_{2}}
\hat{h}^{\mu}_{2\mu} \right) ( \partial_\sigma \partial^\sigma \pi ) .
\label{L-EH1-L-EH2}
\eeqa

We now turn to the matter actions. The matter vierbeins
are given by
\be
e_{\star \mu}^a = s_{\star1}  e^{a}_{1\mu} + s_{\star2}  e^{a}_{2\mu} ,
\label{e-star-def}
\ee
where $\star$ stands for $\rm b$ (baryons) or $\rm d$ (dark matter).
They can be written as
\beq
e_{\star \mu}^a = \hat{e}^a_{\star\nu} \frac{\partial\phi^\nu_\star}{\partial x^{\mu}}
+ \delta\tilde{e}^a_{\star\mu} ,
\label{e-star-a-mu-h-xi}
\eeq
where we introduced
\beq
\hat{e}^a_{\star \mu}  = a_{\star} ( \delta^a_\mu + \hat h^a_{\star\mu} ) ,
\eeq
\be
a_\star \hat{h}_{\star \mu}^a =  s_{\star 1} a_1 \hat{h}^{a}_{1\mu} +
s_{\star 2} a_2 \hat{h}^{a}_{2 \mu} ,
\ee
\beq
\phi^\mu_\star = x^{\mu} + \xi^\mu_\star , \;\;\;
a_\star \xi_\star^\mu =  s_{\star 1} a_1 \xi^\mu_1 + s_{\star 2} a_2 \xi^{\mu}_2 ,
\eeq
and
\beq
\delta\tilde{e}^a_{\star\mu} = - a_\star \hat{h}^a_{\star\nu} \partial_\mu\xi^{\nu}_\star + s_{\star 1} a_1 \hat{h}^a_{1\nu}
\partial_{\mu}\xi^{\nu}_1 + s_{\star 2} a_2 \hat{h}^a_{2\nu} \partial_{\mu}\xi^{\nu}_2 .
\eeq
As compared with Eq.(\ref{e-ell-phi-appendix}), there is an additional term
$\delta\tilde{e}^a_{\star\mu}$
of the form $\hat{h}\partial\xi$ because the matter vierbeins are defined by the
composite expression (\ref{e-star-def}).
Defining the matter Lagrangians as
\beq
L_\star(e_{\star \mu}^a) = \sqrt{-g_\star} {\cal L}_{\star} (g_{\star\mu\nu}) ,
\eeq
we now have
\beq
L_\star(e_{\star \mu}^a) =  L_\star( \hat{e}_{\star \nu}^a  \partial_{\mu}\phi_\star^\nu )
+ \frac{\sqrt{-\bar{g}_\star}}{2} \bar{T}^{\mu\nu}_\star \delta\tilde g_{\star\mu\nu} ,
\eeq
where $\delta\tilde g_{\star\mu\nu}$ is the metric perturbation associated with
$\delta\tilde{e}^a_{\star\mu}$ in Eq.(\ref{e-star-a-mu-h-xi}).
On the other hand, as for the Einstein-Hilbert terms (\ref{coupling-h-pi-1}),
the term $L_\star( \hat{e}_{\star \nu}^a  \partial_{\mu}\phi_\star^\nu )$ gives rise
to the factor
\beq
L_\star(  \hat{e}_{\star \nu}^a  \partial_{\mu}\phi_\star^\nu ) \supset
\sqrt{-\bar{g}_\star} \bar T^{\mu\mu}_\star
\eta_{\mu\mu} a_\star^2 \hat{h}^{\mu}_{\star\mu} \partial_\sigma \xi^{\sigma}_\star .
\label{coupling-h-pi-T-1}
\eeq
Collecting all terms, this gives
\beqa
L_\star(e_{\star \mu}^a) &  \supset & \sqrt{-\bar{g}_\star} \bar T^{\mu\mu}_\star
\eta_{\mu\mu} a_\star [ a_\star \hat{h}^{\mu}_{\star\mu} \partial_\sigma \xi^{\sigma}_\star
-  a_\star \hat{h}^{\mu}_{\star\sigma} \partial_\mu \xi^{\sigma}_\star
\nonumber \\
&& + s_{\star1} a_1 \hat{h}^{\mu}_{1\sigma} \partial_\mu \xi^{\sigma}_1
+ s_{\star2} a_2 \hat{h}^{\mu}_{2\sigma} \partial_\mu \xi^{\sigma}_2 ] .
\eeqa
Using Eq.(\ref{xi-1-2-def}), this yields for the baryonic matter Lagrangian
\beqa
&& L(e^a_{\mu})  \supset  \sqrt{-\bar{g}} \bar T^{\mu\mu}
\eta_{\mu\mu} \left[ \left( \frac{a_1}{s_{\rm d1}} - \frac{a_2}{s_{\rm d2}} \right)
( s_1 a_1 \hat{h}^{\mu}_{1\mu} + s_2 a_2 \hat{h}^{\mu}_{2\mu} ) \right.
\nonumber \\
&& \left. \times ( \partial_\sigma \partial^{\sigma} \pi)
+ \frac{a_1 a_2 (s_1 s_{\rm d1}+s_2 s_{\rm d2})}{s_{\rm d1} s_{\rm d2}}
( \hat{h}^{\mu}_{1\sigma} - \hat{h}^{\mu}_{2\sigma} ) ( \partial_\mu \partial^{\sigma} \pi )
\right] , \nonumber \\
&&
\label{L-baryons-h-pi}
\eeqa
and for the dark matter Lagrangian
\beqa
&& L_{\rm d}(e^a_{\rm d\mu})  \supset  \sqrt{-\bar{g}_{\rm d}} \bar T^{\mu\mu}_{\rm d}
\eta_{\mu\mu} \left[ \left( \frac{a_1}{s_{1}} - \frac{a_2}{s_{2}} \right)
( s_{\rm d1} a_1 \hat{h}^{\mu}_{1\mu} + s_{\rm d2} a_2 \hat{h}^{\mu}_{2\mu} ) \right.
\nonumber \\
&& \left. \times ( \partial_\sigma \partial^{\sigma} \pi)
+ \frac{a_1 a_2 (s_1 s_{\rm d1}+s_2 s_{\rm d2})}{s_{1} s_{2}}
( \hat{h}^{\mu}_{1\sigma} - \hat{h}^{\mu}_{2\sigma} )  ( \partial_\mu \partial^{\sigma} \pi )
\right] . \nonumber \\
&&
\label{L-dark-h-pi}
\eeqa

Collecting (\ref{L-EH1-L-EH2}), (\ref{L-baryons-h-pi}) and (\ref{L-dark-h-pi}),
we find that the terms $\hat{h}\partial^2\pi$ that arise in the total action are
\beqa
L_{\rm EH+matter} & \supset & \left[ \alpha \sqrt{-\bar{g}} \bar{T}^{\mu\mu} a^2
+ \alpha_{\rm d} \sqrt{-\bar{g}_{\rm d}} \bar{T}_{\rm d}^{\mu\mu} a_{\rm d}^2 \right]
\eta_{\mu\mu} \nonumber \\
&& \times ( \hat{h}^{\mu}_{-\sigma} \partial_\mu\partial^\sigma\pi
- \hat{h}^{\mu}_{-\mu} \partial_\sigma \partial^\sigma \pi)
\label{LEH-Lstar-general}
\eeqa
where we have
\beqa
&& \alpha= \frac{a_1 a_2}{a^2} \frac{s_1 s_{\rm d1}+s_2s_{\rm d2}}
{s_{\rm d1}s_{\rm d2}} , \nonumber \\
&& \alpha_{\rm d}= \frac{a_1 a_2}{a_{\rm d}^2} \frac{s_1 s_{\rm d1}+s_2s_{\rm d2}}
{s_{1}s_{2}} ,
\eeqa
and we introduced the metric combination
\be
\hat{h}^{a}_{-\mu} = \hat{h}^{a}_{1\mu} - \hat{h}^{a}_{2\mu} ,
\label{h-_a_mu_def}
\ee
which agrees with Eq.(\ref{h-_def}).

Thus, we find that in all cases, even when the baryonic and dark matter couplings
$s_{\star\ell}$ are different, the St\"uckelberg field $\pi$ only couples to the
same metric combination $\hat{h}_-$.
The $\hat{h}_-\partial^2\pi$ terms in the last set of parentheses in Eq.(\ref{LEH-Lstar-general})
are the same as in Eqs.(\ref{LEH-Lstar-equall}) and (\ref{LEH-Lstar-pi-equall}) below,
and they coincide with the result (\ref{d2S-pi1}) in the main text, where
we only keep the dominant terms with $\partial \gg {\cal H}$.
In particular, by integrating by parts
the terms in  $\pi''$, we can again check that this contribution to the action
can be written in terms of first-order time derivatives only.
Therefore, it does not give rise to Boulware-Deser ghosts.

In the case where the couplings $s_{\rm d\ell}$ and $s_\ell$ are identical,
we can separate the quadratic action in two  sectors $S_+$ and $S_-$,
as explicitly shown in section~\ref{sec:freedom-linear}.
When the baryonic and dark matter metrics are different, we cannot simultaneously
decouple both matter metrics from $h_-$, as we only have two fundamental
metrics $h_1$ and $h_2$, so that $h_{\rm d}$ must be a combination of
$h$ and $h_-$.
This may give rise to a modification of gravity on Hubble scales, although this is the regime where the
derivation presented in this appendix is no longer valid. On small scales,
we have seen in appendix~\ref{sec:non-propagation-Goldstone} that
$\pi$ does not propagate and does not generate a modification of gravity. In the main text, we have described the
 modification of gravity that is seen by the large-scale structures, which  is entirely due
to the fluctuations of the scalar field $\varphi$  whose effect is to generate  a fifth force
as described in section~\ref{sec:linear}.

\subsection{Identical couplings}

In the early-time regime, where $s_{\rm d\ell}=s_\ell$ and $a_{\ell}=s_\ell a$,
we have $\alpha=\alpha_{\rm d} =1$, and Eq.(\ref{LEH-Lstar-general})
simplifies as
\beqa
L_{\rm EH+matter} & \supset & \sqrt{-\bar{g}} ( \bar{T}^{\mu\mu}
+ \bar{T}_{\rm d}^{\mu\mu} ) a^2 \eta_{\mu\mu} \nonumber \\
&& \times ( \hat{h}^{\mu}_{-\sigma} \partial_\mu\partial^\sigma\pi
- \hat{h}^{\mu}_{-\mu} \partial_\sigma \partial^\sigma \pi) . \;\;\;
\label{LEH-Lstar-equall}
\eeqa
Thus, we recover the result of sections~\ref{sec:Quadratic-action} and
\ref{sec:quadratic-action}, obtained from the explicit derivation of the quadratic action,
that the Goldstone boson does not couple to matter at the quadratic order in the
Lagrangian and at the linear level in the equations of motion.
It belongs to the sector $S_-$ of the action and it is only coupled to the graviton
$\hat{h}_-$, which is also decoupled from matter.
Explicitly, this metric coupling reads
\beqa
&& L_{\hat{h}\partial^2\pi} = - a^4 \bar\rho_T ( \hat{h}^{0}_{-i} \partial_i \pi' -
\hat{h}^{0}_{-0} \nabla^2 \pi ) + a^4 \bar{p}_T
\nonumber \\
&& \hspace{0.5cm}  \times ( - \hat{h}^i_{-0} \partial_i \pi' + \hat{h}^{i}_{-j}
\partial_i\partial_j \pi + \hat{h}^{i}_{-i} ( \pi'' - \nabla^2 \pi) ) , \hspace{1cm}
\eeqa
where $\bar{p}_T$ and $\bar{\rho}_T$ are the total pressure and energy densities.
Now using $\hat{h}^{0}_{-0}=\hat{\phi}_-$, $\hat{h}^{0}_{-i}=-\partial_i \hat{V}_-$,
$\hat{h}^{i}_{-0} = \partial_i \hat{V}_-$,
$\hat{h}^{i}_{-j} = -\hat\psi_- \delta^i_j + \partial_i \partial_j \hat{U}_-$,
and $a^2\bar\rho_T= 3 M_{\rm Pl}^2 {\cal H}^2$,
$a^2 \bar{p}_T = - M^2_{\rm Pl} ({\cal H}^2 + 2 {\cal H}')$,
this gives
\beqa
&&\frac{L_{\hat{h}\partial^2\pi}}{a^2M^2_{\rm Pl}}= 3 {\cal H}^2 \hat\phi_- \nabla^2 \pi
+ 2 (2{\cal H}^2 + {\cal H}') (\nabla \hat{V}_-) \cdot (\nabla\pi') \;\;\; \nonumber \\
&&   + ({\cal H}^2+2{\cal H}') [3 \hat\psi_- \pi'' - 2 \hat\psi_- \nabla^2 \pi - (\nabla^2
\hat{U}_-)\pi'' ] .  \;\;\;
\label{LEH-Lstar-pi-equall}
\eeqa
This coincides with the result (\ref{d2S-pi1}) in the main text, when the subdominant terms
have been dropped.
In particular, integrating by parts the terms in  $\pi''$, we recover the fact that the quadratic
action can be written in terms of first-order time derivatives only, without third-
and fourth-order time derivatives left.

\bibliography{ref1}   

\end{document}